\documentclass[aps,pre,preprint,nofootinbib,superscriptaddress]{revtex4-1}

\usepackage{amsmath}
\usepackage{fullpage}
\usepackage{amssymb}
\usepackage{amsthm}
\usepackage{hyperref}
\usepackage{setspace}
\usepackage{graphicx}
\usepackage{bm}
\usepackage{physics}
\usepackage{array}
\usepackage{tabu}
\usepackage{xcolor}

\newcommand\Bstrut{\rule[-2ex]{0pt}{0pt}} 

\usepackage{lineno}

\begin{document}

 \title{Analytic solution of chemical master equations involving gene switching. I: Representation theory and diagrammatic approach to exact solution}

\author{John J. Vastola}
\email[john.vastola.phd@gmail.com]{}
\affiliation{Department of Physics and Astronomy, Vanderbilt University, Nashville, TN, USA}
\affiliation{Quantitative Systems Biology Center, Vanderbilt University, Nashville, TN, USA}

\author{Gennady Gorin}
\affiliation{Division of Chemistry and Chemical Engineering, California Institute of Technology, Pasadena, CA, USA}

\author{Lior Pachter}
\affiliation{Division of Biology and Biological Engineering \& Department of Computing and Mathematical Sciences, California Institute of Technology,
Pasadena, CA, USA}

\author{William R. Holmes}

\affiliation{Department of Physics and Astronomy, Vanderbilt University, Nashville, TN, USA}
\affiliation{Quantitative Systems Biology Center, Vanderbilt University, Nashville, TN, USA}
\affiliation{Department of Mathematics, Vanderbilt University, Nashville, TN, USA}

\date{\today}

\begin{abstract}
The chemical master equation (CME), which describes the discrete and stochastic molecule number dynamics associated with biological processes like transcription, is difficult to solve analytically. It is particularly hard to solve for models involving bursting/gene switching, a biological feature that tends to produce heavy-tailed single cell RNA counts distributions. In this paper, we present a novel method for computing exact and analytic solutions to the CME in such cases, and use these results to explore approximate solutions valid in different parameter regimes, and to compute observables of interest. Our method leverages tools inspired by quantum mechanics, including ladder operators and Feynman-like diagrams, and establishes close formal parallels between the dynamics of bursty transcription, and the dynamics of bosons interacting with a single fermion. We focus on two problems: (i) the chemical birth-death process coupled to a switching gene/the telegraph model, and (ii) a model of transcription and multistep splicing involving a switching gene and an arbitrary number of downstream splicing steps. We work out many special cases, and exhaustively explore the special functionology associated with these problems. This is Part I in a two-part series of papers; in Part II, we explore an alternative solution approach that is more useful for numerically solving these problems, and apply it to parameter inference on simulated RNA counts data.
\end{abstract}

\maketitle

\tableofcontents


\clearpage

\section{Introduction}

This is the first in a series of two papers on solving chemical master equations involving a switching gene, particularly in the context of bursty models of transcription and splicing. This paper describes a novel theoretical approach for solving these problems analytically, and explores the mathematical properties of the solutions we obtain in detail; the second paper \cite{vastolaP2} describes an alternative approach that is more appropriate for efficiently solving these problems on a computer. In other words, this paper tells us how to think about the solutions to these models, while the next paper offers a fast and stable way to work with them in practice---especially in the context of parameter inference.

The chemical master equation (CME) describes how the number of molecules in a system (e.g. RNA in a single cell) changes over time \cite{mcquarrie1967,gillespie1992,gillespie2000,gillespie2007,gillespie2013,fox2017,munsky2018,erban2020,vastolaCLE}. It treats molecule numbers as discrete rather than continuous, and dynamics as stochastic rather than deterministic; especially when there are small numbers of molecules, fluctuations due to the randomness associated with diffusion and molecular interactions become important. 

The equation itself---which can either be described as a differential-difference equation, or as a countably infinite number of coupled ordinary differential equations---is typically challenging to solve both analytically and numerically. Monomolecular reaction networks, treated by Jahnke and Huisinga in a landmark 2007 paper \cite{jahnke2007, vastolaDP}, comprise the largest class of systems for which analytic solutions to the CME are known. But this class does not include systems with many biologically common types of chemical reactions, like protein synthesis and binding, whose presence in a system's reaction list yields CMEs that can at present be solved only in very special cases \cite{schnoerr2017} (see e.g. \cite{mcquarrie1967, laurenzi2000, arslan2008, bokes2012, pendar2013, vastolaDP, beentjes2020, cao2020, cao2020detailed}).

One biological feature that yields CMEs that are hard to solve analytically is \textit{bursting} \cite{robinson2007, robinson2008, di2011, jia2011, oberg2012, kumar_my_2014, kumar2015, amrhein2019, gorin2020}. Bursting refers to the fact that RNA transcription does not seem to happen at a continuous rate in many contexts, but instead in discrete bursts. This is usually modeled by assuming that genes have at least two states that they randomly switch between, and transcribe at different rates when in different states. 

Recent improvements in fluorescence and sequencing-based experimental methods have enabled the identification of sub-molecular features of individual RNA \cite{shah_dynamics_2018, drexler_splicing_2020} in single cells. Specifically, one can detect individual introns, which provides the ability to interrogate the dynamics of RNA splicing---long implicated in the physiological control of gene expression \cite{lehir2003, stamm2005, long2008, wang2014, lim2020}. If the kinetics of these systems are to be understood in terms of a mechanistic model, one must be able to solve CMEs involving both splicing and gene switching, since bursting is known to be one of the salient features. 

In this paper, we present a new and fairly general approach to solving CMEs involving gene switching, and first illustrate our approach by exactly solving a telegraph model-like problem that describes mRNA production and degradation given a gene that randomly switches between two states. This is a particularly useful benchmark, as the problem has been solved before \cite{peccoud1995, raj2006, iyer2009, huang2015, cao2018}. We then use our method to tackle a much more complicated problem: deriving the full steady state distribution of a model of bursty transcription and splicing, assuming a linear path graph and arbitrarily many splicing steps. This path graph model is relatively simple, but plausible and directly applicable to modeling genes that have few introns and a single significant isoform. We validate our work using standard numerical approaches (exact stochastic simulations \cite{gillespie1976, gillespie1977, gillespie2007} and finite state projection \cite{munsky_finite_2006,fox2019}).

The paper is organized as follows. In Sec. \ref{sec:mainresults}, we present our main results and outline the theoretical approach described in later sections. This section is intended to allow readers to appreciate the utility of this new method, and to use the results we have derived with it, without having to enmesh themselves in its technical details. In Sec. \ref{sec:reptheory}, we describe a ladder operator-based approach to solving the chemical birth-death process and pure gene switching, and derive the ladder operators that allow us to effectively reframe the coupled problem. In Sec. \ref{sec:diagrammatic}, we construct a diagrammatic series solution to the coupled problem, and explain how to draw the relevant Feynman-like diagrams. In Sec. \ref{sec:special}, we examine special cases of our series solution, and show how to extract them from the more general result. In Sec. \ref{sec:continuous}, we study the high copy number/continuous concentration limit of our solution, and show that it corresponds to solving a certain Fokker-Planck equation coupled to a switching gene. In Sec. \ref{sec:multistep}, we generalize our theoretical approach to tackle the multistep splicing problem. In Sec. \ref{sec:numerical}, we consider the problem of implementing our solutions  on a computer, and validate them against numerical approaches. Finally, we discuss how our approach could be fruitfully extended and applied to study more general splicing dynamics in Sec. \ref{sec:discussion}.

\section{Main results and outline of approach}
\label{sec:mainresults}

In this section, we summarize our main analytic results. Readers only interested in our results, and not in methodological details, may find this section the most useful. 

\begin{figure}[!htb]
\center{\includegraphics[scale=0.85]{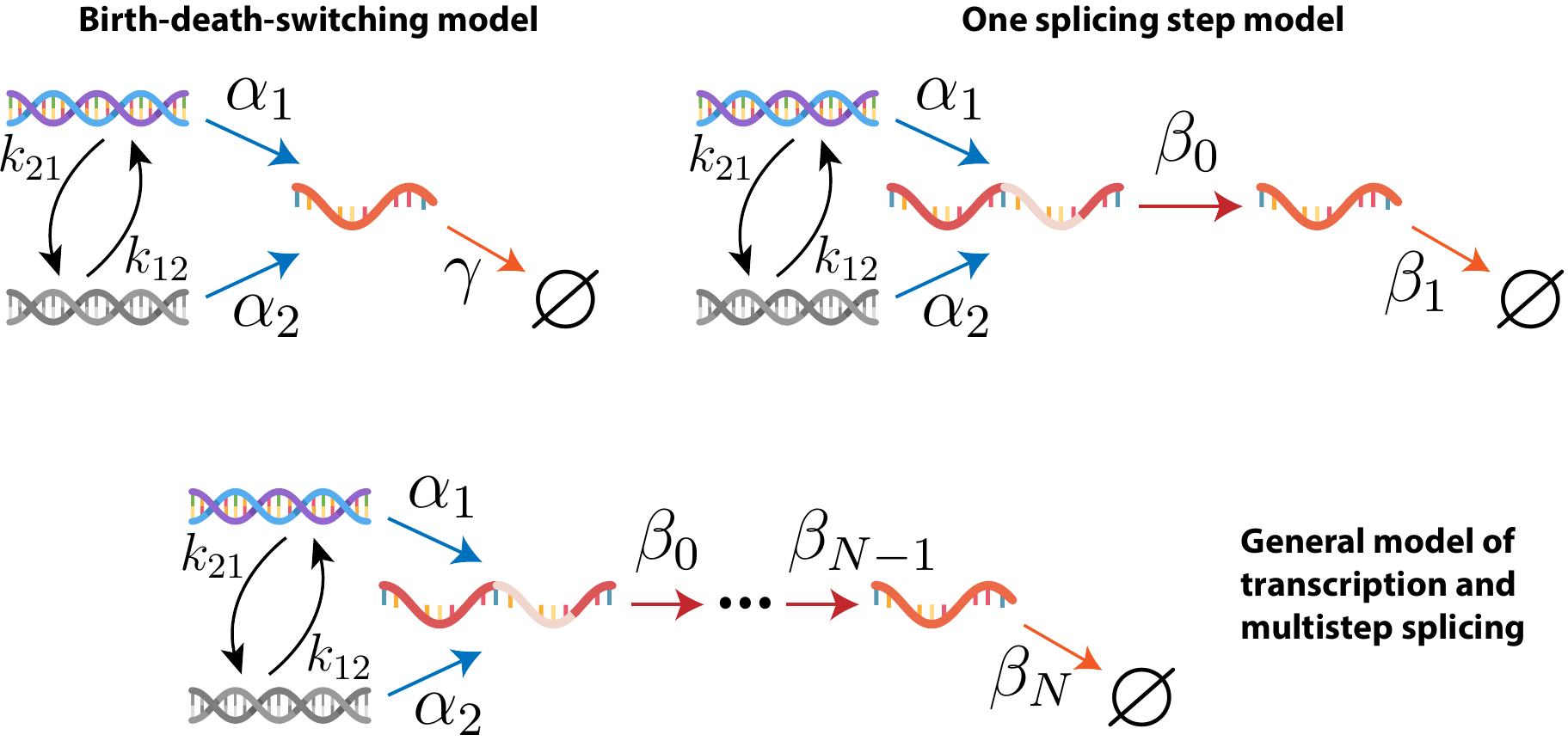}}
\caption{\label{fig:allmodels_cartoon} 
The models we will examine in this paper. The 1 species version is the birth-death-switching model (usually called the telegraph model elsewhere). We generalize it to include some arbitrary number $N$ of downstream splicing steps, so that the most general model has $N+1$ distinct species of RNA overall.}
\end{figure}

We have analytically solved two problems: (i) a telegraph model-like problem involving one RNA species and a switching gene, and (ii) a generalization of it intended to model the dynamics of bursty transcription with multiple downstream splicing steps (Fig \ref{fig:allmodels_cartoon}). The problem statements and main results for each are presented here in their own subsections (with additional, more specialized results in Sec. \ref{sec:diagrammatic} - \ref{sec:multistep}). The third subsection provides a brief guide to the three kinds of qualitative solution behavior one can expect. The last subsection provides a roadmap for the new theoretical methodology explored in the rest of the paper.



\vspace{1in}

\clearpage

\subsection{Chemical birth-death process coupled to a switching gene} \label{sec:results_bds}


\subsubsection{Problem statement}

Let $X$ denote the RNA species, and $G_1$ and $G_2$ the two possible gene states. Using the formalism of the chemical master equation (CME), RNA production and degradation coupled to a switching gene can be modeled using the list of chemical reactions
\begin{equation} \label{eq:bds_rxns}
\begin{split}
G_1 &\xrightarrow{k_{21}} G_2 \hspace{0.8in} G_2 \xrightarrow{k_{12}} G_1 \\
G_1 &\xrightarrow{\alpha_1} G_1 + X \hspace{0.49in} G_2 \xrightarrow{\alpha_2} G_2 + X \\
X &\xrightarrow{\gamma} \varnothing 
\end{split}
\end{equation}
where $k_{21}$ and $k_{12}$ parameterize the rates of gene switching, $\alpha_1$ and $\alpha_2$ parameterize the transcription rate in each gene state, and $\gamma$ parameterizes the RNA's degradation rate. This model is often called the `telegraph model', especially when one of the production rates is zero. For reasons that will become clear in the following sections, we prefer to think of it as two distinct problems coupled together: the chemical birth-death process (describing RNA production and degradation given a constitutively active gene) coupled to a switching gene.

The stochastic dynamics of this system are completely characterized by the probability distribution $P(x, S, t)$, which indicates the probability that the system has $x \in \mathbb{N} := \{ 0, 1, 2, ... \}$ RNA molecules and is in gene state $S \in \{ 1, 2 \}$ at some time $t$ (given some specified initial condition $P(x, S, 0)$ which has no effect on steady state information). The probability distributions $P(x, 1, t)$ and $P(x, 2, t)$ are determined by the CME 
\begin{equation} \label{eq:CME_bds}
\begin{split}
\frac{\partial P(x, 1, t)}{\partial t} =& \ - k_{21} P(x, 1, t) + k_{12} P(x, 2, t) \\ 
& + \alpha_1 \left[ P(x - 1, 1, t) - P(x, 1, t) \right] + \gamma \left[ (x + 1) P(x + 1, 1, t) - x P(x, 1, t) \right] \\
\frac{\partial P(x, 2, t)}{\partial t} =& \ k_{21} P(x, 1, t) - k_{12} P(x, 2, t) \\
&+ \alpha_2 \left[ P(x - 1, 2, t) - P(x, 2, t) \right] + \gamma \left[ (x + 1) P(x + 1, 2, t) - x P(x, 2, t) \right] 
\end{split}
\end{equation}
which is known to be difficult to solve analytically. Heuristically, the difficulty comes from coupling two qualitatively very different kinds of stochastic processes: a discrete switching process on two states, and a biased random walk on a countably infinite molecular number space.

We will not compute $P(x, 1, t)$ or $P(x, 2, t)$ in full generality, but rather the functions
\begin{equation}
\begin{split}
P_{ss}(x) &:= \lim_{t \to \infty} P(x, 1, t) + P(x, 2, t) \\
\psi_{ss}(g) &:= \sum_{x = 0}^{\infty} P_{ss}(x) g^x
\end{split}
\end{equation}
i.e. the steady state probability distribution and steady state (analytic) generating function for a complex variable $g$, both marginalized over gene state. Several distinct objects labeled `generating functions' will appear throughout this paper, so we distinguish this one with the adjective `analytic' to avoid confusion. 


\subsubsection{Steady state results} \label{sec:main_bds_res}

\begin{figure}[!htb]
\center{\includegraphics[scale=1]{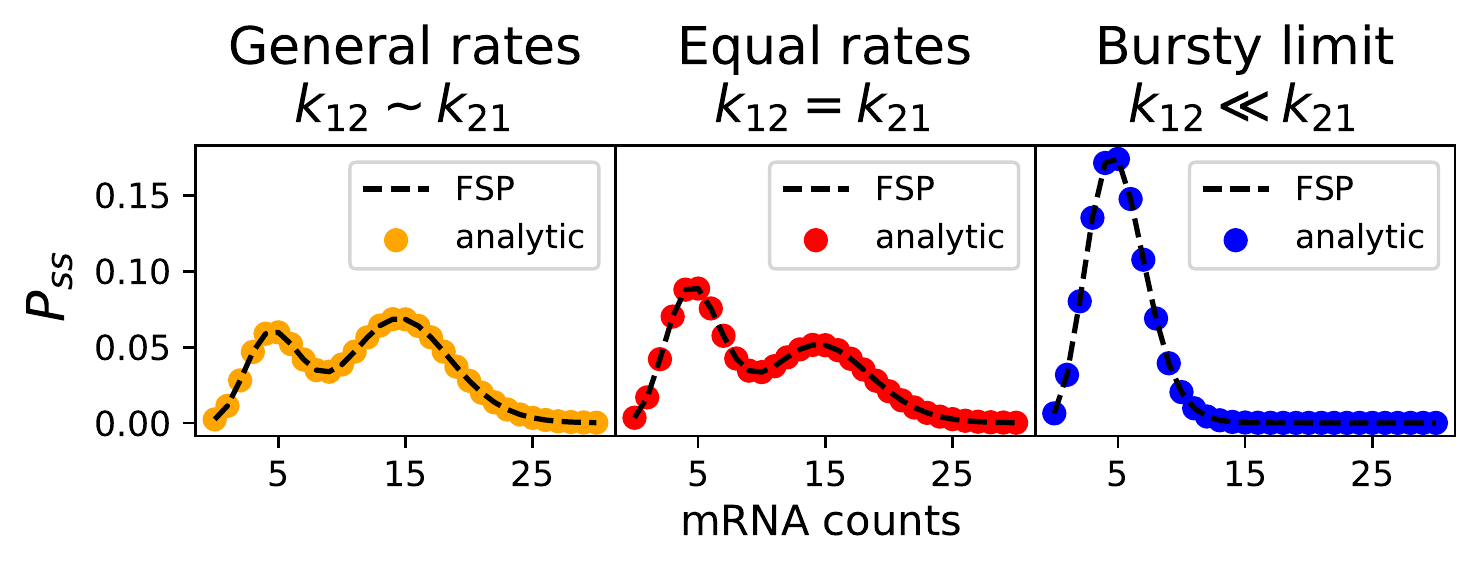}}
\caption{\label{fig:bds_threepanel_valid} 
The birth-death-switching steady state distribution $P_{ss}(x)$ in three representative cases: general rates ($k_{12} \sim k_{21}$), equal rates ($k_{12} = k_{21}$), and very unequal rates ($k_{12} \ll k_{21}$). The black dotted lines correspond to the finite state projection (FSP) result, while the colored dots correspond to the result from our solution approach.}
\label{fig:bds_threepanel}
\end{figure}

\noindent Define the derived quantities
\begin{equation}
\begin{split}
\alpha_{eff} &:= \alpha_1 \left( \frac{k_{12}}{s} \right) + \alpha_2 \left( \frac{k_{21}}{s} \right) \\
\mu &:= \frac{\alpha_1 \left( \frac{k_{12}}{s} \right) + \alpha_2 \left( \frac{k_{21}}{s} \right)}{\gamma} = \frac{\alpha_{eff}}{\gamma} \\
\Delta \alpha &:= \alpha_1 - \alpha_2 \\
c_3 &:=  \frac{k_{12} k_{21}}{s^2} \\ 
c_4 &:= \frac{k_{21} - k_{12}}{s}  \ ,
\end{split}
\end{equation}
where $\alpha_{eff}$ is the effective transcription rate (which takes into account the average amount of time spent in each gene state in the long time limit), $\mu$ is the effective mean molecule number, $\Delta \alpha$ is the production rate difference, and $c_3$ and $c_4$ are non-dimensional coefficients that appear in our solution. Also define the $2 \times 2$ transfer matrix $T$ (c.f. Sec. \ref{sec:transfer}) as
\begin{equation}
T = \begin{pmatrix} T_{00} & T_{01} \\ T_{10}  & T_{11} \end{pmatrix} = \begin{pmatrix} 0 & 1 \\ c_3  & c_4 \end{pmatrix} = \begin{pmatrix} 0 & 1 \\ \frac{k_{12} k_{21}}{s^2}  & \frac{k_{21} - k_{12}}{s} \end{pmatrix} \ .
\end{equation}
Using $C_k(x,\mu)$ to denote the $k$th Charlier polynomial (see Appendix \ref{sec:app_charlier} for the definition and properties of these objects), the exact analytic results for $P_{ss}(x)$ and $\psi_{ss}(g)$ can be written
\begin{equation} \label{eq:bds_sln}
\begin{split}
\frac{P_{ss}(x)}{\text{Poiss}(x, \mu)} &=  1 + \sum_{k = 2}^{\infty} (\Delta \alpha)^k  \ C_k(x, \mu) \sum_{i_1, ..., i_{k-1} = 0, 1} \ \frac{T_{0 i_{k-1}}}{(k \gamma)} \frac{T_{i_{k-1} i_{k-2}}}{[ (k-1)\gamma + i_{k-1} s ]}   \cdots \frac{T_{i_1 0}}{(\gamma + i_1 s)}  \\
&= 1 + (\Delta \alpha)^2 C_2(x, \mu) \cdot  \frac{c_3}{(s + \gamma) 2 \gamma} + (\Delta \alpha)^3 C_3(x, \mu) \cdot \frac{c_3 c_4}{(s + \gamma) (s + 2\gamma) 3 \gamma} + \cdots 
\end{split}
\end{equation}
\begin{equation} \label{eq:1sp_genfunc}
\begin{split}
\frac{\psi_{ss}(g)}{e^{\mu (g-1)}} &=  1 + \sum_{k = 2}^{\infty} \left[ (\Delta \alpha) (g-1) \right]^k  \  \sum_{i_1, ..., i_{k-1} = 0, 1} \ \frac{T_{0 i_{k-1}}}{(k \gamma)} \cdots \frac{T_{i_1 0}}{(\gamma + i_1 s)}  \\
&= 1 +  \frac{[(\Delta \alpha) (g-1)]^2 c_3}{(s + \gamma) 2 \gamma} +  \frac{[(\Delta \alpha) (g-1)]^3  c_3 c_4}{(s + \gamma) (s + 2\gamma) 3 \gamma} + \cdots 
\end{split}
\end{equation}
\noindent where $\text{Poiss}(x,\mu)$ denotes the Poisson distribution
\begin{equation}
\text{Poiss}(x,\mu) := \frac{\mu^x e^{-\mu}}{x!} \ . 
\end{equation}
See Fig. \ref{fig:bds_threepanel_valid}  for plots in three representative cases, and a comparison against numerical results obtained via finite state projection \cite{munsky_finite_2006,fox2019}. Although it is not obvious, our results agree with previously published results on this problem, e.g. results due to Iyer-Biswas et al. \cite{iyer2009}, Huang et al. \cite{huang2015}, and Cao and Grima \cite{cao2018}; see Appendix \ref{sec:app_consistency} for a mathematical proof of this. For special cases and limits of these formulas, see Sec. \ref{sec:special}.

The mean and variance of this distribution are given by
\begin{equation}
\begin{split}
\expval{x} &= \mu \\
\sigma^2_x &= \expval{x^2} - \expval{x}^2 = \mu + \frac{(\Delta \alpha)^2 c_3}{(s + \gamma) \gamma}  \ .
\end{split}
\end{equation}

\clearpage

\subsection{Multistep splicing coupled to a switching gene} \label{sec:results_multistep}

\subsubsection{Problem statement}

Let $X_0$ denote unspliced/nascent RNA, $X_1$ denote RNA that has experienced one splicing step (i.e. one intron has been removed), $X_2$ denote RNA that has experienced two splicing steps, and so on. Assuming there are $N$ splicing steps, let $X_N$ denote the fully processed/mature RNA. As before, let $G_1$ and $G_2$ denote the two possible gene states. Transcription coupled to a switching gene, with multiple downstream splicing steps, can be modeled using the list of chemical reactions
\begin{equation} 
\begin{split}
G_1 &\xrightarrow{k_{21}} G_2 \hspace{0.8in} G_2 \xrightarrow{k_{12}} G_1 \\
G_1 &\xrightarrow{\alpha_1} G_1 + X_0 \hspace{0.44in} G_2 \xrightarrow{\alpha_2} G_2 + X_0 \\
X_0 &\xrightarrow{\beta_0} X_1 \xrightarrow{\beta_1} X_2 \xrightarrow{\beta_2} \cdots \xrightarrow{\beta_{N-1}} X_N \xrightarrow{\beta_N} \varnothing 
\end{split}
\end{equation}
where $k_{21}$ and $k_{12}$ parameterize the rates of gene switching, $\alpha_1$ and $\alpha_2$ parameterize the transcription rate in each gene state, $\beta_i$ parameterizes the rate of the $i$th splicing step (for $0 \leq i < N$), and $\beta_N$ parameterizes the mature RNA's degradation rate. 

\clearpage

Denote the state of the system using $\mathbf{x} := (x_0, x_1, ..., x_N) \in \mathbb{N}^{N+1}$ and $S \in \left\{ 1, 2 \right\}$. Use $\boldsymbol{\epsilon}_j$ to denote the vector with a $1$ in the $j$th place and zeros elsewhere. The CME corresponding to this model reads
\begin{equation} \label{eq:CME_multistep}
\begin{split}
\frac{\partial P(\mathbf{x}, 1, t)}{\partial t} =& \ - k_{21} P(\mathbf{x}, 1, t) + k_{12} P(\mathbf{x}, 2, t) \\ 
& + \alpha_1 \left[ P(\mathbf{x} - \boldsymbol{\epsilon}_0, 1, t) - P(\mathbf{x}, 1, t) \right] \\
& + \sum_{j = 0}^{N-1} \beta_j \left[ (x_j + 1) P(\mathbf{x} + \boldsymbol{\epsilon}_j - \boldsymbol{\epsilon}_{j+1}, 1, t) - x_j P(\mathbf{x}, 1, t) \right] \\
& + \beta_N \left[ (x_N + 1) P(\mathbf{x} + \boldsymbol{\epsilon}_N, 1, t) - x_N P(\mathbf{x}, 1, t) \right] \\
\frac{\partial P(\mathbf{x}, 2, t)}{\partial t} =& \ k_{21} P(\mathbf{x}, 1, t) - k_{12} P(\mathbf{x}, 2, t) \\
& + \alpha_2 \left[ P(\mathbf{x} - \boldsymbol{\epsilon}_0, 2, t) - P(\mathbf{x}, 2, t) \right] \\
& + \sum_{j = 0}^{N-1} \beta_j \left[ (x_j + 1) P(\mathbf{x} + \boldsymbol{\epsilon}_j - \boldsymbol{\epsilon}_{j+1}, 2, t) - x_j P(\mathbf{x}, 2, t) \right] \\
& + \beta_N \left[ (x_N + 1) P(\mathbf{x} + \boldsymbol{\epsilon}_N, 2, t) - x_N P(\mathbf{x}, 2, t) \right] \ .
\end{split}
\end{equation}
For this more complicated many-variable problem, it is helpful to introduce the notation
\begin{equation}
\begin{split}
\mathbf{v}^{\mathbf{x}} &:= v_0^{x_0} \cdots v_N^{x_N} \\
\mathbf{x}! &:= x_0! \cdots x_N! \\
\sum_{\mathbf{x}} &:= \sum_{x_0 = 0}^{\infty} \cdots \sum_{x_N = 0}^{\infty} \\
\int d\mathbf{x} &:= \int dx_0 \cdots \int dx_N  \ .
\end{split}
\end{equation} 
\noindent We aim to compute the steady state probability distribution and (analytic) generating function both marginalized over gene state, i.e. 
\begin{equation}
\begin{split}
P_{ss}(\mathbf{x}) &:= \lim_{t \to \infty} P(\mathbf{x}, 1, t) + P(\mathbf{x}, 2, t) \\
\psi_{ss}(\mathbf{g}) &:= \sum_{\mathbf{x}} P_{ss}(\mathbf{x}) \ \mathbf{g}^\mathbf{x}
\end{split}
\end{equation}
where $\mathbf{g} \in \mathbb{C}^{N+1}$. 

\subsubsection{Steady state results}

We will reuse several derived quantities from Sec. \ref{sec:main_bds_res}. One that must be modified is the effective mean, which in this case is a vector $\boldsymbol{\mu} := (\mu_0, \mu_1, ..., \mu_N)$ whose components are defined as
\begin{equation}
\mu_i = \frac{\alpha_{eff}}{\beta_i} \ .
\end{equation}
We can use it to define the multivariate Poisson distribution 
\begin{equation}
\text{Poiss}(\mathbf{x}, \boldsymbol{\mu}) := \frac{\boldsymbol{\mu}^{\mathbf{x}} e^{- \boldsymbol{\mu} \cdot \mathbf{1}}}{\mathbf{x}!} = \frac{\mu_0^{x_0} e^{- \mu_0}}{x_0!} \cdots \frac{\mu_N^{x_N} e^{- \mu_N}}{x_N!} \ .
\end{equation}
We must also define the vector $\mathbf{q} = (q_0, q_1, ..., q_N)$, and the family of vectors $\mathbf{v}^{(k)}$ (for $0 \leq k \leq N$). The entries of the former are defined via
\begin{equation}
\begin{split}
q_0 &= 1 \\
q_1 &= - \frac{\beta_0}{\beta_1 - \beta_0} \\
q_2 &= \frac{\beta_0 \beta_1}{(\beta_2 - \beta_0) (\beta_2 - \beta_1)} \\
q_j &= (-1)^j \frac{\beta_0 \cdots \beta_{j-1}}{(\beta_j - \beta_{0}) \cdots (\beta_j - \beta_{j-1})}  \hspace{0.25in} ( \ 1 \leq j \leq N \ )
\end{split}
\end{equation}
\noindent while the entries of the latter are defined via
\begin{equation}
v^{(k)}_j = \begin{cases}
0 & j < k \\
1 & j = k \\
\frac{\beta_k \cdots \beta_{j-1}}{(\beta_{k+1} - \beta_k) \cdots (\beta_j - \beta_k)} & j > k \ .
\end{cases}
\end{equation}
Use $V_{\mathbf{n}}(\mathbf{x},\boldsymbol{\mu})$ to denote the multistep orthogonal polynomial associated with the nonnegative integer vector $\mathbf{n} \in \mathbb{N}^{N+1}$; these polynomials are new, and are defined and discussed in Appendix \ref{sec:app_multistep}. Our exact analytic result for $P_{ss}(\mathbf{x})$ is
\begin{equation}
\begin{split}
\frac{P_{ss}(\mathbf{x})}{\text{Poiss}(\mathbf{x}, \boldsymbol{\mu})} =& 1 + \sum_{k = 2}^{\infty}  \sum_{\text{paths } \mathbf{j}} \left(\Delta \alpha\right)^k  \mathbf{q}^{\mathbf{n}} V_{\mathbf{n}}(\mathbf{x}, \boldsymbol{\mu})  \sum_{i_1, ..., i_{k-1} = 0, 1} \ \frac{T_{0 i_{k-1}}}{\left[ \beta_{j_1} + \cdots + \beta_{j_k} \right]} \cdots \frac{T_{i_1 0}}{\left[ \beta_{j_1} + i_1 s \right]}  \ , 
\end{split}
\end{equation}
where the sum requires some explanation. For each $k \geq 2$, we sum over all paths $\mathbf{j}$ of length $k$ with elements in $\left\{0, 1, ..., N \right\}$, i.e. $\mathbf{j} := (j_1, ..., j_k) \in \left\{ 0, 1, ..., N \right\}^k$. The vector $\mathbf{n} := (n_0, n_1, ..., n_N)$ counts the number of times that each integer $i$ appears in a path, i.e.
\begin{equation}
\begin{split}
n_r = \sum_{i \text{ s.t. } j_i = r} 1
\end{split}
\end{equation}
for all $0 \leq r \leq N$. For example, in the case of two splicing steps ($N = 1$), for $k = 1$ there are just two paths: $\mathbf{j}^{(1)} = (0)$ and $\mathbf{j}^{(2)} = (1)$. For $k = 2$, there are four:
\begin{equation}
\begin{split}
\mathbf{j}^{(1)} &= (0, 0) \\
\mathbf{j}^{(2)} &= (0, 1) \\
\mathbf{j}^{(3)} &= (1, 0) \\
\mathbf{j}^{(4)} &= (1, 1) \ .
\end{split}
\end{equation}
In general, for $N$ splicing steps and a set length $k \geq 1$, there will be $(N+1)^k$ distinct paths that must be summed over. The (analytic) generating function is
\begin{equation} \label{eq:multi_analytic_genfunc}
\begin{split}
\frac{\psi_{ss}(\mathbf{g})}{e^{\boldsymbol{\mu} \cdot (\mathbf{g} - \mathbf{1})} } =& 1 + \sum_{k = 2}^{\infty}  \left(\Delta \alpha\right)^k  \sum_{\text{paths } \mathbf{j}^{(k)}}   \ \prod_{m = 0}^N \left[ q_m (\mathbf{g} - \mathbf{1}) \cdot \mathbf{v}^{(m)}  \right]^{n_m} \ \sum_{i_1, ..., i_{k-1} = 0, 1} \ \frac{T_{0 i_{k-1}}}{\left[ \beta_{j_1} + \cdots + \beta_{j_k} \right]} \cdots \frac{T_{i_1 0}}{\left[ \beta_{j_1} + i_1 s \right]}  \\
=& 1 + (\Delta \alpha)^2 \sum_{a = 0}^N \sum_{b = 0}^N \left[ q_a (\mathbf{g} - \mathbf{1}) \cdot \mathbf{v}^{(a)}  \right] \left[ q_b (\mathbf{g} - \mathbf{1}) \cdot \mathbf{v}^{(b)}  \right] \frac{c_3}{(\beta_a + \beta_b) (\beta_a + s)} + \cdots \ .
\end{split}
\end{equation}
For special cases and limits, see Sec. \ref{sec:multistep_special}. The means, variances, and covariances of this distribution are
\begin{equation}
\begin{split}
\langle x_i \rangle &= \mu_i \\
\sigma^2_i &= \expval{x_i^2} - \expval{x_i}^2 = \mu_{i} + (\Delta \alpha)^2 \sum_{a = 0}^N \sum_{b = 0}^N  \frac{2 q_a q_b v^{(a)}_i v^{(b)}_i c_3}{(\beta_a + \beta_b) (\beta_a + s)} \\
\text{Cov}(x_i, x_j) &= \expval{x_i x_j} - \expval{x_i} \expval{x_j} = (\Delta \alpha)^2 \sum_{a = 0}^N \sum_{b = 0}^N  \frac{q_a q_b v^{(a)}_i v^{(b)}_j c_3}{(\beta_a + \beta_b) (\beta_a + s)} \ .
\end{split}
\end{equation}


%
%
%
%

\subsection{Brief guide to qualitative solution behaviors}
\label{sec:brief_guide}

For developing intuition about the behavior of these distributions, it is helpful to keep three edge cases in mind: (i) the Poisson limit, (ii) the Poisson mixture limit, and (iii) the negative binomial limit. Our exact solutions provably reduce to each of these three special cases in various biologically plausible limits (see Sec. \ref{sec:special} and \ref{sec:multistep_special}), and can be viewed as interpolating between them in intermediate parameter regimes.

When the sum of the gene switching rates, $s = k_{12} + k_{21}$, is much larger than the degradation and splicing rates, switching happens so quickly that the RNA only `feels' the effective transcription rate $\alpha_{eff}$. In this parameter regime, the steady state solutions reduce to Poisson distributions---exactly the behavior we would expect if there were no gene switching. 

When $s$ is much smaller than the degradation and splicing rates, switching happens so rarely that the system behaves like a superposition of two birth-death/monomolecular processes: one with transcription rate $\alpha_1$, and the other with transcription rate $\alpha_2$. In this parameter regime, the steady state solutions reduce to the Poisson mixture
\begin{equation}
P_{mix}(\mathbf{x}) = \frac{k_{12}}{s} \text{Poiss}(\mathbf{x}, \boldsymbol{\mu}_1) + \frac{k_{21}}{s} \text{Poiss}(\mathbf{x}, \boldsymbol{\mu}_2)
\end{equation}
where the $k_{12}/s$ and $k_{21}/s$ factors represent the typical fractions of time the system spends in each gene state.

\begin{figure}[!htb]
\center{\includegraphics[scale=1]{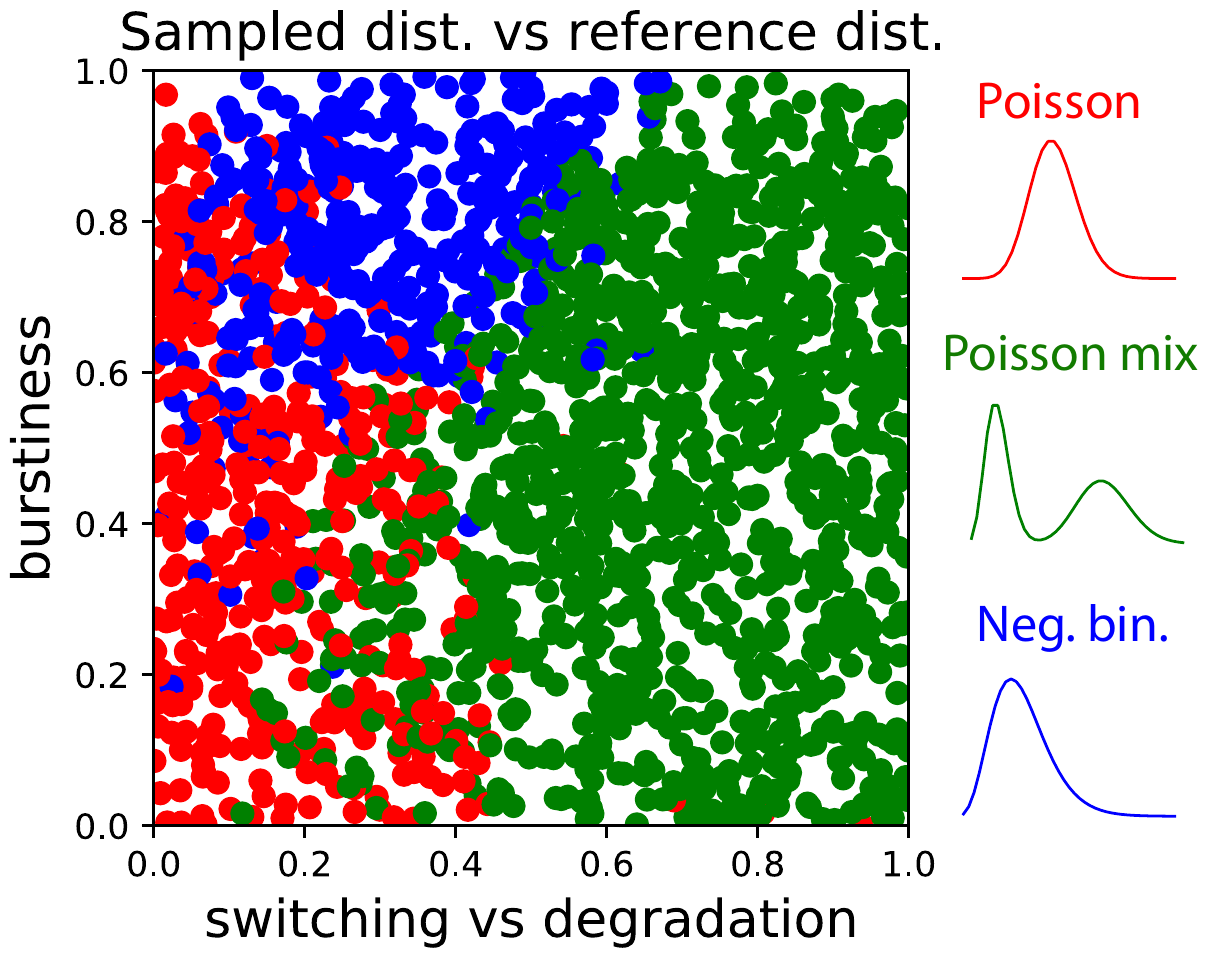}}
\caption{\label{fig:sampling_game} 
The birth-death-switching steady state distribution for many randomly sampled ($N = 2000$) parameter sets compared to three reference distributions (Poisson, Poisson mixture, negative binomial). After sampling a parameter set and computing the true $P_{ss}(x)$, the L2 distance between it and each reference distribution was computed. Each parameter set was plotted in `switching vs degradation'-`burstiness' space and colored according to the closest reference distribution. Red: Poisson, Green: Poisson mixture, Blue: negative binomial.}
\end{figure}

Finally, the bursty limit is defined by the conditions:
\begin{equation}
\begin{split}
\alpha_1 &\gg 1 \\
\alpha_2 &= 0 \\
k_{21} &\gg 1 \hspace{0.25in} k_{21} \gg k_{12}
\end{split}
\end{equation}
where taking $\alpha_1$ and $k_{21}$ large is done while keeping $\alpha_1/k_{21}$ held fixed. This yields a model with burst size $b := \alpha_1/k_{21}$ and burst frequency $k_{12}$, which has a heavy-tailed steady state distribution that is approximately negative binomial \cite{amrhein2019, raj_stochastic_2006}. For example, the birth-death-switching distribution approximately becomes
\begin{equation}
P_{nb}(x) = \binom{x + r - 1}{x} (1 - p)^r p^x
\end{equation}
where
\begin{equation}
\begin{split}
r &:= \frac{k_{12}}{\gamma}  \\
p &:= \frac{b}{1 + b} \ .
\end{split}
\end{equation}
We emphasize again that it will be helpful throughout the paper to think of the somewhat complicated general analytic solutions as interpolating between these three extremes. Indeed, as Fig. \ref{fig:sampling_game} suggests, this can be viewed somewhat literally. If we define the ad-hoc measures
\begin{equation}
\begin{split}
\text{`burstiness'} &:= \frac{\Delta \alpha}{\alpha_1 + \alpha_2} \cdot \frac{k_{12} k_{21}}{(k_{12} + k_{21})^2} \\
\text{`switching vs degradation'} &:= \frac{s}{s + \gamma}
\end{split}
\end{equation}
and compare the exact birth-death-switching solutions for various randomly sampled parameter sets to each of these three limit distributions, we can see that the parameter space can be relatively cleanly divided up into three regions. 

\subsection{Outline of approach to analytic solution} \label{sec:roadmap}

\begin{figure}[!htb]
\center{\includegraphics[scale=1]{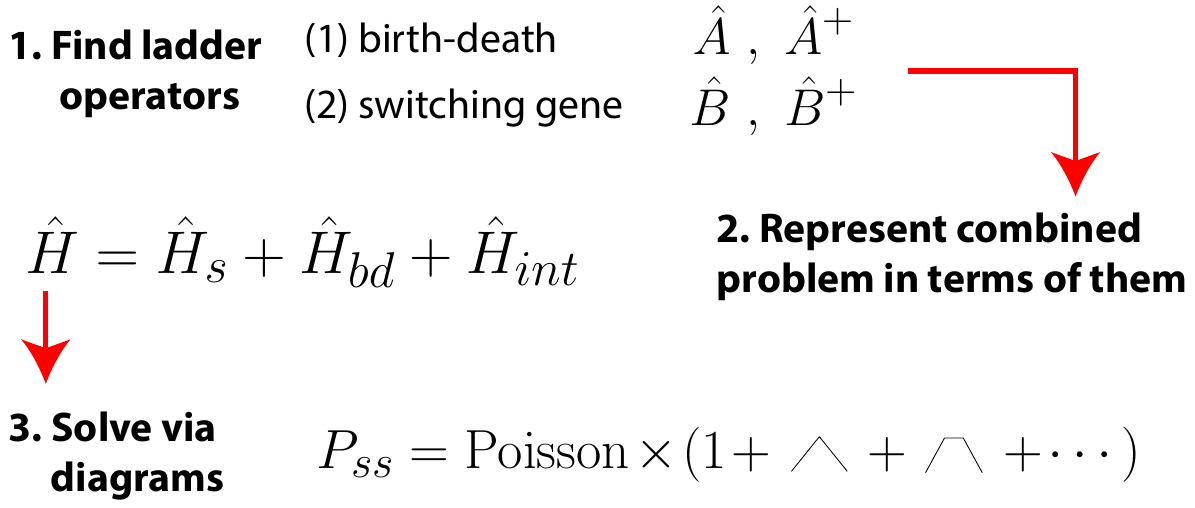}}
\caption{\label{fig:roadmap} 
The three main steps of our theoretical approach. First, we identify ladder operators for the RNA-only problem and the gene-switching-only problem. Then we use those ladder operators to formulate the CME of the full problem as a specific coupling of the individual problems. Finally, we exploit properties of the ladder operators to obtain a series solution, which can be viewed as a sum of Feynman diagrams.}
\end{figure}

The theoretical methodology used to obtain these results is new, and uses ideas inspired by quantum mechanics. Instead of directly trying to solve each CME (Eq. \ref{eq:CME_bds} and Eq. \ref{eq:CME_multistep}), we reframe the problem in terms of states and operators in a certain Hilbert space. In this more abstract space, we must compute the generating function (instead of the probability distribution), whose dynamics are determined by an operator that `looks' like a Hamiltonian. In this way, solving the CME is reduced to solving an abstract Schr\"{o}dinger-like equation. 

But this Schr\"{o}dinger-like equation is, naively, just as difficult to solve as the original CME. The key insight is to split the problem into two separate pieces (the dynamics of RNA by itself, and the dynamics of a switching gene by itself), solve each individual piece in a very particular way, and then pose the original problem as a certain coupling of the individual problems. 

The aforementioned `particular way' is to solve the individual problems in terms of ladder operators, which are mathematically analogous to those used in the algebraic treatment of the quantum harmonic oscillator \cite{griffiths2018} and in quantum field theory \cite{schwartz2014}. The dynamics of RNA by itself can be solved in terms of `bosonic' ladder operators, while the dynamics of a switching gene by itself can be solved in terms of `fermionic' ladder operators. Posing the full coupled problem in terms of these operators makes it amenable to the usual methods of quantum field theory, and enables a diagrammatic approach to computing quantities of interest. It also essentially reduces many computations (e.g. of moments) to straightforward algebra. 

To summarize, our general strategy involves three steps (Fig. \ref{fig:roadmap}):
\begin{enumerate}
\item Find ladder operators for RNA dynamics (assuming a constitutively active gene). Find ladder operators for pure gene switching.
\item Express the Hamiltonian of the coupled problem in terms of these ladder operators.
\item Solve the coupled problem perturbatively/diagrammatically to all orders.
\end{enumerate}
In the following sections, we carry out this plan. We first focus on the birth-death-switching problem because it is less encumbered by notational baggage, and so better illustrates our method. After establishing the method by solving this problem, we go on to solve the multistep splicing problem to demonstrate the power of this new approach.

%



%


\clearpage

\section{Preliminary representation theory}
\label{sec:reptheory}

In this section, we study the representation theory of RNA production and degradation by itself, and gene switching by itself. We will find that the former dynamics can be described by bosonic ladder operators, while the latter dynamics can be described by fermionic ladder operators. In the last subsection, we show how to use these ladder operators to couple the two individual problems and obtain the Hamiltonian operator corresponding to Eq. \ref{eq:CME_bds}.

\subsection{Representation theory of chemical birth-death process}
\label{sec:reptheory_bd}

In this section, we study the chemical birth-death process---a simple stochastic model of RNA being produced and degraded, without any additional features like gene switching or regulation---and in particular how to work with it in terms of ladder operators that make various theoretical features (e.g. the spectrum, the relationship between eigenstates, and computing expectation values) transparent. These ladder operators and the morass of formalism surrounding them will then be used to couple the birth-death process to a switching gene in Sec. \ref{sec:diagrammatic}.

Let $X$ denote the RNA species as before. The chemical birth-death process is defined by the list of reactions
\begin{equation} \label{eq:bd_rxns}
\begin{split}
\varnothing &\xrightarrow{\alpha} X \\
X &\xrightarrow{\gamma} \varnothing 
\end{split}
\end{equation}
where $\alpha$ parameterizes the rate of RNA production and $\gamma$ parameterizes the rate of degradation. For simplicity, we assume both $\alpha$ and $\gamma$ are time-independent. The corresponding CME is 
\begin{equation} \label{eq:CME_bd}
\begin{split}
\frac{\partial P(x, t)}{\partial t} =& \ \alpha \left[ P(x - 1, t) - P(x, t) \right] + \gamma \left[ (x + 1) P(x + 1, t) - x P(x, t) \right]  
\end{split}
\end{equation}
where $P(x, t)$ is the probability that the system has $x \in \left\{ 0, 1, 2, ...\right\}$ RNA molecules at time $t$. 

In order to think about this problem in terms of ladder operators, we must first reframe it in terms of states within a certain Hilbert space, and operators that act on those states. Consider a Hilbert space spanned by the basis kets $\ket{0}, \ket{1}, ...$, where an arbitrary state $\ket{\phi}$ can be written
\begin{equation}
\ket{\phi} := \sum_{x = 0}^{\infty} c(x) \ \ket{x} 
\end{equation}
for some generally complex coefficients $c(x)$. The operators that act on these states will generally be written in terms of the creation and annihilation operators
\begin{equation}
\begin{split}
\hat{\pi} \ket{x} &:= \ket{x + 1} \\
\hat{a} \ket{x} &:= x \ \ket{x - 1}
\end{split}
\end{equation}
which satisfy the familiar commutation relation $[ \hat{a}, \hat{\pi} ] = 1$. In the reframed problem, we work with the generating function
\begin{equation} \label{eq:gf}
\ket{\psi} := \sum_{x = 0}^{\infty} P(x, t) \ \ket{x}
\end{equation}
instead of the probability density $P(x, t)$; by construction, it contains equivalent information, so that knowing one is equivalent to knowing the other. Just as $P(x, t)$ satisfies the CME given by Eq. \ref{eq:CME_bd}, it is easy to show that the generating function $\ket{\psi}$ satisfies the equation of motion
\begin{equation} \label{eq:EOM_bd}
\frac{\partial \ket{\psi}}{\partial t} = \hat{H}_{bd} \ket{\psi}
\end{equation}
where the Hamiltonian operator $\hat{H}_{bd}$ is defined as
\begin{equation}
\hat{H}_{bd} := \alpha \left( \hat{\pi} - 1 \right) + \gamma \left( \hat{a} - \hat{\pi} \hat{a}  \right) \ .
\end{equation}
For technical reasons, it will be slightly more convenient to work in terms of the Grassberger-Scheunert creation operator $\hat{a}^+ := \hat{\pi} - 1$ \cite{grassberger1980, vastolaDP}, in terms of which the Hamiltonian operator reads
\begin{equation}
\hat{H}_{bd} := \alpha \ \hat{a}^+ - \gamma \ \hat{a}^+ \hat{a} \ .
\end{equation}
Now we will proceed with the ladder operator solution of the problem. We are seeking two operators $\hat{A}$ and $\hat{A}^+$ for which the following hold:
\begin{enumerate}
\item \textbf{Hamiltonian decomposition:} $\hat{H}_{bd}$ is proportional to $\hat{A}^+ \hat{A}$.
\item \textbf{Commutation relation:} The pair satisfies $[\hat{A}, \hat{A}^+] = 1$.
\item \textbf{Hermitian conjugates:} There is an inner product with respect to which $\hat{A}$ and $\hat{A}^+$ are Hermitian conjugates.\footnote{This last requirement is optional, and is mainly necessary if one wishes to use ladder operators to derive transient (as opposed to steady state) solutions.}
\end{enumerate}
Although such a pair of operators may not exist for a generic choice of $\hat{H}$, their existence is easy to establish for this problem. Define $\mu := \alpha/\gamma$ (which turns out to be the mean molecule number at steady state), and consider the operators
\begin{equation} \label{eq:bd_lad_def}
\begin{split}
\hat{A}^+ &:= \sqrt{\mu} \ \hat{a}^+ \\
\hat{A} &:= \frac{1}{\sqrt{\mu}} \left( \hat{a} - \mu \right) \ .
\end{split}
\end{equation}
Their fundamental commutation relation $[\hat{A}, \hat{A}^+] = 1$ is easy to verify. In terms of these two operators, the Hamiltonian can be written
\begin{equation}
\hat{H}_{bd} = - \gamma \ \hat{A}^+ \hat{A} \ .
\end{equation}
This decomposition makes it easy to show that the commutation relations
\begin{equation} \label{eq:comm_bd}
\begin{split}
[\hat{A}, \hat{H}_{bd}] &= - \gamma \hat{A} \\
[\hat{A}^+, \hat{H}_{bd}] &= \gamma \hat{A}^+
\end{split}
\end{equation}
hold. They can be used to argue \cite{griffiths2018, vastolaADD} that there is a ground state $\ket{0}$ satisfying
\begin{equation} \label{eq:ground_bd}
\begin{split}
\hat{A} \ket{0} &= 0 \\
\ket{0} &= \sum_{x = 0}^{\infty} P_{ss}(x) \ket{x} = \sum_{x = 0}^{\infty} \frac{\mu^x}{x!} e^{- \mu} \ket{x} \ ,
\end{split}
\end{equation}
where the normalization of $\ket{0}$ was chosen so that its coefficients sum to $1$ (since they correspond to the values of the steady state probability distribution $P_{ss}(x)$), and that the eigenvalues of the Hamiltonian operator are $- E_n$, where
\begin{equation} \label{eq:energies_bd}
E_n = \gamma \ n
\end{equation}
for $n = 0, 1, 2, ...$ Hence, we can write the eigenstates of $\hat{H}_{bd}$ as $\ket{n}$ for $n \in \mathbb{N}$ (e.g. $\ket{0}$ is the ground state, and $\ket{1}$ if the first excited state), so that
\begin{equation}
\hat{H}_{bd} \ket{n} = - \gamma \ n \ket{n} \ .
\end{equation}
Moreover, using the facts that\footnote{This is easy to show by induction, using the fundamental commutation relation between $\hat{A}$ and $\hat{A}^+$.}  $[\hat{A}, (\hat{A}^+)^n] = n (\hat{A}^+)^{n-1}$ and $\hat{A} \ket{0} = 0$, we have 
\begin{equation}
\begin{split}
\hat{H}_{bd} (\hat{A}^+)^n \ket{0} &= - \gamma \hat{A}^+ \hat{A} (\hat{A}^+)^n \ket{0} \\
&= - \gamma \hat{A}^+ \left[ (\hat{A}^+)^n \hat{A} + n  (\hat{A}^+)^{n-1} \right] \ket{0} \\
&= - \gamma \ n (\hat{A}^+)^{n} \ket{0} 
\end{split}
\end{equation}
i.e. that
\begin{equation}
\ket{n} \propto (\hat{A}^+)^n \ket{0} \ .
\end{equation}
In order to fix the normalization of the eigenstates $\ket{n}$, we must decide on an inner product. For this problem, a natural choice turns out to be
\begin{equation} \label{eq:bd_inner}
\braket{x}{y} := \frac{\delta_{x y}}{P_{ss}(x)} = \frac{\delta_{x y} \ x! \ e^{\mu}}{\mu^x }  \ .
\end{equation}
It is straightforward to show that $\hat{A}$ and $\hat{A}^+$ are Hermitian conjugates with respect to this inner product. On the one hand,
\begin{equation}
\begin{split}
\mel{x}{\hat{A}^+}{y} &= \sqrt{\mu} \ \mel{x}{\hat{a}^+ - 1}{y} \\
&= \sqrt{\mu} \ \left[ \ \braket{x}{y + 1} - \braket{x}{y} \ \right] \\
&= \sqrt{\mu} \ \left[ \frac{\delta_{x, y+1} \ x! \ e^{\mu}}{\mu^x }  - \frac{\delta_{x y} \ x! \ e^{\mu}}{\mu^x }  \right] \ .
\end{split}
\end{equation}
On the other hand,
\begin{equation}
\begin{split}
\mel{y}{\hat{A}}{x} &= \frac{1}{\sqrt{\mu}} \ \mel{y}{\hat{a} - \mu}{x} \\
&= \frac{1}{\sqrt{\mu}} \ \left[ \ x \braket{y}{x - 1} - \mu \braket{y}{x} \ \right] \\
&= \frac{1}{\sqrt{\mu}}  \ \left[ x \frac{\delta_{x-1, y} \ (x-1)! \ e^{\mu}}{\mu^{x-1} }  - \mu \frac{\delta_{x y} \ x! \ e^{\mu}}{\mu^x }  \right] \\
&= \sqrt{\mu} \ \left[ \frac{\delta_{x, y+1} \ x! \ e^{\mu}}{\mu^x }  - \frac{\delta_{x y} \ x! \ e^{\mu}}{\mu^x }  \right] \\
&= \mel{x}{\hat{A}^+}{y}  \ .
\end{split}
\end{equation}
Because $\hat{A}$ and $\hat{A}^+$ are Hermitian conjugates, $\hat{H}_{bd}$ is Hermitian, since
\begin{equation}
\left( \hat{H}_{bd} \right)^{\dag} = - \gamma \left( \hat{A}^+ \hat{A} \right)^{\dag} = - \gamma \left( \hat{A} \right)^{\dag} \left( \hat{A}^+ \right)^{\dag} = - \gamma \hat{A}^+ \hat{A} = \hat{H}_{bd} \ .
\end{equation}
This allows us to use a standard argument to show that different eigenstates are orthogonal:
\begin{equation}
\begin{split}
\braket{m}{n} = - \frac{1}{E_n} \mel{m}{\hat{H}_{bd}}{n} = - \frac{1}{E_n} \mel{n}{\hat{H}_{bd}}{m}^* = \frac{E_m}{E_n} \braket{n}{m}^* = \frac{E_m}{E_n} \braket{m}{n}
\end{split}
\end{equation}
which forces $\braket{m}{n} = 0$ since $E_m \neq E_n$. In order to normalize these eigenstates, so that $\braket{n}{n} = 1$ for all $n \in \mathbb{N}$, we note that
\begin{equation}
\braket{n}{n} = \mel{0}{\hat{A}^n (\hat{A}^+)^n}{0} = n! \braket{0}{0} = n! \ .
\end{equation}
This implies the correct normalization for the eigenkets is
\begin{equation} \label{eq:eig_def_bd}
\ket{n} := \frac{( \hat{A}^+ )^n}{\sqrt{n!}} \ \ket{0} \ .
\end{equation}
Using this definition, we can show that
\begin{equation}
\begin{split}
\hat{A}^+ \ket{n} &= \sqrt{n+1} \ket{n+1} \\
\hat{A} \ket{n} &= \sqrt{n} \ket{n-1} \ ,
\end{split}
\end{equation}
properties that will be useful later. 

All of this legwork we put into understanding the eigenstates of $\hat{H}_{bd}$ makes finally solving the equation of motion, Eq. \ref{eq:EOM_bd}, relatively simple. Because the eigenstates as we have defined them are orthonormal, we have the resolution of the identity 
\begin{equation} \label{eq:ROI_bd}
1 = \sum_{n = 0}^{\infty} \ \ket{n} \bra{n} \ .
\end{equation}
The formal solution of Eq. \ref{eq:EOM_bd} is
\begin{equation}
\ket{\psi(t)} = e^{\hat{H}_{bd} (t - t_0)} \ket{\psi_0} 
\end{equation}
where $\ket{\psi_0} = \ket{\psi(t_0)}$ is the initial generating function. Applying Eq. \ref{eq:ROI_bd} to our formal solution, we have
\begin{equation}
\begin{split}
\ket{\psi(t)} &= \sum_{n = 0}^{\infty} e^{\hat{H}_{bd} (t - t_0)} \ket{n} \braket{n}{\psi_0} \\
&= \sum_{n = 0}^{\infty}  e^{- E_n (t - t_0)}  \ket{n} \braket{n}{\psi_0} \\
&= \sum_{n = 0}^{\infty} \braket{n}{\psi_0} \   \ket{n}  \ e^{- \gamma n (t - t_0)} \ .
\end{split}
\end{equation}
We are particularly interested in computing the transition probability $P(x, t; x_0, t_0)$ for arbitrary $x_0 \in \mathbb{N}$, in which case $\ket{\psi_0} = \ket{x_0}$. Denoting the coefficients of $\ket{n}$ by $P_n(x)$, note that
\begin{equation}
\braket{n}{x_0} = \frac{P_n(x_0)}{P_{ss}(x_0)}
\end{equation}
by definition. It can be shown, using the definition of $\ket{n}$ from Eq. \ref{eq:eig_def_bd} and the fact that the coefficients of $\ket{0}$ are known (c.f. Eq. \ref{eq:ground_bd}), that
\begin{equation}
P_n(x) = \sqrt{\frac{\mu^n}{n!}} \ C_n(x, \mu) P_{ss}(x)
\end{equation}
where $C_n(x, \mu)$ are the Charlier polynomials, a family of discrete orthogonal polynomials whose properties are discussed at length in Appendix \ref{sec:app_charlier}. They play a role analogous to the one the Hermite polynomials play in the solution of the quantum harmonic oscillator. Hence, the solution to Eq. \ref{eq:EOM_bd} with $\ket{\psi_0} = \ket{x_0}$ is 
\begin{equation}
\begin{split}
\ket{\psi(t)} &= \sum_{n = 0}^{\infty} \sqrt{\frac{\mu^n}{n!}} \ C_n(x_0, \mu) \   \ket{n}  \ e^{- \gamma n (t - t_0)} \ .
\end{split}
\end{equation}
The corresponding probability distribution $P(x, t; x_0, t_0)$ can be obtained in two ways: (i) by appealing to the definition of $\ket{\psi(t)}$, or (ii) by invoking the Euclidean product. The Euclidean product is another useful inner product, and defined on two basis kets as 
\begin{equation} \label{eq:bd_euclidean}
\braket{x}{y}_{Eu} = \delta_{x, y} 
\end{equation}
and extended to more general kets by linearity.
The expression $\braket{x}{\psi(t)}_{Eu}$ essentially `picks out' the coefficient of $\ket{\psi(t)}$ corresponding to $P(x, t)$; although this is identical to just picking out the $x$ coefficient of the generating function, using this notation makes the mathematical analogy between position-space wave functions and probability distributions more transparent. Finally,
\begin{equation} \label{eq:sln_bd_charlier}
\begin{split}
P(x, t; x_0, t_0) &= \braket{x}{\psi(t)}_{Eu} \\
&= \sum_{n = 0}^{\infty} \sqrt{\frac{\mu^n}{n!}} \ C_n(x_0, \mu) \   \braket{x}{n}_{Eu}  \ e^{- \gamma n (t - t_0)} \\
&= P_{ss}(x) \sum_{n = 0}^{\infty} \frac{\mu^n}{n!} \ C_n(x_0, \mu) C_n(x, \mu)  \ e^{- \gamma n (t - t_0)} \ .
\end{split}
\end{equation}
The more familiar expression for $P(x, t; x_0, t_0)$, given by \cite{jahnke2007, vastolaDP}
\begin{equation} \label{eq:bd_sln}
P(x, t; x_0, t_0) = \sum_{j = 0}^{\min(x_0, x)} \binom{x_0}{j} \ q^{x_0 - j} (1 - q)^j \ \cdot \ \frac{(\mu q)^{x - j} e^{- \mu q}}{(x - j)!} 
\end{equation}
where $q(t) := 1 - e^{- \gamma (t - t_0)}$, can be obtained from Eq. \ref{eq:sln_bd_charlier} using Charlier polynomial properties discussed in Appendix \ref{sec:app_charlier}; for example, one can either use the integral representation given by Eq. \ref{eq:charlier_intrep}, or the Charlier polynomial generating function given by Eq. \ref{eq:charlier_genfunc}. 

Particular features of the solution, like moments, can be obtained by expressing certain operators in terms of ladder operators and then using their algebraic properties to compute the result. For example, suppose we want to compute the mean $\expval{x}$ as a function of time. First, note that
\begin{equation}
\hat{a} \ket{\psi(t)} = \sum_{x = 0}^{\infty} P(x, t) \ x \ket{x - 1} \ ,
\end{equation}
so 
\begin{equation}
\mel{0}{\hat{a}}{\psi(t)} = \sum_{x = 0}^{\infty} P(x, t) x = \expval{x} 
\end{equation}
where we have used the fact that the inner product of the vacuum state with any basis ket $\ket{x}$ is $\braket{0}{x} = 1$. Now, since $\hat{a} = \sqrt{\mu} \ \hat{A} + \mu$, we can compute
\begin{equation}
\begin{split}
\expval{x} &= \mel{0}{(\sqrt{\mu} \ \hat{A} + \mu)}{\psi(t)} \\
&= \sum_{n = 0}^{\infty} \sqrt{\frac{\mu^n}{n!}} \ C_n(x_0, \mu) \   \mel{0}{(\sqrt{\mu} \ \hat{A} + \mu)}{n}  \ e^{- \gamma n (t - t_0)} \\
&= \sum_{n = 0}^{\infty} \sqrt{\frac{\mu^n}{n!}} \ C_n(x_0, \mu) \   \left[ \sqrt{\mu} \mel{0}{\hat{A}}{n} + \mu \braket{0}{n} \right] \ e^{- \gamma n (t - t_0)} \\
&= \sum_{n = 0}^{\infty} \sqrt{\frac{\mu^n}{n!}} \ C_n(x_0, \mu) \   \left[ \sqrt{\mu n} \braket{0}{n-1} + \mu \braket{0}{n}\right] \ e^{- \gamma n (t - t_0)} \\
&= \sum_{n = 0}^{\infty} \sqrt{\frac{\mu^n}{n!}} \ C_n(x_0, \mu) \   \left[ \sqrt{\mu n} \ \delta_{n, 1} + \mu \ \delta_{n, 0} \right] \ e^{- \gamma n (t - t_0)} \\
&= \left( x_0 - \mu \right) e^{- \gamma (t - t_0)} + \mu 
\end{split}
\end{equation}
where we have used the fact that $C_1(x_0, \mu) = (x_0/\mu) - 1$. 

Parenthetically, we may note that the use of these ladder operators to understand the birth-death process evokes striking parallels with the quantum harmonic oscillator; the analogy is even more striking in the continuous limit, where Hermite polynomials replace the Charlier polynomials (see Sec. \ref{sec:continuous} and \cite{vastolaADD}). In some sense, a single chemical species being randomly produced and degraded behaves like a free boson. More generally, we will see in Sec. \ref{sec:multistep} that $N$ chemical species being produced and degraded behave like $N$ free bosons.

\clearpage

\subsection{Representation theory of pure gene switching}
\label{sec:reptheory_gene}
In this section, we will consider the stochastic dynamics of a switching gene by itself. A careful study of the switching gene uncovers ladder operators that will be useful when coupling switching dynamics to the birth-death process in the next section.

Consider a gene with two states, which we will label as $G_1$ and $G_2$. Our list of chemical reactions is
\begin{equation} \label{eq:gene_rxns}
\begin{split}
G_1 &\xrightarrow{k_{21}} G_2 \\
G_2 &\xrightarrow{k_{12}} G_1 
\end{split}
\end{equation}
and the corresponding CME reads
\begin{equation}
\begin{split}
\frac{\partial P(1, t)}{\partial t} =& \ - k_{21} P(1, t) + k_{12} P(2, t) \\ 
\frac{\partial P(2, t)}{\partial t} =& \ k_{21} P(1, t) - k_{12} P(2, t) 
\end{split}
\end{equation}
where $P(1, t)$ is the probability that the gene is in state $1$ at time $t$, and $P(2, t)$ is the probability that the gene is in state $2$ at time $t$. If we define the state vector 
\begin{equation}
\vec{P}(t) := \begin{pmatrix} P(1, t) \\ P(2, t) \end{pmatrix} \ ,
\end{equation} 
then we can write the CME in the form
\begin{equation}
\dot{\vec{P}} = \hat{H}_s \vec{P} \ ,
\end{equation}
where the switching Hamiltonian $\hat{H}_s$ is defined to be
\begin{equation} \label{eq:Hs}
\hat{H}_s := \begin{pmatrix} - k_{21} & k_{12} \\ k_{21} & - k_{12}  \end{pmatrix} \ .
\end{equation}
One important property of the matrix $\hat{H}_s$ is that it is \textit{infinitesimal stochastic} (see \cite{baez2018} for more discussion of this property), which means that $1^T \hat{H}_s = 0^T$, or equivalently that the columns of $\hat{H}_s$ sum to zero. This property guarantees probability conservation:
\begin{equation}
\frac{d}{dt} \left( 1^T \vec{P} \right) = 1^T \frac{d \vec{P}}{dt} = 1^T \hat{H}_s \vec{P} = 0^T \vec{P} = 0 \ .
\end{equation}
It is easy enough to find the exact time-dependent solution to this system by diagonalizing $\hat{H}_s$. Its eigenvalues are $\lambda_0 = 0$ and $\lambda_1 = - s$, where $s := k_{12} + k_{21}$ is the gene switching rate sum. The corresponding eigenvectors are
\begin{equation}
\vec{v}_0 = \begin{pmatrix} \frac{k_{12}}{k_{12} + k_{21}} \\ \frac{k_{21}}{k_{12} + k_{21}} \end{pmatrix} \ \vec{v}_1 = \begin{pmatrix} 1 \\ -1 \end{pmatrix} \ .
\end{equation}
Hence, the exact solution is
\begin{equation}
\vec{P}(t) = \vec{v}_0 + c \ \vec{v}_1 \ e^{- s t}
\end{equation}
where the constant $c$ depends upon one's initial condition. 


A more interesting task is to try and express $\hat{H}_s$ in terms of ladder operators that will make its solution structure more explicit. Recall that the anticommutator of two matrices $M_1$ and $M_2$ is defined as $\{ M_1, M_2 \} := M_1 M_2 + M_2 M_1$. We are seeking two operators $\hat{B}$ and $\hat{B}^+$ for which the following hold:
\begin{enumerate}
\item \textbf{Hamiltonian decomposition:} $\hat{H}_{s}$ is proportional to $\hat{B}^+ \hat{B}$.
\item \textbf{Commutation relations:} The pair satisfies $\{ \hat{B}, \hat{B} \} = \{ \hat{B}^+, \hat{B}^+ \} = 0$ and $\{ \hat{B}, \hat{B}^+ \} = 1$. Here, $1$ is shorthand for the $2\times2$ identity matrix.
\item \textbf{Hermitian conjugates:} There is an inner product with respect to which $\hat{B}$ and $\hat{B}^+$ are Hermitian conjugates.
\end{enumerate}
\clearpage
\noindent By trial and error\footnote{For a more principled route to determining them, see Appendix \ref{sec:app_moreswitch}.}, we can find that the operators
\begin{equation}
\begin{split}
\hat{B} := \begin{pmatrix} \frac{k_{12} k_{21}}{s^2} & - \frac{k_{12}^2}{s^2} \\ \frac{k_{21}^2}{s^2} & - \frac{k_{12} k_{21}}{s^2}  \end{pmatrix} \\
\hat{B}^+ := \begin{pmatrix} 1 & 1 \\ -1 & -1  \end{pmatrix}
\end{split}
\end{equation}
suffice. In terms of them, the switching Hamiltonian $\hat{H}_s$ can be written
\begin{equation}
\hat{H}_s = - s \ \hat{B}^+ \hat{B} \ .
\end{equation}
One nice feature of $\hat{B}$ and $\hat{B}^+$ is that their action on the eigenvectors of $\hat{H}_s$ is particularly simple: 
\begin{equation} \label{eq:B_effect}
B \vec{v}_0 = 0 \hspace{1in} B \vec{v}_1 = \vec{v}_0
\end{equation}
\begin{equation} \label{eq:Bplus_effect}
B^+ \vec{v}_0 = \vec{v}_1 \hspace{1in} B^+ \vec{v}_1 = 0 \ .
\end{equation}
These are properties that we will repeatedly exploit later. One can define a natural inner product on $\mathbb{R}^2$ using the symmetric and positive-definite weight matrix
\begin{equation}
\hat{W} = 
\begin{pmatrix}
1 + \frac{k_{21}^2}{s^2} & 1 - \frac{k_{12} k_{21}}{s^2} \\ 1 - \frac{k_{12} k_{21}}{s^2} & 1 + \frac{k_{12}^2}{s^2}
\end{pmatrix} \ .
\end{equation}
Note that, since $k_{12}$ and $k_{21}$ are nonnegative, $0 \leq (k_{12} k_{21})/s^2 \leq 1/4$, which makes the entries of $\hat{W}$ nonnegative. For more discussion on the appropriateness of this weight matrix, along with some mathematical motivation for considering these kinds of ladder operators, see Appendix \ref{sec:app_moreswitch}. For now, assuming our choice of $\hat{W}$ is well-motivated, we define the inner product of two vectors $\vec{z}_1, \vec{z}_2 \in \mathbb{R}^2$ via
\begin{equation}
\langle \vec{z}_1, \vec{z}_2 \rangle := \vec{z}_1^T W \vec{z}_2 \ .
\end{equation} 
With respect to this inner product, $\hat{B}$ and $\hat{B}^+$ are Hermitian conjugates (see Appendix \ref{sec:app_moreswitch} for the mathematical details), i.e. we have $\hat{W} \hat{B} = (\hat{B}^+)^T \hat{W}$ and $\hat{W} \hat{B}^+ = (\hat{B})^T \hat{W}$.

Recalling that the eigenvectors of $\hat{H}_s$ and $\hat{H}^T_s$ are orthogonal with respect to the usual dot product offers another way to view this inner product. That is, given the eigenvectors $\vec{w}_0$ and $\vec{w}_1$ (which correspond to the eigenvalues $\lambda_0 = 0$ and $\lambda_1 = - s$ since $\hat{H}_s$ and $\hat{H}^T_s$ have the same eigenvalues), we have $\vec{v}_i^T W = \vec{w}_i^T$. For completeness' sake, we note that
\begin{equation}
\vec{w}_0 = \begin{pmatrix} 1 \\ 1 \end{pmatrix}  \ \vec{w}_1 = \begin{pmatrix} \frac{k_{21}}{k_{12} + k_{21}} \\ - \frac{k_{12}}{k_{12} + k_{21}} \end{pmatrix} \ .
\end{equation}
Because applying $\vec{w}_0^T$ on the left corresponds to summing over each gene state, we will end up using it to marginalize over gene state in Sec. \ref{sec:special}. 
 
It will be helpful to recast this inner product in terms of bra-ket notation; to that end, we identify the eigenvectors $\vec{v}_0$ and $\vec{v}_1$ with $\ket{0}$ and $\ket{1}$, $\vec{w}_0$ and $\vec{w}_1$ with $\bra{0}$ and $\bra{1}$, and note that
\begin{equation} \label{eq:switch_ortho}
\begin{split}
\braket{0}{1} &= 0  \\
\braket{0}{0} &= \braket{1}{1} = 1 
\end{split}
\end{equation}
i.e. they constitute an orthonormal basis. 

It is interesting to note that the switching gene behaves much like a single spin-1/2 fermion. Given that many RNA species being produced and degraded behave like many free bosons (see Sec. \ref{sec:multistep}), it is tempting to speculate that the dynamics of switching between many gene states corresponds to the dynamics of many free fermions. As we discuss at the end of Appendix \ref{sec:app_moreswitch}, this generically does not seem to be true, at least in a straightforward generalization of our approach.

\subsection{Coupling a switching gene to the birth-death process}
\label{sec:reptheory_couple}
We would like to couple a switching gene to the chemical birth-death process; in particular, we would like to make the production rate dependent on the current gene state. This means that, instead of the production rate being a constant parameter $\alpha$, we can view it as an operator 
\begin{equation}
\hat{\alpha} := \begin{pmatrix} \alpha_1 & 0 \\ 0 & \alpha_2 \end{pmatrix}
\end{equation}
where $\alpha_1$ and $\alpha_2$ are the distinct production rates we first introduced in Eq. \ref{eq:bds_rxns}. We stress that we \textit{do not} assume anything about $\alpha_1$ or $\alpha_2$ (e.g. that they are distinct, or that one is significantly larger than the other), even though these assumptions apply in the biologically interesting case of a bursty gene. 

This section's central trick is to note that the matrix $\hat{\alpha}$ can be written in terms of ladder operators, since the set $\{ 1, \hat{B}, \hat{B}^+, \hat{B}^+ \hat{B} \}$ forms a basis for the space of all real $2\times2$ matrices. We will use this fact to write
\begin{equation}
\begin{pmatrix} \alpha_1 & 0 \\ 0 & \alpha_2 \end{pmatrix} = r_1 + r_2 \hat{B} + r_3 \hat{B}^+ + r_4 \hat{B}^+ \hat{B}  
\end{equation}
for some coefficients $r_1$, $r_2$, $r_3$, and $r_4$. To determine these coefficients, we recall from the previous subsection the facts that (i) the actions of $\hat{B}$ and $\hat{B}^+$ on the eigenvectors of $\hat{H}_s$ are easy to compute (c.f. Eq. \ref{eq:B_effect} and Eq. \ref{eq:Bplus_effect}), and (ii) the eigenvectors of $\hat{H}_s$ are orthonormal with respect to the inner product we defined (c.f. Eq. \ref{eq:switch_ortho}). For example,
\begin{equation}
\begin{split}
\hat{\alpha} \ket{0} &= r_1 \ket{0} + r_2 \hat{B} \ket{0} + r_3 \hat{B}^+ \ket{0} + r_4 \hat{B}^+ \hat{B} \ket{0} \\
&= r_1 \ket{0} + r_3 \ket{1} \\
\implies r_1 &= \mel{0}{\hat{\alpha}}{0} \ . 
\end{split}
\end{equation}
Similarly,
\begin{equation}
\begin{split}
r_2 &= \mel{0}{\hat{\alpha}}{1}  \\
r_3 &= \mel{1}{\hat{\alpha}}{0}  \\
r_4 &= \mel{1}{\hat{\alpha}}{1} - r_1  \ .
\end{split}
\end{equation}
It should be noted that, although we are interested here in the case where the production rate depends on the current gene state in the sense specified above, this method can in principle be used to tackle many possible generalizations of this problem. 

Our particular coefficients read
\begin{equation}
\begin{split}
r_1 &=  \frac{\alpha_1 k_{12} + \alpha_2 k_{21}}{s} \\
r_2 &= \alpha_1 - \alpha_2 \\
r_3 &= (\alpha_1 - \alpha_2) \frac{k_{12} k_{21}}{s^2} \\
r_4 &= (\alpha_1 - \alpha_2) \ \frac{k_{21} - k_{12}}{s}  \ .
\end{split}
\end{equation}
If we define the effective production rate
\begin{equation}
\alpha_{eff} := \frac{\alpha_1 k_{12} + \alpha_2 k_{21}}{s} \ ,
\end{equation}
then we can write the matrix $\hat{\alpha}$ in terms of ladder operators as
\begin{equation}
\hat{\alpha} = \alpha_{eff} + (\alpha_1 - \alpha_2) \left[ B + \frac{k_{12} k_{21}}{s^2} B^+ + \frac{k_{21} - k_{12}}{s}  B^+ B \right] \ .
\end{equation}
With this done, we can write out the Hamiltonian operator for our full, coupled problem\footnote{This can be derived from the full CME, Eq. \ref{eq:CME_bds}.}. It reads
\begin{equation}
\begin{split}
\hat{H} &= - s \hat{B}^+ \hat{B} + \hat{\alpha} \  \hat{a}^+  - \gamma \ \hat{a}^+ \hat{a} \\
&= - s \hat{B}^+ \hat{B} - \gamma \hat{A}^+ \hat{A} + (\alpha_1 - \alpha_2) \ \hat{a}^+ \left[ \hat{B} + \frac{k_{12} k_{21}}{s^2} \hat{B}^+ + \frac{k_{21} - k_{12}}{s}  \hat{B}^+ \hat{B} \right] \\
&= - s \hat{B}^+ \hat{B} - \gamma \hat{A}^+ \hat{A} + \frac{(\alpha_1 - \alpha_2)}{\sqrt{\mu}} \hat{A}^+ \left[ \hat{B} + \frac{k_{12} k_{21}}{s^2} \hat{B}^+ + \frac{k_{21} - k_{12}}{s}  \hat{B}^+ \hat{B} \right] \\
&= \hat{H}_s + \hat{H}_{bd} + \hat{H}_{int} 
\end{split}
\end{equation}
where we define the interaction Hamiltonian $\hat{H}_{int}$ as
\begin{equation}
\hat{H}_{int} := \frac{(\alpha_1 - \alpha_2)}{\sqrt{\mu}} \hat{A}^+ \left[ \hat{B} + \frac{k_{12} k_{21}}{s^2} \hat{B}^+ + \frac{k_{21} - k_{12}}{s}  \hat{B}^+ \hat{B} \right]  
\end{equation}
along with the shorthand notations
\begin{equation}
\begin{split}
\mu &:= \frac{\alpha_{eff}}{\gamma} \\
\hat{A}^+ &= \sqrt{\mu} \left( a^+ - 1 \right) \\
\hat{A} &= \frac{1}{\sqrt{\mu}} \left( a - \mu \right)  \ .
\end{split}
\end{equation}
The parameter $\mu$ that appears in this Hamiltonian is actually the effective mean---a sort of average of what the mean would be given the typical amount of time the system spends in each gene state. It should not be conflated with $\alpha_1/\gamma$ or $\alpha_2/\gamma$ in what follows.

Interestingly, $\hat{H}_{int}$ looks like an interaction term with interaction strength proportional to $(\alpha_1 - \alpha_2)$, the difference between the two production rates. In particular, since the $\hat{A}$ operators are `bosonic', and the $\hat{B}$ operators are `fermionic', our full Hamiltonian looks like it describes one boson interacting with one fermion. When the production rates are equal (i.e. when $\alpha_1 = \alpha_2$), the interaction term disappears, and the birth-death process and the gene switching dynamics become uncoupled. This makes sense---from the point of view of the birth-death process, it doesn't matter what the current gene state is as long as it doesn't affect the production rate.

It is natural to speculate that the Hamiltonian describing a birth-death process coupled to a gene with $N$ gene states would correspond to one boson interacting with $N-1$ fermions (which also possibly interact with each other). Coupling additional birth-death processes (for example, representing gene-gene interactions or downstream products of the original RNA) may look like having more bosons. It is not completely clear what the representation theory of $\hat{H}_s$ would look like in these more general cases, or even if a ladder operator view would be appropriate. Regrettably, the representation theory of infinitesimal stochastic matrices seems not particularly well-studied; for example, it is not clear to the authors that infinitesimal stochastic matrices (or even restricted classes of them, like irreducible infinitesimal stochastic matrices) have a spectral theorem that would permit eigenvector decompositions similar to the one we have used here. 

\clearpage





\section{Diagrammatic approach to exact solution}
\label{sec:diagrammatic}

In this section, we use the ladder operators identified in the previous section to develop a diagrammatic approach to obtaining the exact steady state solution of CMEs involving switching (and in particular, of Eq. \ref{eq:CME_bds}). 

\subsection{Constructing eigenstates of full Hamiltonian}

Consider a state formed by naively combining the eigenstates of the two original problems\footnote{The birth-death problem has a countably infinite number of eigenstates, which can be indexed by a natural number $n$, and the switching gene has two, which we label using $0$ and $1$. Note that these two eigenstates are distinct from the two gene states, which are labeled using $1$ and $2$.}, which we will denote by $\ket{n; g}$, and which satisfies
\begin{equation}
\begin{split}
\hat{H}_s \ket{n; 0} &= 0 \\
\hat{H}_s \ket{n; 1} &= - s \ket{n; 1} \\
\hat{H}_{bd} \ket{n; g} &= - \gamma n \ket{n; g}  \ .
\end{split}
\end{equation}
The action of our ladder operators on this state is
\begin{equation}
\begin{split}
\hat{B} \ket{n; 0} &= \hat{B}^+ \ket{n; 1} = 0 \\
\hat{B}^+ \ket{n; 0} &= \ket{n; 1} \\
\hat{B} \ket{n; 1} &= \ket{n; 0} \\
\hat{A} \ket{n; g} &= \sqrt{n} \ket{n - 1; g} \\
\hat{A}^+ \ket{n; g} &= \sqrt{n + 1} \ket{n + 1; g} \ .
\end{split}
\end{equation}
All of this means that we can compute the action of $\hat{H}$ on these states. It is
\begin{equation}
\begin{split}
\hat{H} \ket{n; 0} &= - \gamma n \ket{n; 0} + \frac{(\alpha_1 - \alpha_2)}{\sqrt{\mu}} \frac{k_{12} k_{21}}{s^2} \sqrt{n + 1} \ket{n + 1; 1} \\
\hat{H} \ket{n; 1} &= - (s + \gamma n) \ket{n; 1} + \frac{(\alpha_1 - \alpha_2)}{\sqrt{\mu}} \sqrt{n+1} \left[ \ket{n+1; 0} + \frac{k_{21} - k_{12}}{s}  \ket{n+1; 1} \right] \ .
\end{split}
\end{equation}
From the above, it should be clear that our states $\ket{n; g}$ are `almost' eigenstates of $\hat{H}$, if it were not for the term proportional to $(\alpha_1 - \alpha_2)$ due to the interaction $\hat{H}_{int}$ between the bosonic and fermionic ladder operators. If $\alpha_1$ is very close to $\alpha_2$, this term would be small, and it may be reasonable to approximately consider the states $\ket{n; g}$ as eigenstates.

But this suggests something interesting. Consider the action of $\hat{H}$ on the `next' states, indexed by $n+1$:
\begin{equation}
\begin{split}
\hat{H} \ket{n + 1; 0} &= - \gamma (n+1) \ket{n + 1; 0} + \frac{(\alpha_1 - \alpha_2)}{\sqrt{\mu}} \frac{k_{12} k_{21}}{s^2} \sqrt{n + 2} \ket{n + 2; 1} \\
\hat{H} \ket{n + 1; 1} &= - [ s + \gamma (n+1) ] \ket{n + 1; 1} + \frac{(\alpha_1 - \alpha_2)}{\sqrt{\mu}} \sqrt{n+2} \left[ \ket{n+2; 0} + \frac{k_{21} - k_{12}}{s}  \ket{n+2; 1} \right] \ .
\end{split}
\end{equation}
Note that
\begin{equation}
\begin{split}
& \hat{H} \left\{ \ket{n; 0} + \frac{(\alpha_1 - \alpha_2)}{\sqrt{\mu}} \frac{k_{12} k_{21}}{s^2 (s + \gamma)} \sqrt{n + 1} \ket{n + 1; 1} \right\} \\
=& - \gamma n \ \left\{ \ket{n; 0} + \frac{(\alpha_1 - \alpha_2)}{\sqrt{\mu}} \frac{k_{12} k_{21}}{s^2 (s + \gamma)} \sqrt{n + 1} \ket{n + 1; 1} \right\} \\
&+  \left[ \frac{(\alpha_1 - \alpha_2)}{\sqrt{\mu}} \right]^2 \frac{k_{12} k_{21}}{s^2 (s + \gamma)} \sqrt{(n+1)(n+2)} \left[ \ket{n+2; 0} + \frac{k_{21} - k_{12}}{s}  \ket{n+2; 1} \right]
\end{split}
\end{equation}
which is to even \textit{better} approximation an eigenstate of $\hat{H}$ when $\alpha_1$ is close to $\alpha_2$. It seems like this process may be continued to obtain an infinite series in powers of $(\alpha_1 - \alpha_2)$---and indeed it can. To show how it can, first let us ease notation by writing
\begin{equation}
\begin{split}
d &:= \frac{\alpha_1 - \alpha_2}{\sqrt{\mu}} \\
c_3 &:=  \frac{k_{12} k_{21}}{s^2} \\
c_4 &:= \frac{k_{21} - k_{12}}{s}
\end{split}
\end{equation}
so that the interaction Hamiltonian $\hat{H}_{int}$ can be written as
\begin{equation} \label{eq:Hint_notation}
\hat{H}_{int} = d \ \hat{A}^+ \left[ \hat{B} + c_3 \hat{B}^+ + c_4 \hat{B}^+ \hat{B} \right] \ .
\end{equation}
If we write our result for the eigenstate $\ket{n}$ of the full coupled problem (which we label as such since it has eigenvalue $- \gamma n$) as
\begin{equation} \label{eq:np0_exp}
\begin{split}
\ket{n} &:= \ket{n; 0} + d \sqrt{n+1} \left\{ \ q^0_{1, 0} \ket{n + 1; 0} + q^0_{1, 1} \ket{n+1; 1} \ \right\} + \cdots  \\
&= \ket{n; 0} + \sum_{k = 1}^{\infty} d^k \sqrt{(n+1)_k} \left\{ \ q^0_{k, 0} \ket{n + k; 0} + q^0_{k, 1} \ket{n + k; 1} \ \right\}
\end{split}
\end{equation}
for some coefficients $q^0_{k, g}$ determined by the procedure above, with $(n+1)_k := (n + 1) \cdots (n + k)$, we find that the coefficients satisfy the recurrence relations
\begin{equation} \label{eq:rec_q0}
\begin{split}
k \gamma \ q^0_{k, 0} &= q^0_{k - 1, 1} \\
(k \gamma + s) \ q^0_{k, 1} &= c_3 \ q^0_{k - 1, 0} + c_4 \ q^0_{k - 1, 1}   
\end{split}
\end{equation}
for all $k \geq 1$, with the initial conditions $q^0_{0,0} = 1$ and $q^0_{0,1} = 0$. 

Similarly, we can construct eigenstates $\ket{n + s/\gamma}$ (which we label as such since they have eigenvalues $- \gamma \left( n + s/\gamma \right)$ ) as an infinite series
\begin{equation}
\begin{split}
\ket{n + s/\gamma} &:= \ket{n; 1} + d \sqrt{n+1} \left\{ \ q^1_{1, 0} \ket{n + 1; 0} + q^1_{1, 1} \ket{n+1; 1} \ \right\} + \cdots  \\
&= \ket{n; 1} + \sum_{k = 1}^{\infty} d^k \sqrt{(n+1)_k} \left\{ \ q^1_{k, 0} \ket{n + k, 0} + q^1_{k, 1} \ket{n + k; 1} \ \right\}
\end{split}
\end{equation}
whose coefficients are determined by the recurrence relations
\begin{equation} \label{eq:rec_q1}
\begin{split}
( k \gamma - s ) \ q^1_{k, 0} &= q^1_{k - 1, 1} \\ 
k \gamma \ q^1_{k, 1} &= c_3 \ q^1_{k - 1, 0} + c_4 \ q^1_{k - 1, 1}  
\end{split}
\end{equation}
for all $k \geq 1$, with the initial conditions $q^1_{0,0} = 0$ and $q^1_{0,1} = 1$. 

The above recurrences have closed form solutions, but their expressions are somewhat cumbersome; fortunately, there is an interesting diagrammatic interpretation of the coefficients $q^0_{k, g}$ and $q^1_{k, g}$. We will explore these points in the next two subsections.

For now, let us stop and note that we have established that there are eigenstates of $\hat{H}$ with eigenvalues $- \gamma \ n - s \ g$ for all $n \in \mathbb{N}$ and $g \in \{ 0, 1 \}$. Although it is not clear how to prove this mathematically, it is likely that this is the \textit{complete} collection of eigenstates, so that the full spectrum of $\hat{H}$ is given by
\begin{equation}
E_{n, g} = \gamma \ n + s \ g
\end{equation}
for $n \in \mathbb{N}$ and $g \in \{ 0, 1 \}$. One nice feature of this result is that it offers a natural way to understand previous observations regarding the relative importance of different time scales in this problem. For example, Iyer-Biswas et al. \cite{iyer2009} point out (in our notation) that $\text{min}(\gamma, s)$ determines the time scale of relaxation to steady state. This makes sense, because $- \text{min}(\gamma, s)$ is precisely the smallest nonzero eigenvalue.


\subsection{The transfer matrix} \label{sec:transfer}

Closed form solutions to the recurrence relations given by Eq. \ref{eq:rec_q0} and \ref{eq:rec_q1} can be written in terms of a certain $2 \times 2$ matrix, which we will call the transfer matrix $T$. First, consider that Eq. \ref{eq:rec_q0} can be rewritten in the form
\begin{equation}
\begin{pmatrix} q^0_{k,0} \\ q^0_{k,1} \end{pmatrix} = \begin{pmatrix} 0 & \frac{1}{k \gamma} \\ \frac{c_3}{k \gamma + s} & \frac{c_4}{k \gamma + s} \end{pmatrix} \begin{pmatrix} q^0_{k-1,0} \\ q^0_{k-1,1} \end{pmatrix} = \begin{pmatrix} \frac{1}{k \gamma} & 0 \\ 0 & \frac{1}{k \gamma + s} \end{pmatrix} \begin{pmatrix} 0 & 1 \\ c_3 & c_4 \end{pmatrix} \begin{pmatrix} q^0_{k-1,0} \\ q^0_{k-1,1} \end{pmatrix} \ .
\end{equation}
Similarly, Eq. \ref{eq:rec_q1} can be rewritten in the form
\begin{equation}
\begin{pmatrix} q^1_{k,0} \\ q^1_{k,1} \end{pmatrix} = \begin{pmatrix} \frac{1}{k \gamma - s} & 0 \\ 0 & \frac{1}{k \gamma} \end{pmatrix} \begin{pmatrix} 0 & 1 \\ c_3 & c_4 \end{pmatrix} \begin{pmatrix} q^1_{k-1,0} \\ q^1_{k-1,1} \end{pmatrix} \ .
\end{equation}
This motivates defining the matrix $T$ whose entries are
\begin{equation}
\begin{split}
T_{00} &= 0 \\
T_{01} &= 1 \\
T_{10} &= c_3 = \frac{k_{12} k_{21}}{s^2} \\
T_{11} &= c_4 = \frac{k_{21} - k_{12}}{s} \ ,
\end{split}
\end{equation}
i.e.
\begin{equation}
T = \begin{pmatrix} T_{00} & T_{01} \\ T_{10}  & T_{11} \end{pmatrix} = \begin{pmatrix} 0 & 1 \\ c_3  & c_4 \end{pmatrix} = \begin{pmatrix} 0 & 1 \\ \frac{k_{12} k_{21}}{s^2}  & \frac{k_{21} - k_{12}}{s} \end{pmatrix} \ .
\end{equation}
Keeping in mind the initial condition, we can now straightforwardly compute the coefficients $q^0_{k,0}$ via
\begin{equation}
\begin{split}
\begin{pmatrix} q^0_{k,0} \\ q^0_{k,1} \end{pmatrix} &= \begin{pmatrix} 0 & \frac{1}{k \gamma} \\ \frac{c_3}{k \gamma + s} & \frac{c_4}{k \gamma + s} \end{pmatrix} \cdots \begin{pmatrix} 0 & \frac{1}{\gamma} \\ \frac{c_3}{\gamma + s} & \frac{c_4}{\gamma + s} \end{pmatrix}  \begin{pmatrix} 1 \\ 0 \end{pmatrix} \ .
\end{split}
\end{equation}
In other words, $q^0_{k,0}$ is the $00$ (upper left) entry of a $k$-fold product of $2 \times 2$ matrices, and $q^0_{k,1}$ is the $10$ (bottom left) entry. By the same argument, $q^1_{k, 0}$ and $q^1_{k,1}$ are the $01$ and $11$ entries of a $k$-fold product. Explicitly, 
\begin{equation}
\begin{split}
q^0_{k, 0} (0\to0 \text{ in } k \text{ steps}) &= \sum_{i_1, ..., i_{k-1} = 0, 1} \ \frac{T_{0 i_{k-1}}}{(k \gamma)} \cdots \frac{T_{i_m i_{m-1}}}{(m \gamma + i_m s)} \cdots \frac{T_{i_2 i_1}}{(2 \gamma + i_2 s)} \frac{T_{i_1 0}}{(\gamma + i_1 s)} \\
q^0_{k, 1} (0\to1 \text{ in } k \text{ steps}) &= \sum_{i_1, ..., i_{k-1} = 0, 1} \ \frac{T_{1 i_{k-1}}}{(k \gamma)} \cdots \frac{T_{i_m i_{m-1}}}{(m \gamma + i_m s)} \cdots \frac{T_{i_2 i_1}}{(2 \gamma + i_2 s)} \frac{T_{i_1 0}}{(\gamma + i_1 s)} \\
q^1_{k, 0} (1\to0 \text{ in } k \text{ steps}) &= \sum_{i_1, ..., i_{k-1} = 0, 1} \ \frac{T_{0 i_{k-1}}}{(k \gamma - s)} \cdots \frac{T_{i_m i_{m-1}}}{(m \gamma - s + i_m s)} \cdots  \frac{T_{i_1 1}}{(\gamma - s + i_1 s)} \\
q^1_{k, 1} (1\to1 \text{ in } k \text{ steps}) &= \sum_{i_1, ..., i_{k-1} = 0, 1} \ \frac{T_{1 i_{k-1}}}{(k \gamma)} \cdots \frac{T_{i_m i_{m-1}}}{(m \gamma - s + i_m s)} \cdots \frac{T_{i_1 1}}{(\gamma - s + i_1 s)} \ .
\end{split}
\end{equation}
These are the cumbersome expressions referred to in the previous subsection. In the next subsection, we offer a diagrammatic interpretation of them.

\clearpage

\subsection{Feynman rules}

It turns out that we can associate with each coefficient $q^a_{k, b}$ a diagram with $k$ lines and $k + 1$ vertices. We obtain the following Feynman rules for computing the Hilbert space/bra-ket representation of the eigenstates.


\subsubsection{Hilbert space Feynman rules}

\noindent In order to compute the coefficient of $\ket{n + k; g}$ in the infinite series expansion of $\ket{n}$ (c.f. Eq. \ref{eq:np0_exp}), draw all valid diagrams going from $0$ to $g \in \left\{0,1\right\}$ in $k$ steps according to the following rules, and add the numbers corresponding to each diagram. 
\begin{enumerate}
\item \textbf{Set up grid}: Write out positions $0$ through $k$ from left to right. Draw two parallel horizontal lines above these labels to denote the $0$ and $1$ gene eigenstates. The diagram will consist of $k+1$ vertices, each located at a horizontal $\{0, ..., k\}$ position and vertical gene eigenstate (lower or upper) position, and lines connecting those vertices.
\item \textbf{Draw lines}: Place the first vertex at horizontal position $0$ and on the bottom row. Fill in the following positions from left to right. There are three possible moves: (i) if at $0$, you must next go to $1$; (ii) if at $1$, you can go to $0$ next; (iii) if at $1$, you can stay at $1$. If $g = 0$, the last vertex must be on the bottom row. If $g = 1$, the last vertex must be on the top row. 
\item \textbf{Numerical factors}: Associate each move/line with a numerical factor. In particular, associate the move from position $m-1$ to position $m$ with the factor:
\begin{itemize}
\item $0 \to 1$ \textit{flip}: $\displaystyle \frac{c_3}{s + m \gamma}$
\item $1 \to 1$ \textit{stay}: $\displaystyle \frac{c_4}{s + m \gamma}$
\item $1 \to 0$ \textit{flip}: $\displaystyle \frac{1}{m \gamma}$
\end{itemize}
\item \textbf{Tack on generic factors}: Multiply the numbers associated with each line together, along with the generic factors $d^k \sqrt{(n+1) \cdots (n+k)}$, to get the number corresponding to the diagram you drew.
\end{enumerate}
The steps are displayed in the context of a specific example in Fig. \ref{fig:feynman_howto}. See Tables \ref{table:feynman_examples0} and \ref{table:feynman_examples1} in Appendix \ref{sec:app_tables} for the values of many low-order Feynman diagrams. 

Analogous rules can be written for computing $q^1_{k,g}$: one must start on the upper row instead of the lower row, and the denominators associated with step 3 are slightly different, since we would have to make the substitutions $s + m \gamma \to m \gamma$ and $m \gamma \to m \gamma - s$.


\begin{figure}[!htb]
\center{\includegraphics[scale=1.2]{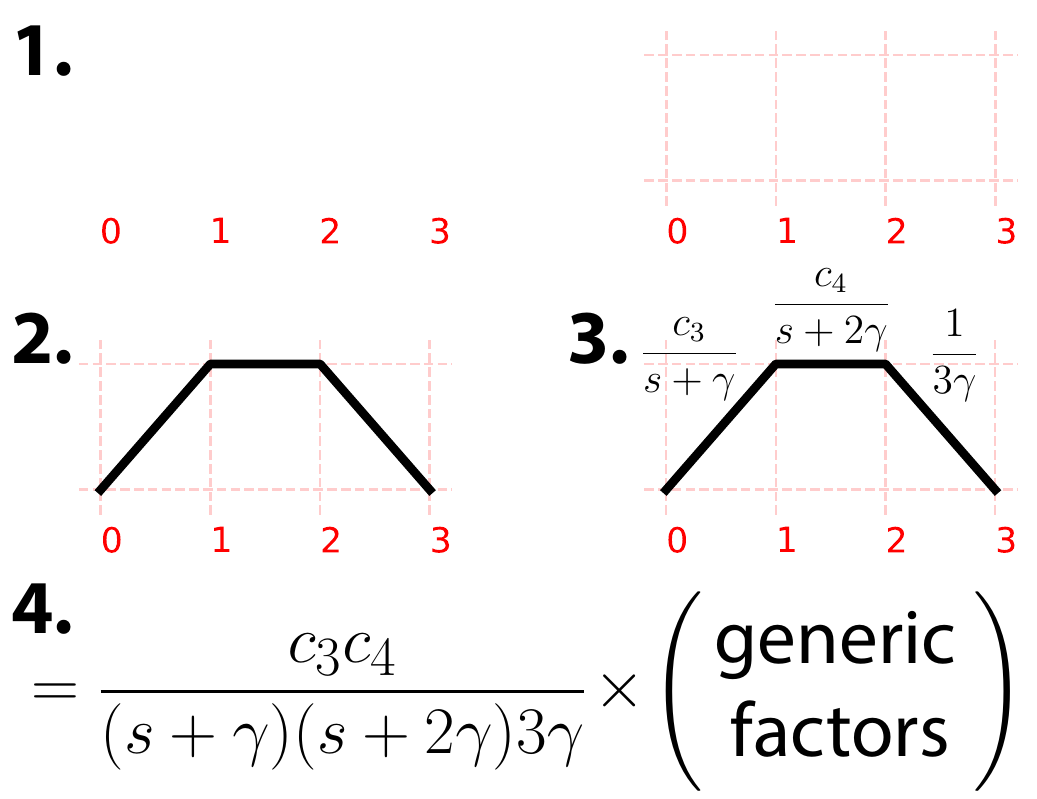}}
\caption{\label{fig:feynman_howto} 
How to draw a Feynman diagram. Here, we illustrate the process using the unique diagram that goes from $0$ to $0$ in three steps. Step 1: Draw the grid. Step 2: Draw the lines for your diagram. Step 3: Associate each line with numerical factors. Step 4: Multiply the numbers for each line together, and tack on generic factors.}
\end{figure}

\clearpage

\section{Special cases and limits}
\label{sec:special}

In this section, we examine various special cases of the result we found in the previous section. Of particular biological relevance are the steady state probability distribution and various special cases of it, including the limiting distributions described earlier in Sec. \ref{sec:brief_guide}.

To summarize our work from the previous section, we found that $\hat{H}$ has eigenstates $\ket{n}$ (where $n \in \mathbb{N}$) with energies $E_n = \gamma n$ which can be written in terms of the naive eigenstates $\ket{n; g}$ via
\begin{equation}
\begin{split}
\ket{n} &= \ket{n; 0} + \sum_{k = 1}^{\infty} \sum_{i_k = 0, 1} d^k \sqrt{(n+1)_k} \ q^0_{k, i_k} \ket{n + k; i_k} \\
&= \ket{n; 0} + \sum_{k = 1}^{\infty} d^k \sqrt{(n+1)_k} \sum_{i_1, ..., i_{k} = 0, 1} \ \frac{T_{i_k i_{k-1}}}{(k \gamma + i_k s)} \cdots \frac{T_{i_1 0}}{(\gamma + i_1 s)} \ket{n + k; i_k} \ ,
\end{split}
\end{equation}
and eigenstates $\ket{n + s/\gamma}$ with energies $E_n = \gamma n + s$ that can be written in terms of naive eigenstates as
\begin{equation}
\begin{split}
\ket{n + s/\gamma} &= \ket{n; 1} + \sum_{k = 1}^{\infty} \sum_{i_k = 0, 1} d^k \sqrt{(n+1)_k}  \ q^1_{k, i_k} \ket{n + k; i_k} \\
&= \ket{n; 1} + \sum_{k = 1}^{\infty} d^k \sqrt{(n+1)_k} \sum_{i_1, ..., i_{k} = 0, 1} \ \frac{T_{i_k i_{k-1}}}{(k \gamma - s + i_k s)} \cdots \frac{T_{i_1 1}}{(\gamma - s + i_1 s)} \ket{n + k; i_k} \ .
\end{split}
\end{equation}
For comparison with experiment, we are most interested in a special eigenstate: the eigenstate with eigenvalue $0$, which corresponds to the steady state probability distribution. Specializing the above, we have the `vacuum state' 
\begin{equation} \label{eq:vac_state}
\begin{split}
\ket{0} &= \ket{0; 0} + \sum_{k = 1}^{\infty} d^k \sqrt{k!} \sum_{i_1, ..., i_{k} = 0, 1} \ \frac{T_{i_k i_{k-1}}}{(k \gamma + i_k s)} \cdots \frac{T_{i_1 0}}{(\gamma + i_1 s)} \ket{k; i_k} \ .
\end{split}
\end{equation}

\clearpage

\subsection{Steady state probability distribution}

We can extract what we \textit{really} care about---the probability distribution corresponding to Eq. \ref{eq:vac_state}---by either picking out coefficients, or invoking the Euclidean product as in Sec. \ref{sec:reptheory_bd}. For the states $\ket{x; S}$ (where $x$ represents molecule number and $S \in \{1, 2\}$ represents gene state, rather than gene eigenstate), from which the naive eigenkets $\ket{n; g}$ can be constructed, the Euclidean product can be defined via
\begin{equation}
\braket{x_1; S_1}{x_2; S_2}_{Eu} := \delta_{x_1, x_2} \delta_{S_1, S_2}
\end{equation}
so that
\begin{equation}
\braket{x; S}{n; g}_{Eu} = \sqrt{\frac{\mu^n}{n!}} \ C_n(x, \mu) \text{Poiss}(x,\mu) (\vec{v}_g)_{S} \ .
\end{equation}
Either way, we find
\begin{equation}
\begin{split}
\frac{P_{ss}(x, \vec{S})}{\text{Poiss}(x,\mu)} &=  \vec{v}_0 + \sum_{k = 1}^{\infty} (\Delta \alpha)^k  \ C_k(x, \mu) \sum_{i_1, ..., i_{k} = 0, 1} \ \frac{T_{i_k i_{k-1}}}{(k \gamma + i_k s)} \cdots \frac{T_{i_1 0}}{(\gamma + i_1 s)}  \ \vec{v}_{i_k}  
\end{split}
\end{equation}
where $\Delta \alpha := \alpha_1 - \alpha_2$ is the production rate difference. We can marginalize this result over gene state by applying $\vec{w}_0^T = (1, 1)^T$ on the left; doing so, we obtain
\begin{equation} \label{eq:bds_Pss_marg}
\begin{split}
\frac{P_{ss}(x)}{\text{Poiss}(x,\mu)} &=  1 + \sum_{k = 2}^{\infty} (\Delta \alpha)^k  \ C_k(x, \mu) \sum_{i_1, ..., i_{k-1} = 0, 1} \ \frac{T_{0 i_{k-1}}}{(k \gamma)} \cdots \frac{T_{i_1 0}}{(\gamma + i_1 s)}    \ .
\end{split}
\end{equation}
The corresponding result for the (analytic) generating function, which is straightforward to compute from this formula, is presented in Sec. \ref{sec:results_bds}. If we are only interested in the steady state distribution marginalized over gene state, we can modify the Feynman rules we presented earlier to compute it instead of the Hilbert space coefficients. These can be viewed as the Feynman rules for `molecular number space' rather than Hilbert space.

\clearpage

\subsubsection{Molecule number space Feynman rules}

\begin{figure}[!htb]
\center{\includegraphics[scale=1.3]{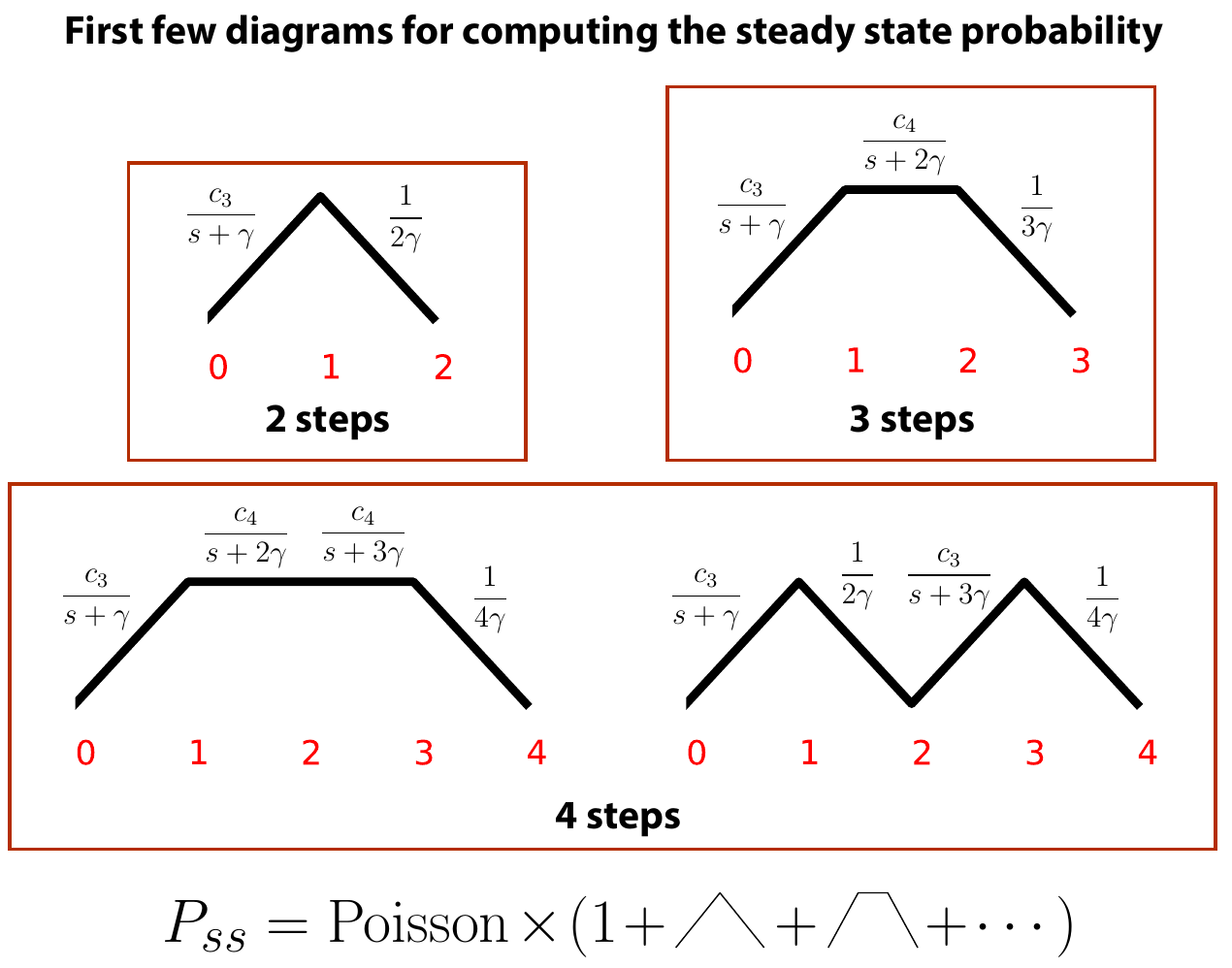}}
\caption{\label{fig:feynman_firstfew} 
First few Feynman diagrams contributing to $P_{ss}(x)$, with the numerical factors corresponding to each line shown explicitly. The overall result for $P_{ss}(x)$ is obtained by multiplying the sum of all diagrams by a Poisson distribution.}
\end{figure}

\noindent In order to compute the terms of order $(\Delta \alpha)^k$ in the infinite series expansion of $P_{ss}(x)/\text{Poiss}(x,\mu)$ (c.f. Eq. \ref{eq:bds_Pss_marg}), draw all valid diagrams going from $0$ to $0$ in $k$ steps according to the following rules, and add the numbers corresponding to each diagram. 
\begin{enumerate}
\item \textbf{Set up grid}: Write out positions $0$ through $k$ from left to right. Draw two parallel horizontal lines above these labels to denote the $0$ and $1$ gene eigenstates. The diagram will consist of $k+1$ vertices, each located at a horizontal $\{0, ..., k\}$ position and vertical gene eigenstate (lower or upper) position, and lines connecting those vertices.
\item \textbf{Draw lines}: Start on the bottom row ($0$). There are three possible moves: (i) if at $0$, you must next go to $1$; (ii) if at $1$, you can go to $0$ next; (iii) if at $1$, you can stay at $1$. The last vertex must be on the bottom row.
\item \textbf{Numerical factors}: Associate each move/line with a numerical factor. In particular, associate the move from position $m-1$ to position $m$ with the factor:
\begin{itemize}
\item $0 \to 1$ \textit{flip}: $\displaystyle \frac{c_3}{s + m \gamma}$
\item $1 \to 1$ \textit{stay}: $\displaystyle \frac{c_4}{s + m \gamma}$
\item $1 \to 0$ \textit{flip}: $\displaystyle \frac{1}{m \gamma}$
\end{itemize}
\item \textbf{Tack on generic factors}: Multiply the numbers associated with each line together, along with the generic factors $(\Delta \alpha)^k C_k(x,\mu)$, to get the number corresponding to the diagram you drew.
\end{enumerate}
It appears that the appropriate correlation function for this problem is given by the function
\begin{equation}
\mel{0}{\hat{A}^2}{0} = \frac{\sigma_x^2 - \mu}{\mu} = \frac{(\Delta \alpha)^2 c_3}{(s + \gamma) \gamma} \leq \frac{(\Delta \alpha)^2}{4 \gamma^2}
\end{equation}
where the inner product is defined analogously to before (c.f. Eq. \ref{eq:bd_inner}) on molecule number basis kets via
\begin{equation}
\braket{x}{y} := \frac{\delta_{x y}}{P_{ss}(x)} \ .
\end{equation}
This function measures how much the steady state distribution deviates from a Poisson distribution; it corresponds to the first nontrivial $0 \to 0$ diagram, which can be seen in the upper left corner of Fig. \ref{fig:feynman_firstfew}. It may be possible to view this as the propagator for some sort of quasiparticle (a transcripton?), but the advantages of such a view are not completely clear. In the next few sections, we examine biologically relevant special cases and limits of Eq. \ref{eq:bds_Pss_marg}.

\clearpage

\subsection{Equal switching rates}

\begin{figure}[!htb]
\center{\includegraphics[scale=1]{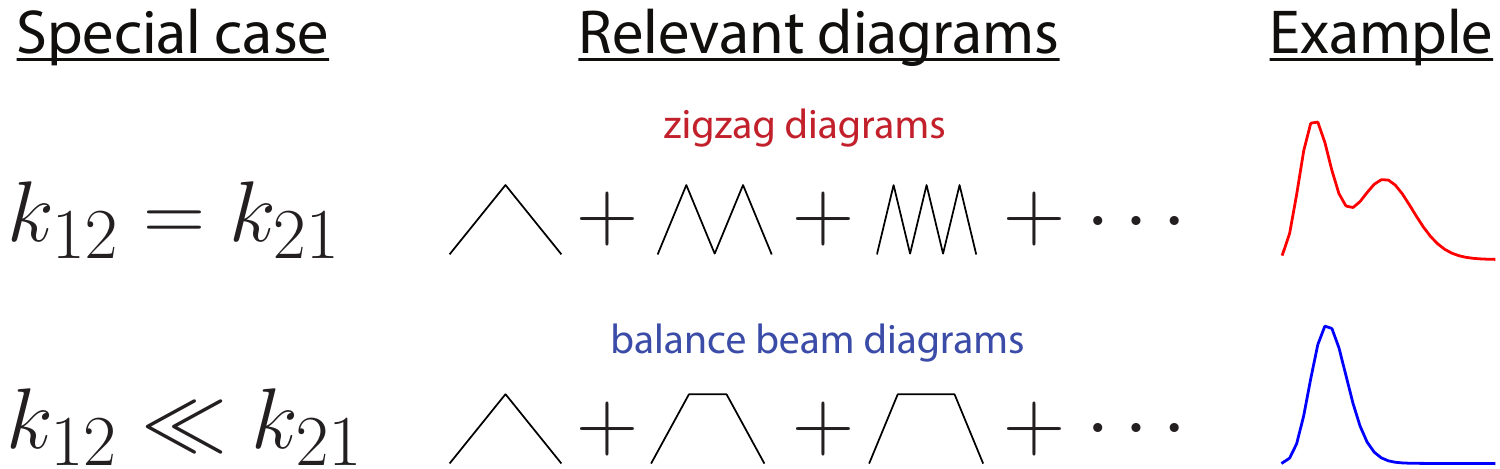}}
\caption{\label{fig:feynman_special} 
Dominant Feynman diagrams in two special cases. When the gene switching rates are equal ($k_{12} = k_{21}$), only the `zigzag' diagrams are nonzero. When the gene switching rates are very unequal (e.g. $k_{12} \ll k_{21}$), the `balance beam' diagrams contribute the leading terms.}
\end{figure}

In the special case of equal switching rates, where $k_{12} = k_{21}$, the exact solution simplifies considerably. In this regime, we have
\begin{equation}
\begin{split}
c_3 &= \frac{k_{12} k_{21}}{s^2} = \frac{1}{4} \\
c_4 &=  \frac{k_{21} - k_{12}}{s} = 0
\end{split}
\end{equation}
which means that all Feynman diagrams involving a $1 \to 1$ `stay' line vanish. This leaves only diagrams of the form $0 \to 1$, $0 \to 1 \to 0$, $0 \to 1 \to 0 \to 1$, and so on, i.e. the `zigzag' diagrams (see Fig. \ref{fig:feynman_special}). In particular, for $k$ even we have
\begin{equation}
\begin{split}
q^0_{k, 0} &= \frac{c_3^{k/2}}{(s + \gamma) 2 \gamma (s + 3 \gamma) \cdots [s + (k-1)\gamma] k \gamma} \\
q^0_{k, 1} &= 0 \\
q^1_{k, 0} &= 0 \\
q^1_{k, 1} &= \frac{c_3^{k/2}}{(\gamma - s)2\gamma(3\gamma - s) \cdots[(k-1)\gamma - s]k\gamma} 
\end{split}
\end{equation}
and for $k$ odd we have
\begin{equation}
\begin{split}
q^0_{k, 0} &= 0 \\
q^0_{k, 1} &= \frac{c_3^{(k+1)/2}}{(s + \gamma) 2 \gamma (s + 3 \gamma) \cdots [(k-1)\gamma](s + k \gamma)} \\
q^1_{k, 0} &= \frac{c_3^{(k-1)/2}}{(\gamma - s)2\gamma(3\gamma - s)\cdots(k-1)\gamma(k\gamma - s)} \\
q^1_{k, 1} &= 0  \ .
\end{split}
\end{equation}
\clearpage
\noindent This allows us to write explicit formulas like
\begin{equation}
\begin{split}
\ket{n} =& \ket{n, 0} + \sum_{k \text{ even} > 0}^{\infty} \frac{d^k \sqrt{(n+1)_k} \ c_3^{k/2}}{(s + \gamma) 2 \gamma (s + 3 \gamma) \cdots [s + (k-1)\gamma] k \gamma} \ket{n + k; 0} \\
&+ \sum_{k \text{ odd} }^{\infty}  \frac{d^k \sqrt{(n+1)_k} \ c_3^{(k+1)/2}}{(s + \gamma) 2 \gamma (s + 3 \gamma) \cdots [(k-1)\gamma](s + k \gamma)} \ket{n + k; 1} \ .
\end{split}
\end{equation}
The corresponding formula for the steady state probability distribution marginalized over gene state is
\begin{equation}
\frac{P_{ss}(x)}{\text{Poiss}(x,\mu)}  =  1 + \sum_{k = 2, 4, ...}^{\infty} (\Delta \alpha)^k  \frac{c_3^{k/2}}{(s + \gamma) 2 \gamma (s + 3 \gamma) \cdots [s + (k-1)\gamma] k \gamma} C_k(x, \mu)  \ .
\end{equation}
Rewriting it in terms of Gamma functions, we have the formula
\begin{equation}
\frac{P_{ss}(x)}{\text{Poiss}(x,\mu)} = \Gamma\left( \frac{s}{2\gamma} + \frac{1}{2} \right) \sum_{m = 0}^{\infty} \left[  \frac{\Delta \alpha}{4 \gamma}   \right]^{2m}  \frac{ C_{2m}(x, \mu)}{m! \ \Gamma\left( \frac{s}{2\gamma} + \frac{1}{2} + m \right)} \ ,
\end{equation}
which makes it easy to compute that the steady state (analytic) generating function is
\begin{equation}
\psi_{ss}(g) = e^{\mu(g-1)} \Gamma\left( \frac{s}{2\gamma} + \frac{1}{2} \right) \left[ \frac{4\gamma}{\Delta \alpha (g-1)} \right]^{\frac{s}{2\gamma} - \frac{1}{2}} I_{\frac{s}{2\gamma} - \frac{1}{2}}\left( \frac{\Delta \alpha}{2 \gamma} (g-1)  \right) 
\end{equation}
where $I_{\nu}(z)$ is the modified Bessel function of the first kind. 

In the case of equal switching rates, for sufficiently different $\alpha_1$ and $\alpha_2$ the distribution is generically bimodal with unequally sized peaks. In the limit of very large $\alpha_1$ and $\alpha_2$ (i.e. the continuous concentration limit discussed in Sec. \ref{sec:continuous}), this asymmetry disappears, and $P_{ss}$ becomes completely symmetric about the effective mean $\mu$ (see Fig. \ref{fig:continuous_fig}). 
 
\subsection{Very unequal switching rates} \label{sec:very_unequal}

Suppose that $k_{12}$ and $k_{21}$ are very different---as is often the case in practice with bursty transcription, where one gene state is considered `active' and the other `inactive'. Without loss of generality, let $k_{21} \gg k_{12}$, so that the $1 \to 2$ transition happens much more readily than the $2 \to 1$ transition. Define $r := k_{12}/k_{21}$ and note $r \ll 1$. In this regime, we are interested in finding the steady state probability distribution to first order in $r$. Note that
\begin{equation}
\begin{split}
c_3 &= \frac{k_{12} k_{21}}{(k_{12} + k_{21})^2} = \frac{r}{1 + 2 r + r^2} \approx r \\
c_4 &= \frac{k_{21} - k_{12}}{k_{21} + k_{12}} = \frac{1 - r}{1 + r} \approx 1 - 2 r
\end{split}
\end{equation}
to first order approximation in $r$. Hence, $c_3 \sim r$ and $c_4 \sim 1$ at leading order. Then the dominant diagrams contributing to the steady state probability sum have first step $0 \to 1$, all intermediate steps $1 \to 1$, and last step $1 \to 0$. 

There exists such a diagram for all $k \geq 2$, and its value is always (to first order in $r$) proportional to $r$. We will call these `balance beam' diagrams (see Fig. \ref{fig:feynman_special} for a visual). Then we have
\begin{equation}
\begin{split}
\sum_{i_1, ..., i_{k-1} = 0, 1} \ \frac{T_{0 i_{k-1}}}{(k \gamma)} \cdots \frac{T_{i_1 0}}{(\gamma + i_1 s)}  &\approx  \frac{T_{0 1}}{(k \gamma)} \cdots \frac{T_{11}}{(2 \gamma + s)} \frac{T_{1 0}}{(\gamma + s)} \\
&\approx \frac{r}{\gamma^k} \ \frac{1}{[(s/\gamma) + 1][(s/\gamma) + 2]\cdots[(s/\gamma) + (k-1)] k}  \\
&= \frac{r}{\gamma^k} \ \frac{(s/\gamma)}{\left( \frac{s}{\gamma} \right)_k k}  \ ,
\end{split}
\end{equation}
so that our approximate formula for $P_{ss}(x)$ in the very unequal switching rates regime reads
\begin{equation} \label{eq:pss_almost_nb}
\begin{split}
\frac{P_{ss}(x)}{\text{Poiss}(x,\mu)} &\approx  1 + \frac{s r}{\gamma} \sum_{k = 2}^{\infty} \left( \frac{\Delta \alpha}{\gamma} \right)^k  \ C_k(x, \mu) \ \frac{1}{\left( \frac{s}{\gamma} \right)_k k}   \ .
\end{split}
\end{equation}
Like the negative binomial distribution, this distribution is somewhat heavy-tailed. This extra variance is thought to be one of the more dramatic consequences of gene switching for empirical single cell RNA counts distributions. 

It should also be noted that this `dominant' diagram kind of analysis recalls what one often does in field theory, e.g. in studies of condensed matter systems \cite{mattuck1992}. This is one of the advantages of this Feynman diagram way of thinking: various approximations can be justified by observing in some regime that certain diagrams are dominant, and others can be ignored. 

The very unequal switching rates limit is closely related to the \textit{bursty} / negative binomial limit we mentioned in Sec. \ref{sec:brief_guide}. Recall that the relevant parameter conditions are
\begin{equation}
\begin{split}
\alpha_2 &= 0 \\
\alpha_1 \to \infty, & \ k_{21} \to \infty \\
b &:= \frac{\alpha_1}{k_{21}} \text{ held fixed} \\
\omega &:= \frac{k_{12}}{\gamma} \text{ held fixed }
\end{split}
\end{equation}
where $b$ is called the burst size and $\omega$ is called the burst frequency. Unfortunately, because this limit requires keeping $\alpha_1/k_{21}$ held fixed, we cannot simply take the relevant limit of Eq. \ref{eq:pss_almost_nb}. Instead, we would have to start again with Eq. \ref{eq:bds_Pss_marg}. 

It turns out that a somewhat slicker way to recover the negative binomial limit is to use the hypergeometric form of the (analytic) generating function, which we derive from Eq. \ref{eq:bds_Pss_marg} in Appendix \ref{sec:app_consistency}. For $\alpha_2 = 0$, it reads
\begin{equation} \label{eq:no_a2}
\psi_{ss}(g) = \sum_{n = 0}^{\infty} \frac{\left[ \frac{\alpha_1}{\gamma} (g-1) \right]^n}{n!} \frac{\left( \frac{k_{12}}{\gamma} \right)_n}{\left( \frac{s}{\gamma} \right)_n} \ .
\end{equation}
First, note that
\begin{equation}
\begin{split}
\left( \frac{s}{\gamma} \right)_n &= \frac{(k_{12} + k_{21})}{\gamma} \left(  \frac{(k_{12} + k_{21})}{\gamma} + 1 \right) \cdots \left( \frac{(k_{12} + k_{21})}{\gamma} + n - 1 \right) \\
&\approx \frac{k_{21}}{\gamma} \left(  \frac{k_{21}}{\gamma} + 1 \right) \cdots \left( \frac{k_{21}}{\gamma} + n - 1 \right) \\
&\approx \left( \frac{k_{21}}{\gamma} \right)^n 
\end{split}
\end{equation}
in the large $k_{21}$ limit. Taking $\alpha_1$ and $k_{21} $ large while keeping $b$ and $\omega$ held fixed, 
\begin{equation}
\begin{split}
\psi_{ss}(g) &\approx \sum_{n = 0}^{\infty} \frac{\left[ \frac{\alpha_1}{\gamma} (g-1) \right]^n}{n!} \frac{\left( \frac{k_{12}}{\gamma} \right)_n}{\left( \frac{k_{21}}{\gamma} \right)^n} \\
&= \sum_{n = 0}^{\infty} \frac{\left[ b (g-1) \right]^n}{n!} \left( \omega \right)_n \\
&= \left[ \frac{1}{1 - b ( g - 1 )} \right]^{\omega}
\end{split}
\end{equation}
which is precisely the generating function of a negative binomial distribution with $p = b/(1 + b)$ and $r = \omega$.

\subsection{Switching much faster than degradation}
\label{sec:switching_faster_deg}

If gene switching tends to happen much more frequently than degradation, so that $s \gg \gamma$, we can approximate the Feynman denominators $(m \gamma + i_m s) \approx i_m s$ where $i_m = 1$, yielding an asymptotic series in powers of $1/s$: 
\begin{equation} \label{eq:bds_Pss_fastswitch}
\begin{split}
\frac{P_{ss}(x)}{\text{Poiss}(x,\mu)} &\approx  1 + \frac{1}{s} \left[ (\Delta \alpha)^2 C_2 \frac{c_3}{2 \gamma} \right] + \frac{1}{s^2} \left[ (\Delta \alpha)^3 C_3 \frac{c_3 c_4}{3 \gamma} +  (\Delta \alpha)^4 C_4 \frac{c_3^2}{8 \gamma^2}  \right] + \cdots \ . 
\end{split}
\end{equation}
In the $s/\gamma \to \infty$ limit, the distribution becomes exactly Poisson with mean $\mu$. Intuitively, gene switching happens so fast in this regime that the gene product only `feels' the effective transcription rate given by $\alpha_{eff}$, and the problem reduces to the familiar chemical birth-death process.

\clearpage

\subsection{Switching much slower than degradation}
\label{sec:slow_switching}

If degradation tends to happen much more frequently than gene switching, so that $\gamma \gg s$, we can approximate the Feynman denominators $(m \gamma + i_m s) \approx m \gamma$, yielding
\begin{equation} \label{eq:pmix_approx_base}
\begin{split}
\frac{P_{ss}(x)}{\text{Poiss}(x,\mu)} &\approx  1 + \sum_{k = 1}^{\infty} \left[ \frac{\Delta \alpha}{\gamma} \right]^k \frac{1}{k!}  \ C_k(x, \mu) \sum_{i_1, ..., i_{k-1} = 0, 1} \ T_{0 i_{k-1}} \cdots T_{i_1 0} \\
&\stackrel{00}{=} \sum_{k=0}^{\infty} \left[ \frac{\Delta \alpha}{\gamma} T \right]^k \frac{1}{k!}  \ C_k(x, \mu)
\end{split}
\end{equation}
where the notation $\stackrel{00}{=}$ means that the LHS is the $00$ (upper left) entry of the $2 \times 2$ matrix described by the RHS. We can sum this in closed form using the known formula for the generating function of the Charlier polynomials (c.f. Eq. \ref{eq:charlier_genfunc} in Appendix \ref{sec:app_charlier}) to obtain the beautiful approximate formula
\begin{equation}
\frac{P_{ss}(x)}{\text{Poiss}(x,\mu)} \stackrel{00}{\approx} \left[ 1 + \frac{\Delta \alpha}{\gamma \mu} T \right]^{x} e^{- (\Delta \alpha) T/\gamma} \ .
\end{equation}
Recall from Sec. \ref{sec:brief_guide} that, when $\gamma \gg s$, we expect the steady state probability distribution to look like a mixture of two Poisson distributions (with means $\alpha_1/\gamma$ and $\alpha_2/\gamma$). Because switching is slow compared to the dynamics of production and degradation, on short time scales the system looks like a birth-death process; the effect of switching is to move from one Poisson distribution to the other.  

Although this formula appears superficially different, it turns out to be \textit{exactly the same} as a Poisson mixture. To see why, note that we can write
\begin{equation}
T = \begin{pmatrix} 0 & 1 \\ f (1 - f) & 1 - 2 f  \end{pmatrix}
\end{equation}
where $f := k_{12}/s$. Diagonalizing $T$, we find that
\begin{equation}
T = \begin{pmatrix} 1 & 1 \\ - f & 1 - f \end{pmatrix} \begin{pmatrix} - f & 0 \\ 0 & 1 - f \end{pmatrix} \begin{pmatrix} 1 - f & - 1 \\ f & 1 \end{pmatrix} \ .
\end{equation}
This representation makes it easy to take powers of $T$, which we can use to evaluate Eq. \ref{eq:pmix_approx_base} in a different way. We have
\begin{equation}
\left( T^k \right)_{00} = f (1 - f)^k + (1 - f) (-f)^k 
\end{equation}
which means
\begin{equation}
\begin{split}
\frac{P_{ss}(x)}{\text{Poiss}(x,\mu)} &\approx  \sum_{k=0}^{\infty} \left[ \frac{\Delta \alpha}{\gamma} \right]^k \left\{ \ f (1 - f)^k + (1 - f) (-f)^k \ \right\} \frac{1}{k!}  \ C_k(x, \mu) \\
&=  f \sum_{k=0}^{\infty} \left[ \frac{\Delta \alpha}{\gamma} (1 - f) \right]^k  \frac{1}{k!}  \ C_k(x, \mu) + (1 - f) \sum_{k=0}^{\infty} \left[ \frac{\Delta \alpha}{\gamma} (- f) \right]^k  \frac{1}{k!}  \ C_k(x, \mu) \\
&=  f \left( 1 + \frac{(1 - f) \frac{\Delta \alpha}{\gamma}}{\mu} \right)^x e^{- (1 - f) \frac{\Delta \alpha}{\gamma}} + (1 - f) \left( 1 - \frac{f \frac{\Delta \alpha}{\gamma}}{\mu} \right)^x e^{f \frac{\Delta \alpha}{\gamma}} 
\end{split}
\end{equation}
where we used the generating function of the Charlier polynomials (Eq. \ref{eq:charlier_genfunc}) in the last step. Simplifying, we obtain
\begin{equation}
\begin{split}
P_{ss}(x) &\approx \frac{\mu^x}{x!} e^{- \mu} \left[ \ \frac{k_{12}}{s} \left( 1 + \frac{\frac{k_{21}}{s} \frac{\Delta \alpha}{\gamma}}{\mu} \right)^x e^{- \frac{k_{21}}{s} \frac{\Delta \alpha}{\gamma}} + \frac{k_{21}}{s} \left( 1 - \frac{\frac{k_{12}}{s} \frac{\Delta \alpha}{\gamma}}{\mu} \right)^x e^{\frac{k_{12}}{s} \frac{\Delta \alpha}{\gamma}}  \  \right] \\
&= \frac{k_{12}}{s} \frac{\left( \alpha_1/\gamma \right)}{x!} e^{- \alpha_1/\gamma} + \frac{k_{21}}{s} \frac{\left( \alpha_2/\gamma \right)}{x!} e^{- \alpha_2/\gamma}  
\end{split}
\end{equation}
i.e. a Poisson mixture. 


\clearpage

\section{The continuous concentration limit}
\label{sec:continuous}

Remarkably, by solving the birth-death-switching CME we have also solved a qualitatively very different problem almost for free: the problem of coupling an Ornstein-Uhlenbeck-like process to a switching gene, which corresponds to the continuous concentration limit of the discrete birth-death process. The same ladder operator/diagrammatic approach we have just followed works, with only minor changes. 

The chemical birth-death process with additive noise \cite{vastolaADD} is governed by the Fokker-Planck equation
\begin{equation} \label{eq:FP_bda}
\frac{\partial P(x,t)}{\partial t} = - \frac{\partial}{\partial x} \left[ (\alpha - \gamma x) P(x,t) \right] + \frac{\sigma^2}{2} \frac{\partial^2 P(x,t)}{\partial x^2} 
\end{equation}
where $P(x, t)$ is the probability that the concentration of $X$ molecules in the system is $x \in \mathbb{R}$ at time $t$, and where $\sigma > 0$ is called the additive noise coefficient. It is equivalent to the Ornstein-Uhlenbeck process up to a simple change of variables, which is most transparently seen by noting that its stochastic trajectories follow the stochastic differential equation (SDE)
\begin{equation}
\dot{x} = \alpha - \gamma x + \sigma \eta(t)
\end{equation}
with $\eta(t)$ a Gaussian white noise term. It is biologically interesting because, in the regime where typical molecule numbers are large, it is reasonable to approximate the discrete molecule number dynamics of the chemical birth-death process as the continuous concentration dynamics described by this Fokker-Planck equation\footnote{Negative concentrations are allowed but are overwhelmingly improbable in the regime where this continuous model is used, i.e. assuming $\alpha/\gamma \gg 1$ so that typical molecule numbers are large.}.

If we coupled this process to a switching gene, we would obtain the Fokker-Planck equations
\begin{equation} \label{eq:FP_switching}
\begin{split}
\frac{\partial P(x, 1, t)}{\partial t} &= - k_{21} P(x, 1, t) + k_{12} P(x, 2, t) - \frac{\partial}{\partial x} \left[ (\alpha_1 - \gamma x) P(x, 1, t) \right] + \frac{\sigma^2}{2} \frac{\partial^2 P(x, 1, t)}{\partial x^2} \\
\frac{\partial P(x, 2, t)}{\partial t} &= k_{21} P(x, 1, t) - k_{12} P(x, 2, t) - \frac{\partial}{\partial x} \left[ (\alpha_2 - \gamma x) P(x, 2, t) \right] + \frac{\sigma^2}{2} \frac{\partial^2 P(x, 2, t)}{\partial x^2} 
\end{split}
\end{equation}
with $P(x, i, t)$ denoting the probability that the system has concentration $x \in \mathbb{R}$ and gene state $i \in \{ 1, 2 \}$ at time $t$. 

Analogously to before, we can first consider the problem without a switching gene (Eq. \ref{eq:FP_bda}), and define the generating function via
\begin{equation}
\ket{\phi} = \int_{-\infty}^{\infty} dx \ P(x, t) \ket{x} \ .
\end{equation}
If we introduce operators $\hat{x}, \hat{p}$ that act as
\begin{equation}
\begin{split}
\hat{x} \ket{\phi} &:= \int_{-\infty}^{\infty} dx \ x \ c(x) \ket{x} \\
 \hat{p} \ket{\phi} &:= \int_{-\infty}^{\infty} dx \ - \frac{\partial c(x)}{\partial x} \ket{x} 
\end{split}
\end{equation}
on a general state $\ket{\phi}$, we can write the equation of motion satisfied by the generating function in the form
\begin{equation}
\frac{\partial \ket{\psi}}{\partial t} = \hat{H}_{bda} \ket{\psi} 
\end{equation}
where the Hamiltonian operator is defined as
\begin{equation}
\hat{H}_{bda} = \hat{p} \left( \alpha - \gamma \hat{x} \right) + \frac{\sigma^2}{2} \ \hat{p}^2 \ .
\end{equation}
We can go on to derive ladder operators
\begin{equation}
\begin{split}
\hat{A} &:= \frac{1}{\sqrt{\sigma^2/2\gamma}} \left[ \hat{x} - \mu - \frac{\sigma^2}{2 \gamma} \hat{p} \right] \\
\hat{A}^+ &:= \sqrt{\frac{\sigma^2}{2 \gamma}} \ \hat{p} 
\end{split}
\end{equation}
whose properties match those discussed in Sec. \ref{sec:reptheory_bd}, which includes that we can write the Hamiltonian in terms of them as
\begin{equation}
\hat{H} = - \gamma \ \hat{A}^+ \hat{A} \ .
\end{equation}
With that done, we have ladder operators for the continuous birth-death problem, as well as the ones derived in Sec. \ref{sec:reptheory_gene} for the switching gene problem. We can pose the coupled problem in terms of them, as in Sec. \ref{sec:reptheory_couple}, via
\begin{equation}
\begin{split}
\hat{H} &= - s \hat{B}^+ \hat{B} + \hat{p} \left[ \hat{\alpha} - \gamma \hat{x} \right]  + \frac{\sigma^2}{2} \hat{p}^2 \\
&= - s \hat{B}^+ \hat{B} - \gamma \hat{A}^+ \hat{A} + (\alpha_1 - \alpha_2) \ \hat{p} \ \left[ \hat{B} + \frac{k_{12} k_{21}}{s^2} \hat{B}^+ + \frac{k_{21} - k_{12}}{s}  \hat{B}^+ \hat{B} \right] \\
&= - s \hat{B}^+ \hat{B} - \gamma \hat{A}^+ \hat{A} + \frac{(\alpha_1 - \alpha_2)}{\sqrt{\sigma^2/2\gamma}} \ \hat{A}^+ \ \left[ \hat{B} + \frac{k_{12} k_{21}}{s^2} \hat{B}^+ + \frac{k_{21} - k_{12}}{s}  \hat{B}^+ \hat{B} \right] \\
&= \hat{H}_s + \hat{H}_{bda} + \hat{H}_{int} \ .
\end{split}
\end{equation}
As before, we can simplify notation by defining the quantities
\begin{equation}
\begin{split}
d &:= \frac{\alpha_1 - \alpha_2}{\sqrt{\sigma^2/2\gamma}} \\
c_3 &:=  \frac{k_{12} k_{21}}{s^2} \\
c_4 &:= \frac{k_{21} - k_{12}}{s}
\end{split}
\end{equation}
so that the interaction Hamiltonian $\hat{H}_{int}$ can be written as
\begin{equation}
\hat{H}_{int} = d \ \hat{A}^+ \left[ \hat{B} + c_3 \hat{B}^+ + c_4 \hat{B}^+ \hat{B} \right] \ .
\end{equation}
But this is precisely the same interaction Hamiltonian we solved earlier (c.f. Eq. \ref{eq:Hint_notation})! Because the method from here on out works in Hilbert space, rather than molecule number space or concentration space, we can go on derive Feynman rules of exactly the same form. 

\begin{figure}[!htb]
\center{\includegraphics[scale=1.]{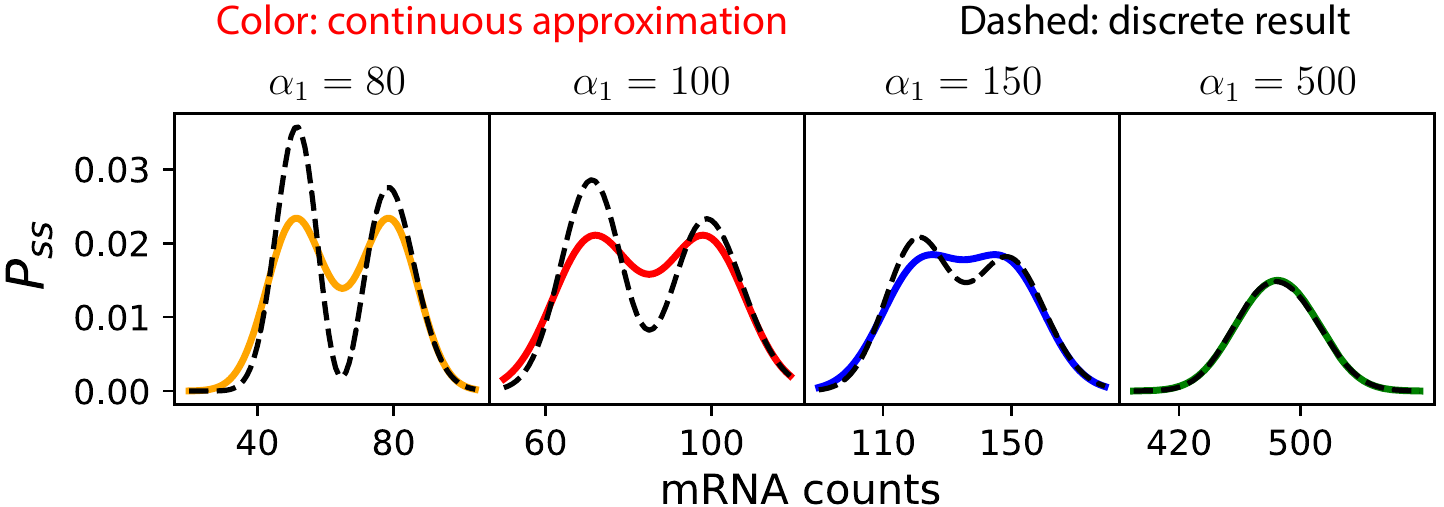}}
\caption{\label{fig:continuous_fig} 
Comparison of the continuous approximation to the discrete birth-death-switching result for equal switching rates as $\alpha_1$ is varied. $\alpha_2 = \alpha_1 - 30$, $\gamma = 1$, $k_{12} = k_{21} = 0.1$. The continuous result is symmetric about the effective mean $\mu$ in all cases; the discrete result approaches the continuous result as $\alpha_1$ becomes large.}
\end{figure}

We must do something slightly different only when we go from the generating function to the probability distribution. The analogue to the Euclidean product (c.f. Eq. \ref{eq:bd_euclidean}) is 
\begin{equation}
\braket{\phi_1}{\phi_2}_{Eu} := \int dx \ c_1^*(x) c_2(x) 
\end{equation}
and the relevant results (given no gene switching) can be summarized as
\begin{equation}
\begin{split}
\braket{x}{\psi}_{Eu} &= P(x, t) \\
\braket{x}{n}_{Eu} &= \sqrt{\frac{1}{2^n n!}} H_n\left( \sqrt{\frac{\gamma}{\sigma^2}} (x - \mu) \right) \cdot \text{Gauss}\left(x, \mu, \frac{\sigma^2}{2 \gamma} \right) \\
\text{Gauss}(x, m, s^2) &:= \frac{1}{\sqrt{2 \pi s^2}} e^{- \frac{(x - m)^2}{2 s^2}} 
\end{split}
\end{equation}
where $H_n$ denotes the $n$th Hermite polynomial. Continuing as before, we find\footnote{See \cite{thomas2015} for another interesting series expansion involving the Hermite polynomials. Thomas and Grima study higher-order terms in the system size expansion of the CME; the Hermite polynomials seem to appear essentially due to fact that (most) arbitrary CMEs can be locally approximated as a multivariate birth-death process with additive noise.}
\begin{equation} \label{eq:bdas_Pss}
\begin{split}
\frac{P_{ss}(x, \vec{g})}{\text{Gauss}\left(x, \mu, \frac{\sigma^2}{2 \gamma} \right)} &=  \vec{v}_0 + \sum_{k = 1}^{\infty} \left[ \frac{\Delta \alpha}{\sqrt{\sigma^2/\gamma}} \right]^k  \ H_k\left( \sqrt{\frac{\gamma}{\sigma^2}} (x - \mu) \right) \sum_{i_1, ..., i_{k} = 0, 1} \ \frac{T_{i_k i_{k-1}}}{(k \gamma + i_k s)} \cdots \frac{T_{i_1 0}}{(\gamma + i_1 s)}  \ \vec{v}_{i_k} \ .
\end{split}
\end{equation}
If we marginalize over gene state, we obtain
\begin{equation} \label{eq:bdas_Pss_marg}
\begin{split}
\frac{P_{ss}(x)}{\text{Gauss}\left(x, \mu, \frac{\sigma^2}{2 \gamma} \right)} &=  1 + \sum_{k = 2}^{\infty} \left[ \frac{\Delta \alpha}{\sqrt{\sigma^2/\gamma}} \right]^k  \ H_k\left( \sqrt{\frac{\gamma}{\sigma^2}} (x - \mu) \right) \sum_{i_1, ..., i_{k-1} = 0, 1} \ \frac{T_{0 i_{k-1}}}{(k \gamma)} \cdots \frac{T_{i_1 0}}{(\gamma + i_1 s)}   \ .
\end{split}
\end{equation}
One easily notes that the only difference between these results and the ones for the discrete birth-death process are the substitutions
\begin{equation}
\begin{split}
\text{Poiss}(x, \mu) &\to \text{Gauss}\left(x, \mu, \frac{\sigma^2}{2 \gamma} \right) \\
C_k(x, \mu) &\to \left[ \frac{1}{\sqrt{\sigma^2/\gamma}} \right]^k \ H_k\left( \sqrt{\frac{\gamma}{\sigma^2}} (x - \mu) \right) \ .
\end{split}
\end{equation}
Because the birth-death process with additive noise (with the specific additive noise constant $\sigma = \sqrt{2 \alpha}$) is the large $\mu$ limit of the discrete birth-death process (see Fig. \ref{fig:continuous_fig}), there is an alternative way to get this result. One merely needs to know that
\begin{equation}
C_k(x, \mu) \xrightarrow{\mu \gg 1} \left[ \frac{1}{\sqrt{2 \mu}} \right]^k H_k\left( \frac{x - \mu}{\sqrt{2 \mu}} \right)
\end{equation}
so that Eq. \ref{eq:bds_Pss_marg} becomes Eq. \ref{eq:bdas_Pss_marg} (with $\sigma = \sqrt{2 \alpha}$). 

It is interesting to note that problems of similar form appear elsewhere. For example, by specializing the parameter values of this problem, one finds that Eq. \ref{eq:bdas_Pss} and Eq. \ref{eq:bdas_Pss_marg} can exactly describe the solution of an active matter run-and-tumble motion problem \cite{gm2020}. The precise correspondence is described in Table \ref{table:runandtumble} below.
\begin{table}[ht!]
\centering
 \begin{tabular}{| c | c |} 
 \hline
Gene regulation problem & Run-and-tumble problem \\ [0.5ex] 
 \hline
Concentration $x$ & Position $x$ \\ 
Gene state $i \in \{1, 2\}$ & Right-moving or left-moving \\
$P(x, 1, t)$ and $P(x, 2, t)$ & $P_{\phi}(x, t)$ and $P_{\psi}(x, t)$ \\
Transcription rates $\alpha_1, \alpha_2$ & Self-propulsion speeds $\frac{w}{2}, - \frac{w}{2}$ \\ 
Degradation rate $\gamma$ & Spring constant $k$ \\
Gene switching rates $k_{12}, k_{21}$ & Direction switching rates $\frac{\alpha}{2}, \frac{\alpha}{2}$ \\
Effective mean concentration $\mu = \frac{\alpha_1 k_{12} + \alpha_2 k_{21}}{\gamma (k_{12} + k_{21})}$ & Effective mean position $\mu = 0$ \\
 \hline
 \end{tabular}
 \caption{The parameter correspondence between the continuous birth-death-switching problem considered here, and the run-and-tumble problem considered by Garcia-Millan and Pruessner \cite{gm2020}.}
\label{table:runandtumble}
\end{table}

\clearpage

\section{Generalization to multistep splicing}
\label{sec:multistep}

Fortuitously, our approach to solving the birth-death-switching problem---a problem that has been solved before \cite{iyer2009, huang2015, cao2018}, albeit using very different methods---straightforwardly generalizes, so that we can solve a significantly more complicated problem that has not been solved before. In this section, we consider more realistic transcription dynamics, involving an arbitrary number of splicing steps that occur in a fixed order, coupled to a switching gene. 

Our plan of attack is the same as the one detailed in Sec. \ref{sec:roadmap}. The ladder operators for pure gene switching are the same as before, so we will first find the ladder operators for the dynamics of transcription and multistep splicing given a constitutively active gene. The list of reactions and CME for the coupled problem are as stated in Sec. \ref{sec:results_multistep} (c.f. Eq. \ref{eq:CME_multistep}).

\subsection{Representation theory of multistep splicing}

Assuming a constitutively active gene, the reaction list for transcription and multistep splicing (with $N$ splicing steps that occur in a fixed order) reads
\begin{equation} \label{eq:splice_rxns}
\begin{split}
\varnothing &\xrightarrow{\alpha} X_0  \\
X_0 &\xrightarrow{\beta_0} X_1 \\
X_1 &\xrightarrow{\beta_1} X_2 \\
& \ \vdots  \\
X_{N-1} &\xrightarrow{\beta_{N-1}} X_N \\
X_N &\xrightarrow{\beta_N} \varnothing  
\end{split}
\end{equation}
where $\alpha$ parameterizes the transcription rate, $\beta_i$ parameterizes the rate of the $i$th splicing step (for $0 \leq i < N$), and $\beta_N$ parameterizes the mature RNA's degradation rate. For simplicity, we assume that the $\beta_i$ are all distinct. Denote the state of this system using the vector $\mathbf{x} := (x_0, x_1, ..., x_N) \in \mathbb{N}^N$. If we use $\boldsymbol{\epsilon}_j$ to denote the vector with a $1$ in the $j$th place and zeros elsewhere, we can write the CME for this system as
\begin{equation} 
\begin{split}
\frac{\partial P(\mathbf{x}, t)}{\partial t} =&  \ \alpha \left[ P(\mathbf{x} - \boldsymbol{\epsilon}_0, t) - P(\mathbf{x}, t) \right] \\
& + \sum_{j = 0}^{N-1} \beta_j \left[ (x_j + 1) P(\mathbf{x} + \boldsymbol{\epsilon}_j - \boldsymbol{\epsilon}_{j+1}, t) - x_j P(\mathbf{x}, t) \right] \\
& + \beta_N \left[ (x_N + 1) P(\mathbf{x} + \boldsymbol{\epsilon}_N, t) - x_N P(\mathbf{x}, t) \right] \ .
\end{split}
\end{equation}
As before, we will reframe the problem in terms of a generating function in a certain Hilbert space, where the dynamics are completely determined by a Hamiltonian operator. In terms of the Grassberger-Scheunert \cite{grassberger1980, vastolaDP} creation operators ($\hat{a}_i^+ := \hat{\pi}_i - 1$), the Hamiltonian operator $\hat{H}_{ms}$ of the multistep splicing problem reads
\begin{equation}
\begin{split}
\hat{H}_{ms} &:= \alpha \ \hat{a}_0^+ + \beta_0 \left( \hat{a}_1^+ - \hat{a}_0^+ \right) \hat{a}_0 + \beta_1 \left( \hat{a}_2^+ - \hat{a}_1^+ \right) \hat{a}_1 + \cdots + \beta_{N-1} \left( \hat{a}_N^+ - \hat{a}_{N-1}^+ \right) \hat{a}_{N-1} - \beta_N \hat{a}_N^+ \hat{a}_N \\
&= \alpha \ \hat{a}_0^+ + \sum_{j = 0}^{N-1} \beta_j \left( \hat{a}_{j+1}^+ - \hat{a}_j^+ \right) \hat{a}_j  - \beta_N \hat{a}_N^+ \hat{a}_N \ .
\end{split}
\end{equation}
We are seeking operators satisfying the commutation relations (c.f. Eq. \ref{eq:comm_bd})
\begin{equation} \label{eq:comm_bd_H}
\begin{split}
[\hat{A}, \hat{H}_{ms}] &= - \lambda \hat{A} \\
[\hat{A}^+, \hat{H}_{ms}] &= \lambda \hat{A}^+
\end{split}
\end{equation}
where $\lambda > 0$ is some constant with units of inverse time. A reasonable strategy is to compute many commutators by hand, and combine them by trial and error in order to construct the desired operators. Some commutator results that are useful for this purpose are:
\begin{equation}
\begin{split}
[\hat{a}_j^+, \hat{H} ] &= \beta_j \ \hat{a}_j^+ - \beta_j \ \hat{a}_{j+1}^+ \hspace{1 in} j = 0, ..., N - 1 \\
[\hat{a}_N^+, \hat{H} ] &= \beta_N \ \hat{a}_N^+  \\
[\hat{a}_j, \hat{H} ] &= - \beta_j \ \hat{a}_j + \beta_{j-1} \ \hat{a}_{j-1} \hspace{1 in} j = 1, ..., N \\
[\hat{a}_0, \hat{H} ] &= \alpha - \beta_0 \hat{a}_0 = - \beta_0 \ \left( \hat{a}_0 - \frac{\alpha}{\beta_0} \right)  \ .
\end{split}
\end{equation}
From the above, it is immediately clear that $\hat{a}^+_N$ is an up ladder operator (with constant $\beta_N$), and $\hat{a}_0 - (\alpha/\beta_0)$ is a down ladder operator (with constant $\beta_0$). 

To find the rest, we can note that $\hat{a}_{N-1}^+$ is an up ladder operator up to a correction involving $\hat{a}_N^+$; making that correction allows us to find an up ladder operator (with constant $\beta_{N-1}$) that is a linear combination of $\hat{a}_{N-1}^+$ and $\hat{a}_N^+$. Similarly, we can find an up ladder operator (with constant $\beta_{N-2}$) that is a linear combination of $\hat{a}_{N-2}^+$, $\hat{a}_{N-1}^+$, and $\hat{a}_N^+$. In general, we have up ladder operators $\hat{A}_i^+$ (for $i = 0, ..., N$), where
\begin{equation}
\begin{split}
\hat{A}_N^+ &:= \hat{a}_N^+ \\
\hat{A}_{N-1}^+ &:= \hat{a}_{N-1}^+ + \frac{\beta_{N-1}}{\beta_N - \beta_{N-1}} \ \hat{a}_N^+ \\
\hat{A}_{N-2}^+ &:= \hat{a}_{N-2}^+ + \frac{\beta_{N-2}}{\beta_{N-1} - \beta_{N-2}} \ \hat{a}_{N-1}^+ + \frac{\beta_{N-1} \ \beta_{N-2}}{(\beta_{N-1} - \beta_{N-2}) (\beta_N - \beta_{N-2})} \ \hat{a}_N^+ \\
\hat{A}_i^+ &:= \hat{a}_i^+ + \sum_{j = i + 1}^N \frac{\beta_i \cdots \beta_{j-1}}{(\beta_j - \beta_i) \cdots (\beta_{i+1} - \beta_i)} \ \hat{a}_j^+ \ \hspace{1in} i = 0, 1, ..., N .
\end{split}
\end{equation}
One can apply the same argument to derive the down ladder operators, starting from $\hat{A}_0$ and constructing $\hat{A}_N$ last. We obtain
\begin{equation}
\begin{split}
\hat{A}_0 &:= \hat{a}_0 - \frac{\alpha}{\beta_0} \\
\hat{A}_{1} &:= \hat{a}_{1} - \frac{\beta_{0}}{\beta_1 - \beta_0} \ \hat{a}_0 + \frac{\alpha \ \beta_0}{\beta_1 (\beta_1 - \beta_0)} \\
\hat{A}_{2} &:= \hat{a}_{2} - \frac{\beta_{1}}{\beta_{2} - \beta_{1}} \ \hat{a}_{1} + \frac{\beta_{0} \ \beta_{1}}{(\beta_{2} - \beta_{1}) (\beta_2 - \beta_{0})} \ \hat{a}_0 - \frac{\alpha \ \beta_0 \ \beta_1}{\beta_2 (\beta_2 - \beta_0) (\beta_2 - \beta_1)} \\
\hat{A}_i &:= \hat{a}_i + \frac{(-1)^{i+1} \alpha \ \beta_0 \cdots \beta_{i-1}}{\beta_i (\beta_i - \beta_{i-1}) \cdots (\beta_i - \beta_0)} + \sum_{j = 0}^{i - 1} (-1)^{i - j}\frac{\beta_j \cdots \beta_{i-1}}{(\beta_i - \beta_{i-1}) \cdots (\beta_{i} - \beta_j)} \  \hat{a}_j \  .
\end{split}
\end{equation}
These ladder operators satisfy the important commutation properties:
\begin{equation}
\begin{split}
[ \hat{A}_j, \hat{H}_{ms} ] &= - \beta_j \ \hat{A}_j  \\
[ \hat{A}_j^+, \hat{H}_{ms} ] &= \beta_j \ \hat{A}_j^+  \\
[ \hat{A}_i, \hat{A}_j^+ ] &= \delta_{i, j}
\end{split}
\end{equation}
for all $i, j = 0, 1, ..., N$. We can write the Hamiltonian in terms of them as
\begin{equation}
\hat{H}_{ms} = - \beta_0 \ \hat{A}_0^+ \hat{A}_0 - \beta_1 \ \hat{A}_1^+ \hat{A}_{1} - \cdots -  \beta_N \ \hat{A}_N^+ \hat{A}_N \ ,
\end{equation}
which looks like the Hamiltonian of $(N+1)$ uncoupled harmonic oscillators or non-interacting bosons. The energies of this Hamiltonian are
\begin{equation}
E_{n_0, n_1, ..., n_N} = \beta_0 \ n_0 + \beta_1 \ n_1 + \cdots + \beta_N \ n_N 
\end{equation}
with each $n_j \in \mathbb{N}$. Using ladder operators or other methods, it is clear that the ground state $\ket{\mathbf{0}}$ (which has energy $E_{0, ..., 0} = 0$) corresponds to the steady state probability distribution
\begin{equation} \label{eq:poiss_general}
\text{Poiss}(\mathbf{x}, \boldsymbol{\mu}) := \frac{\boldsymbol{\mu}^{\mathbf{x}} e^{- \boldsymbol{\mu} \cdot \mathbf{1}}}{\mathbf{x}!} = \frac{\mu_0^{x_0} e^{- \mu_0}}{x_0!} \cdots \frac{\mu_N^{x_N} e^{- \mu_N}}{x_N!}
\end{equation}
where the components of the vector $\boldsymbol{\mu}$ are $\mu_i = \alpha/\beta_i$ for all $0 \leq i \leq N$. 

In what follows, it will be helpful to consider simple cases in order to get intuition for the more general case. The `toy' case of one splicing step ($N = 1$, two distinct species) has ladder operators
\begin{equation}
\begin{split}
\hat{A}_0 &:= \hat{a}_0 - \frac{\alpha}{\beta_0} \\
\hat{A}_{1} &:= \hat{a}_{1} - \frac{\beta_{0}}{\beta_1 - \beta_0} \ \hat{a}_0 + \frac{\alpha \ \beta_0}{\beta_1 (\beta_1 - \beta_0)} \\
\hat{A}_{0}^+ &:= \hat{a}_{0}^+ + \frac{\beta_{0}}{\beta_1 - \beta_{0}} \ \hat{a}_1^+ \\
\hat{A}_1^+ &:= \hat{a}_1^+  \ .
\end{split}
\end{equation}
\clearpage
\noindent Properties like the canonical commutation relations and ladder operator decomposition of the Hamiltonian are straightforward to verify in this case. The case of two splicing steps ($N = 2$, three distinct species) has ladder operators
\begin{equation}
\begin{split}
\hat{A}_0 &:= \hat{a}_0 - \frac{\alpha}{\beta_0} \\
\hat{A}_{1} &:= \hat{a}_{1} - \frac{\beta_{0}}{\beta_1 - \beta_0} \ \hat{a}_0 + \frac{\alpha \ \beta_0}{\beta_1 (\beta_1 - \beta_0)} \\
\hat{A}_{2} &:= \hat{a}_{2} - \frac{\beta_{1}}{\beta_{2} - \beta_{1}} \ \hat{a}_{1} + \frac{\beta_{0} \ \beta_{1}}{(\beta_{2} - \beta_{1}) (\beta_2 - \beta_{0})} \ \hat{a}_0 - \frac{\alpha \ \beta_0 \ \beta_1}{\beta_2 (\beta_2 - \beta_0) (\beta_2 - \beta_1)} \\
\hat{A}_{0}^+ &:= \hat{a}_{0}^+ + \frac{\beta_{0}}{\beta_{1} - \beta_{0}} \ \hat{a}_{1}^+ + \frac{\beta_{0} \ \beta_{1}}{(\beta_{1} - \beta_{0}) (\beta_2 - \beta_{0})} \ \hat{a}_2^+ \\
\hat{A}_{1}^+ &:= \hat{a}_{1}^+ + \frac{\beta_{1}}{\beta_2 - \beta_{1}} \ \hat{a}_2^+ \\
\hat{A}_2^+ &:= \hat{a}_2^+ \ .
\end{split}
\end{equation}
The reader may notice that, in these ladder operators, there are no analogues of the $\sqrt{\mu}$ factors that appear in Eq. \ref{eq:bd_lad_def}. Those factors appeared earlier in order to make $\hat{A}$ and $\hat{A}^+$ Hermitian conjugates with respect to a certain inner product. Because we will only solve for the steady state probability distribution, such an inner product, along with guaranteeing the up and down ladder operators are Hermitian conjugates of one another, is not necessary. Furthermore, it is not clear what the `correct' choice of inner product is in this case, because choosing one directly analogous to Eq. \ref{eq:bd_inner} (but using Eq. \ref{eq:poiss_general} instead of a one-dimensional Poisson distribution) does not seem to work. 

\clearpage

\subsection{Diagrammatic approach to exact solution}

We will now use our ladder operators to reframe the coupled problem as in Sec. \ref{sec:reptheory_couple}, and proceed with a diagrammatic approach to the exact solution as in \ref{sec:diagrammatic}. The Hamiltonian of the coupled problems reads
\begin{equation}
\begin{split}
\hat{H} &= - s \hat{B}^+ \hat{B} + \hat{\alpha} \ \hat{a}_0^+ + \sum_{j = 0}^{N-1} \beta_j \left( \hat{a}_{j+1}^+ - \hat{a}_j^+ \right) \hat{a}_j  - \beta_N \hat{a}_N^+ \hat{a}_N \\
&= - s \hat{B}^+ \hat{B} - \sum_{j = 0}^{N} \beta_j \hat{A}^+_j \hat{A}_j + (\Delta \alpha) \ \hat{a}_0^+ \left[ B + c_3 B^+ + c_4 B^+ B \right] \ .
\end{split}
\end{equation}
All we need to do in order to proceed is to express $\hat{a}_0^+$ in terms of our ladder operators. In the case of one splicing step, this is easy:
\begin{equation}
\hat{a}_0^+ = \hat{A}_0^+ - \frac{\beta_0}{\beta_1 - \beta_0} \hat{A}_1^+
\end{equation}
In case of two splicing steps:
\begin{equation}
\hat{a}_0^+ = \hat{A}_0^+ - \frac{\beta_0}{\beta_1 - \beta_0} \hat{A}_1^+ + \frac{\beta_0 \beta_1}{(\beta_2 - \beta_0) (\beta_2 - \beta_1)} \hat{A}_2^+ \ .
\end{equation}
In general, for $N$ splicing steps, we have
\begin{equation}
\hat{a}_0^+ = \hat{A}_0^+ + \sum_{j=1}^N  (-1)^j \frac{\beta_0 \cdots \beta_{j-1}}{(\beta_j - \beta_{0}) \cdots (\beta_j - \beta_{j-1})} \ \hat{A}_j^+ \ .
\end{equation}
To ease notation, we can write
\begin{equation}
\hat{a}_0^+ = \sum_{j=0}^N q_j \hat{A}_j^+
\end{equation}
\clearpage
\noindent and define the vector $\mathbf{q} = (q_0, ..., q_N)$ via
\begin{equation}
\begin{split}
q_0 &= 1 \\
q_1 &= - \frac{\beta_0}{\beta_1 - \beta_0} \\
q_2 &= \frac{\beta_0 \beta_1}{(\beta_2 - \beta_0) (\beta_2 - \beta_1)} \\
q_j &= (-1)^j \frac{\beta_0 \cdots \beta_{j-1}}{(\beta_j - \beta_{0}) \cdots (\beta_j - \beta_{j-1})}  \ ( \ 1 \leq j \leq N \ ) \ .
\end{split}
\end{equation}
Hence, we have
\begin{equation}
\begin{split}
\hat{H} &= - s \hat{B}^+ \hat{B} - \sum_{j = 0}^{N} \beta_j \hat{A}^+_j \hat{A}_j + (\Delta \alpha) \sum_{j=0}^N q_j \hat{A}_j^+ \left[ B + c_3 B^+ + c_4 B^+ B \right] \\
&= \hat{H}_s + \hat{H}_{ms} + \hat{H}_{int}
\end{split}
\end{equation}
where the interaction Hamiltonian $\hat{H}_{int}$ is defined to be
\begin{equation}
\hat{H}_{int} = (\Delta \alpha) \sum_{j=0}^N q_j \hat{A}_j^+ \left[ B + c_3 B^+ + c_4 B^+ B \right] \ .
\end{equation}
As before, we will work in terms of the naive eigenstates $\ket{\mathbf{n}; g} = \ket{n_0, n_1, ..., n_N; g}$. The action of our ladder operators on these states is easily computed. We seek to calculate the `ground state', which we will write as
\begin{equation}
\ket{\bar{0}} := \ket{0 + \cdots + 0} = \ket{\mathbf{0}; 0} + \sum_{\mathbf{n} \neq \mathbf{0}} \sum_{g = 0, 1} \mathbf{q}^{n} \sqrt{\mathbf{n}!} \ Q(\mathbf{n}, g) \ket{\mathbf{n}; g}
\end{equation}
for some coefficients $Q(\mathbf{n}, g)$. Start by observing that
\begin{equation}
\hat{H} \ket{\mathbf{0}; 0} = (\Delta \alpha) c_3 \sum_{j = 0}^N q_j  \ket{\boldsymbol{\epsilon}_j; 1} \ .
\end{equation}
Next,
\begin{equation}
\begin{split}
\hat{H} \ket{\boldsymbol{\epsilon}_i; 1} =& \  - (s + \beta_i) \ket{\boldsymbol{\epsilon}_i; 1} + (\Delta \alpha) q_i \sqrt{2} \left[ \  \ket{2\boldsymbol{\epsilon}_i; 0} + c_4 \ket{2\boldsymbol{\epsilon}_i; 1} \ \right] \\
& + (\Delta \alpha) \sum_{j \neq i} q_j \left[ \ \ket{\boldsymbol{\epsilon}_i + \boldsymbol{\epsilon}_j; 0} + c_4 \ket{\boldsymbol{\epsilon}_i + \boldsymbol{\epsilon}_j; 1} \ \right] \ .
\end{split}
\end{equation}
This means that, to second order in $(\Delta \alpha)$, the vacuum state is
\begin{equation}
\begin{split}
\ket{\bar{0}} \approx& \ \ket{\mathbf{0}; 0} + (\Delta \alpha) \sum_{i = 0}^N q_i \frac{T_{10}}{s + \beta_i} \ket{\boldsymbol{\epsilon}_i; 1} \\
& + (\Delta \alpha)^2 \sqrt{2} \sum_{i = 0}^N (q_i)^2 \left[ \frac{T_{01} T_{10}}{2 \beta_i (s + \beta_i)} \ket{2\boldsymbol{\epsilon}_i; 0} + \frac{T_{11} T_{10}}{(s + 2 \beta_i)(s + \beta_i)} \ket{2\boldsymbol{\epsilon}_i; 1} 
 \right] \\
 & + (\Delta \alpha)^2 \sum_{i \neq j} q_i q_j \left[ \frac{T_{01} T_{10} \ket{\boldsymbol{\epsilon}_i+\boldsymbol{\epsilon}_j; 0}}{(\beta_j + \beta_i) (s + \beta_i)}  + \frac{T_{11} T_{10} \ket{\boldsymbol{\epsilon}_i+\boldsymbol{\epsilon}_j; 1}}{(s + \beta_j + \beta_i)(s + \beta_i)}  \right] \ .
\end{split}
\end{equation}
The qualitative difference between this problem and the one-dimensional problem we solved earlier, at least as far as this diagrammatic approach goes, is quickly becoming clear. Earlier, we had to think about paths through gene eigenstate space (e.g. one diagram might correspond to $0 \to 1 \to 0$) when writing down our solution, but the `path' through RNA eigenstate space was simple. In the case of the vacuum state, it went $0 \to 1 \to 2 \to \cdots$, with each increase corresponding to another application of the up ladder operator $\hat{A}^+$. 

But in this case, each time we increment the RNA eigenstate number vector, we have $(N+1)$ choices of up ladder operator. For example, among the second order terms there is one that corresponds to first applying $\hat{A}_0^+$, and then applying $\hat{A}_1^+$. The order of traversal matters, because this is distinct from the term whose path involves first applying $\hat{A}_1^+$, and then $\hat{A}_0^+$---as one can see from the Feynman diagram denominators.  

Continuing this procedure, we find that
\begin{equation} \label{eq:multistep_vac}
\begin{split}
\ket{\bar{0}} =& \ \ket{\mathbf{0}; 0} + \sum_{k = 1}^{\infty}  \sum_{\text{paths } \mathbf{j}} (\Delta \alpha)^k \mathbf{q}^{\mathbf{n}} \sqrt{\mathbf{n}!} \sum_{i_1, ..., i_{k}} \ \frac{T_{i_k i_{k-1}}}{\left[ \beta_{j_1} + \cdots + \beta_{j_k} + i_k s \right]} \cdots \frac{T_{i_1 0} \ket{\mathbf{n}; i_k}}{\left[ \beta_{j_1} + i_1 s \right]}  
\end{split}
\end{equation}
where the sum over paths should be understood as follows. For each $k \geq 1$, we sum over all paths $\mathbf{j}$ of length $k$ with elements in $\left\{0, 1, ..., N \right\}$, i.e. $\mathbf{j} := (j_1, ..., j_k) \in \left\{ 0, 1, ..., N \right\}^k$. The vector $\mathbf{n} := (n_0, n_1, ..., n_N)$ counts the number of times that each integer $i$ appears in a path, i.e.
\begin{equation}
\begin{split}
n_r = \sum_{i \text{ s.t. } j_i = r} 1
\end{split}
\end{equation}
for all $0 \leq r \leq N$. The Hilbert space Feynman rules we obtain are almost the same as before, up to this change involving considering paths through RNA eigenstate space. 

\subsubsection{Hilbert space Feynman rules (multistep version)}

\noindent In order to compute the coefficient of $\ket{\mathbf{n}; g}$ in the infinite series expansion of $\ket{\bar{0}}$ (c.f. Eq. \ref{eq:multistep_vac}), draw all valid diagrams going from $0$ to $g \in \left\{0,1\right\}$ in $k = n_0 + \cdots n_N$ steps according to the following rules, and add the numbers corresponding to each diagram. 
\begin{enumerate}
\item \textbf{Set up grid}: Write out positions $0$ through $k$ from left to right. Draw two parallel horizontal lines above these labels to denote the $0$ and $1$ gene eigenstates. The diagram will consist of $k+1$ vertices, each located at a horizontal $\{0, ..., k\}$ position and vertical gene eigenstate (lower or upper) position, and lines connecting those vertices.
\item \textbf{Draw lines}: Place the first vertex at horizontal position $0$ and on the bottom row. Fill in the following positions from left to right. There are three possible moves: (i) if at $0$, you must next go to $1$; (ii) if at $1$, you can go to $0$ next; (iii) if at $1$, you can stay at $1$. If $g = 0$, the last vertex must be on the bottom row. If $g = 1$, the last vertex must be on the top row. 
\item \textbf{Numerical factors}: Associate each move/line with a numerical factor. In particular, associate the move from position $m-1$ to position $m$ with the factor:
\begin{itemize}
\item $0 \to 1$ \textit{flip}: $\displaystyle \frac{c_3}{s + \beta_{j_1} + \cdots + \beta_{j_m}}$
\item $1 \to 1$ \textit{stay}: $\displaystyle \frac{c_4}{s + \beta_{j_1} + \cdots + \beta_{j_m}}$
\item $1 \to 0$ \textit{flip}: $\displaystyle \frac{1}{\beta_{j_1} + \cdots + \beta_{j_m}}$
\end{itemize}
\item \textbf{Sum over RNA paths}: Write down all possible paths on $\{0, 1, ..., N\}^k$ of length $k$ in which each integer $i$ appears exactly $n_i$ times. Evaluate your diagram for all of these paths (by multiplying the numbers associated with each line together) and sum the contributions due to each path.
\item \textbf{Tack on generic factors}: Multiply in the generic factors $(\alpha)^k \mathbf{q}^{\mathbf{n}} \sqrt{\mathbf{n}!}$ to get the number corresponding to the diagram you drew.
\end{enumerate}

\subsection{Steady state probability distribution} \label{sec:multistep_Pss}

\begin{figure}[!htb]
\center{\includegraphics[scale=1.1]{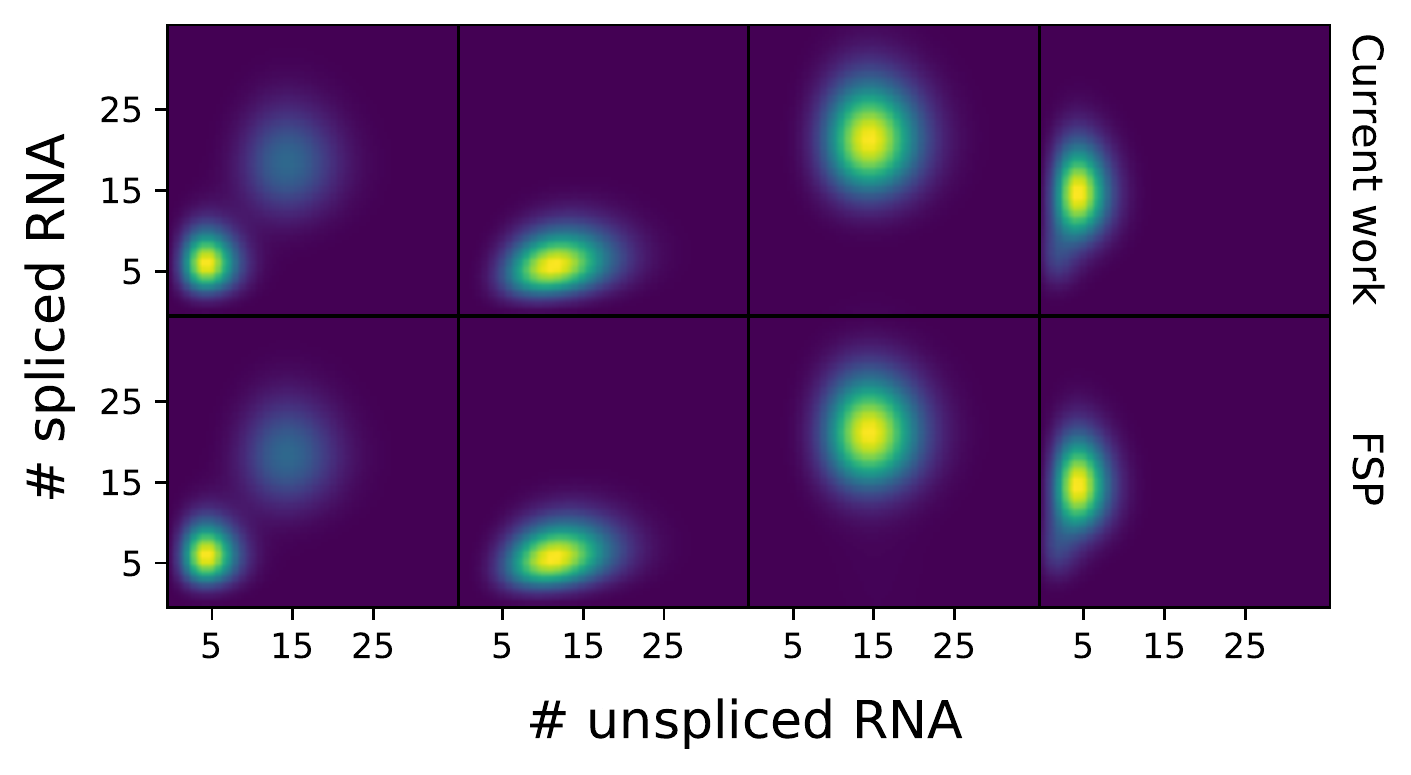}}
\caption{\label{fig:2sp_valid_FSP} 
Comparison of the 2 species/1 splicing step (i.e. $N = 1$) result described by Eq. \ref{eq:multi_Pss_marg} against finite state projection (FSP) in a variety of parameter conditions. Unspliced RNA corresponds to $x_0$, while spliced RNA corresponds to $x_1$. In these cases, our formula matches orthogonal numerical results quite well.}
\end{figure}

Let us determine the steady state probability distribution corresponding to Eq. \ref{eq:multistep_vac}. As before, we can either invoke the Euclidean product or manually pick out coefficients in molecule number space.  For the states $\ket{\mathbf{x}; S}$ (where $\mathbf{x}$ represents molecule number and $S \in \{1, 2\}$ represents gene state, rather than gene eigenstate), from which the naive eigenkets $\ket{\mathbf{n}; g}$ can be constructed, the Euclidean product can be defined via
\begin{equation}
\braket{\mathbf{x}_1; S_1}{\mathbf{x}_2; S_2}_{Eu} := \delta_{\mathbf{x}_1, \mathbf{x}_2} \delta_{S_1, S_2}
\end{equation}
so that
\begin{equation}
\braket{\mathbf{x}; S}{\mathbf{n}; g}_{Eu} = \frac{1}{\sqrt{\mathbf{n}!}} \ V_{\mathbf{n}}(\mathbf{x}, \boldsymbol{\mu}) \text{Poiss}(\mathbf{x},\boldsymbol{\mu}) \ (\vec{v}_g)_{S} 
\end{equation}
where $V_{\mathbf{n}}(\mathbf{x}, \boldsymbol{\mu})$ denotes a novel generalization of the Charlier polynomials that we define and discuss in Appendix \ref{sec:app_multistep}. Using this, we have
\begin{equation}
\begin{split}
\frac{P_{ss}(\mathbf{x}, \vec{S})}{\text{Poiss}(\mathbf{x}, \boldsymbol{\mu})} = \vec{v}_0 + \sum_{k = 1}^{\infty}  \sum_{\text{paths } \mathbf{j}} \left(\Delta \alpha\right)^k  \mathbf{q}^{\mathbf{n}} V_{\mathbf{n}}(\mathbf{x}, \boldsymbol{\mu})  \sum_{i_1, ..., i_{k}} \ \frac{T_{i_k i_{k-1}}}{\left[ \beta_{j_1} + \cdots + \beta_{j_k} + i_k s \right]} \cdots \frac{T_{i_1 0} \ \vec{v}_{i_k}}{\left[ \beta_{j_1} + i_1 s \right]}  \ . 
\end{split}
\end{equation}
Marginalizing over gene state,
\begin{equation} \label{eq:multi_Pss_marg}
\begin{split}
\frac{P_{ss}(\mathbf{x})}{\text{Poiss}(\mathbf{x}, \boldsymbol{\mu})} = 1 + \sum_{k = 2}^{\infty}  \sum_{\text{paths } \mathbf{j}} \left(\Delta \alpha\right)^k  \mathbf{q}^{\mathbf{n}} V_{\mathbf{n}}(\mathbf{x}, \boldsymbol{\mu})  \sum_{i_1, ..., i_{k-1}} \ \frac{T_{0 i_{k-1}}}{\left[ \beta_{j_1} + \cdots + \beta_{j_k} \right]} \cdots \frac{T_{i_1 0}}{\left[ \beta_{j_1} + i_1 s \right]}  \ . 
\end{split}
\end{equation}
See Fig. \ref{fig:2sp_valid_FSP} for some representative distributions in the $N = 1$ case, and how they compare to numerical results obtained by finite state projection \cite{munsky_finite_2006,fox2019}. It is interesting to note that this result is the same as our one species solution (c.f. Eq. \ref{eq:bds_Pss_marg}) up to the following replacements:
\begin{equation}
\begin{split}
\sum_{k = 2}^{\infty} &\xrightarrow{} \sum_{k = 2}^{\infty}  \sum_{\text{paths } \mathbf{j}}  \\
d^k &\xrightarrow{} (\Delta \alpha)^k \left( q_0 \right)^{n_0} \cdots \left( q_N \right)^{n_N} \\
k \gamma &\xrightarrow{} \beta_{j_1} + \cdots + \beta_{j_k} \\
\text{Poiss}(x, \mu) &\xrightarrow{} \text{Poiss}(\mathbf{x}, \boldsymbol{\mu}) \\
C_k(x, \mu) &\xrightarrow{} V_{n_0, ..., n_N}(\mathbf{x}, \boldsymbol{\mu}) \ .
\end{split}
\end{equation}
We can write down our last set of Feynman rules for computing this steady state probability distribution directly. 

\subsubsection{Molecule number space Feynman rules (multistep version)}

\noindent In order to compute the terms of order $(\Delta \alpha)^k$ in the infinite series expansion of $P_{ss}(\mathbf{x})/\text{Poiss}(\mathbf{x}, \boldsymbol{\mu})$ (c.f. Eq. \ref{eq:multi_Pss_marg}), draw all valid diagrams going from $0$ to $0$ in $k$ steps according to the following rules, and add the numbers corresponding to each diagram. 
\begin{enumerate}
\item \textbf{Set up grid}: Write out positions $0$ through $k$ from left to right. Draw two parallel horizontal lines above these labels to denote the $0$ and $1$ gene eigenstates. The diagram will consist of $k+1$ vertices, each located at a horizontal $\{0, ..., k\}$ position and vertical gene eigenstate (lower or upper) position, and lines connecting those vertices.
\item \textbf{Draw lines}: Place the first vertex at horizontal position $0$ and on the bottom row. Fill in the following positions from left to right. There are three possible moves: (i) if at $0$, you must next go to $1$; (ii) if at $1$, you can go to $0$ next; (iii) if at $1$, you can stay at $1$. If $g = 0$, the last vertex must be on the bottom row. If $g = 1$, the last vertex must be on the top row. 
\item \textbf{Numerical factors}: Associate each move/line with a numerical factor. In particular, associate the move from position $m-1$ to position $m$ with the factor:
\begin{itemize}
\item $0 \to 1$ \textit{flip}: $\displaystyle \frac{c_3}{s + \beta_{j_1} + \cdots + \beta_{j_m}}$
\item $1 \to 1$ \textit{stay}: $\displaystyle \frac{c_4}{s + \beta_{j_1} + \cdots + \beta_{j_m}}$
\item $1 \to 0$ \textit{flip}: $\displaystyle \frac{1}{\beta_{j_1} + \cdots + \beta_{j_m}}$
\end{itemize}
\item \textbf{Sum over RNA paths + generic factors}: Write down all possible paths on $\{0, 1, ..., N\}^k$ of length $k$. Evaluate your diagram for all of these paths (by multiplying the numbers associated with each line together) and sum the contributions due to each path with the weights $(\Delta \alpha)^k \mathbf{q}^{\mathbf{n}} V_{\mathbf{n}}(\mathbf{x}, \boldsymbol{\mu})$ where $\mathbf{n} = (n_0, ..., n_N)$ counts the number of times each integer $i$ appears in a given path.   
\end{enumerate}

\clearpage

\subsection{Special cases} \label{sec:multistep_special}

In this section, we will specialize our result for the steady state probability distribution in various ways, just as we did earlier for the one species problem in Sec. \ref{sec:special}. 

If we want to marginalize over $x_1, ..., x_N$, leaving only the distribution for the number $x_0$ of nascent RNA, we need the formula
\begin{equation}
\sum_{x_1, ..., x_N} V_{\mathbf{n}}(\mathbf{x}, \boldsymbol{\mu}) \ \text{Poiss}(\mathbf{x},\boldsymbol{\mu}) = (\delta_{n_1, 0} \cdots \delta_{n_N, 0}) \ C_{n_0}(x_0, \mu_0) \ \text{Poiss}(x_0, \mu_0) \ .
\end{equation}
Only the terms with $n_1 = \cdots = n_N = 0$ survive, converting the remaining polynomials to Charlier polynomials and eliminating any paths other than the ones with all zeros. We end up with
\begin{equation} 
\begin{split}
\frac{P_{ss}(x_0)}{\text{Poiss}(x_0, \mu_0)} = 1 + \sum_{k = 2}^{\infty} \left(\Delta \alpha\right)^k  C_{n_0}(x_0, \mu_0)  \sum_{i_1, ..., i_{k-1}} \ \frac{T_{0 i_{k-1}}}{\left( k \beta_0 \right)} \cdots \frac{T_{i_1 0}}{\left( \beta_{0} + i_1 s \right)}  
\end{split}
\end{equation}
i.e. the result is the same as what we found in the one species problem (c.f. Eq. \ref{eq:bds_Pss_marg}). This makes sense, because the dynamics downstream of the nascent RNA species $X_0$ should have no impact on its distribution. More generally, if we were to marginalize over $x_{m+1}, ..., x_N$, leaving only $x_0, ..., x_{m}$, the distribution would match the distribution for the multistep problem with $m$ splicing steps. 

Meanwhile, if we want to marginalize over $x_0, ..., x_{N-1}$, leaving only the distribution for the number $x_N$ of mature/fully processed RNA, we need the formula
\begin{equation}
\sum_{x_0, ..., x_{N-1}} V_{\mathbf{n}}(\mathbf{x}, \boldsymbol{\mu}) \ \text{Poiss}(\mathbf{x},\boldsymbol{\mu}) =  C_{n_N}(x_N, \mu_N) \ \text{Poiss}(x_N, \mu_N) \prod_{k = 0}^N \left[ v^{(k)}_N \right]^{n_k} \ .
\end{equation}
\clearpage
\noindent Since all of the $v^{(k)}_N$ are nonzero, no paths vanish this time. We have
\begin{equation} 
\begin{split}
\frac{P_{ss}(x_N)}{\text{Poiss}(x_N, \mu_N)} = & \  1 + \sum_{k = 2}^{\infty}  \sum_{\text{paths } \mathbf{j}} \left(\Delta \alpha\right)^k  \mathbf{q}^{\mathbf{n}} C_{n_N}(x_N, \mu_N) \times \\
& \times \prod_{k = 0}^N \left[ v^{(k)}_N \right]^{n_k} \sum_{i_1, ..., i_{k-1}} \ \frac{T_{0 i_{k-1}}}{\left[ \beta_{j_1} + \cdots + \beta_{j_k} \right]} \cdots \frac{T_{i_1 0}}{\left[ \beta_{j_1} + i_1 s \right]}  \ . 
\end{split}
\end{equation}
Qualitatively, this formula indicates that the distribution of the mature RNA species $X_N$ \textit{is} affected by all of the preceding splicing steps---including how many steps there were, and the associated rates. This observation is less trivial than it at first seems, since the marginal steady state distribution of $X_N$ for a constitutively active gene (which is a Poisson distribution) is completely independent of upstream splicing dynamics. 

Let us move on to special parameter regimes of the full joint distribution. In the case of very unequal switching rates ($k_{21} \gg k_{12}$), we can argue as in Sec. \ref{sec:very_unequal} that the `balance beam' Feynman diagrams contribute the most to first order in the switching rate ratio $r := k_{12}/k_{21}$. The end result is analogous:
\begin{equation}
\begin{split}
\frac{P_{ss}(\mathbf{x})}{\text{Poiss}(\mathbf{x}, \boldsymbol{\mu})} =& \  1 + \sum_{k = 2}^{\infty}  \sum_{\text{paths } \mathbf{j}} \left(\Delta \alpha\right)^k  \mathbf{q}^{\mathbf{n}} V_{\mathbf{n}}(\mathbf{x}, \boldsymbol{\mu}) \times \\
& \times \frac{r}{\left( \beta_{j_1} + \cdots + \beta_{j_k} \right) \left( \beta_{j_1} + \cdots + \beta_{j_{k-1}} + s \right) \cdots \left( \beta_{j_1} + s \right)}  \ . 
\end{split}
\end{equation}
If switching is much faster than all of the $\beta_i$ (for $0 \leq i \leq N$), we can again write an asymptotic series in $1/s$:
\begin{equation}
\frac{P_{ss}(\mathbf{x})}{\text{Poiss}(\mathbf{x}, \boldsymbol{\mu})} \approx 1 + \frac{1}{s} (\Delta \alpha)^2 c_3 \sum_{i, j}   \frac{\mathbf{q}^{\mathbf{n}} V_{\mathbf{n}}(\mathbf{x}, \boldsymbol{\mu})}{\beta_i + \beta_j}  +  \mathcal{O}\left( \frac{1}{s^2} \right)
\end{equation}
where $\mathbf{n} := \boldsymbol{\epsilon}_i + \boldsymbol{\epsilon}_j$. Finally, if switching is much slower than all of the $\beta_i$, we can approximate the Feynman denominators as $(\beta_{j_1} + \cdots \beta_{j_m} + i_m s) \approx (\beta_{j_1} + \cdots \beta_{j_m})$ and use the combinatorial result\footnote{This can be proved by induction.} that
\begin{equation}
\sum_{\text{paths } \mathbf{j} \ , \ \mathbf{n} \text{ fixed}} \frac{1}{\left( \beta_{j_1} + \cdots + \beta_{j_k} \right) \cdots \left( \beta_{j_1} + \beta_{j_2} \right) \left( \beta_{j_1} \right)} = \frac{1}{\mathbf{n}!} \frac{1}{\beta_0^{n_0} \cdots \beta_N^{n_N}}
\end{equation}
where we consider only paths of length $k$ with the same sum vector $\mathbf{n}$. Then our steady state probability becomes
\begin{equation}
\begin{split}
\frac{P_{ss}(\mathbf{x})}{\text{Poiss}(\mathbf{x}, \boldsymbol{\mu})} = 1 + \sum_{\mathbf{n}} \frac{1}{\mathbf{n}!} \prod_{j = 0}^N \left[ \frac{(\Delta \alpha) q_j}{\beta_j} T \right]^{n_j} V_{\mathbf{n}}(\mathbf{x}, \boldsymbol{\mu})   \ . 
\end{split}
\end{equation}
Using the formula for the generating function of these orthogonal polynomials (c.f. Eq. \ref{eq:multi_genfunc} in Appendix \ref{sec:app_multistep}), we can sum this in closed form to obtain
\begin{equation}
\frac{P_{ss}(\mathbf{x})}{\text{Poiss}(\mathbf{x}, \boldsymbol{\mu})} \stackrel{00}{\approx}  \prod_{j = 0}^N \left[ 1 + \frac{(\Delta \alpha)}{\mu_j} T \sum_k v^{(k)}_j \frac{q_k}{\beta_k}  \right]^{x_j} e^{- (\Delta \alpha) T \sum_k v^{(k)}_j \frac{q_k}{\beta_k} } \ .
\end{equation}
Using a straightforward generalization of the argument from Sec. \ref{sec:slow_switching}, we can show that this interesting-looking formula is exactly equal to a Poisson mixture, i.e. that
\begin{equation} \label{eq:my_pmix}
P_{ss}(\mathbf{x}) \approx \frac{k_{12}}{s} \text{Poiss}(\mathbf{x}, \boldsymbol{\mu}_1) + \frac{k_{21}}{s} \text{Poiss}(\mathbf{x}, \boldsymbol{\mu}_2)
\end{equation}
in this limit. 


\clearpage

\section{Numerical implementation and validation}
\label{sec:numerical}

In this section, we explain how to implement the formulas presented in the previous sections efficiently, and discuss typical numerical behavior. 

The most important thing to note is that, while our results are formally correct (see e.g. Appendix \ref{sec:app_consistency} for a mathematical proof of the agreement between our birth-death-switching result and the previously known result), they are likely not very useful for numerically implementing the solution to these problems in large swaths of parameter space. In particular, they are frequently numerically unstable, with the more general formulas presented in the previous section somewhat more unstable than the birth-death-switching formulas. 

\begin{figure}[h]
\centering
\includegraphics[width=\columnwidth]{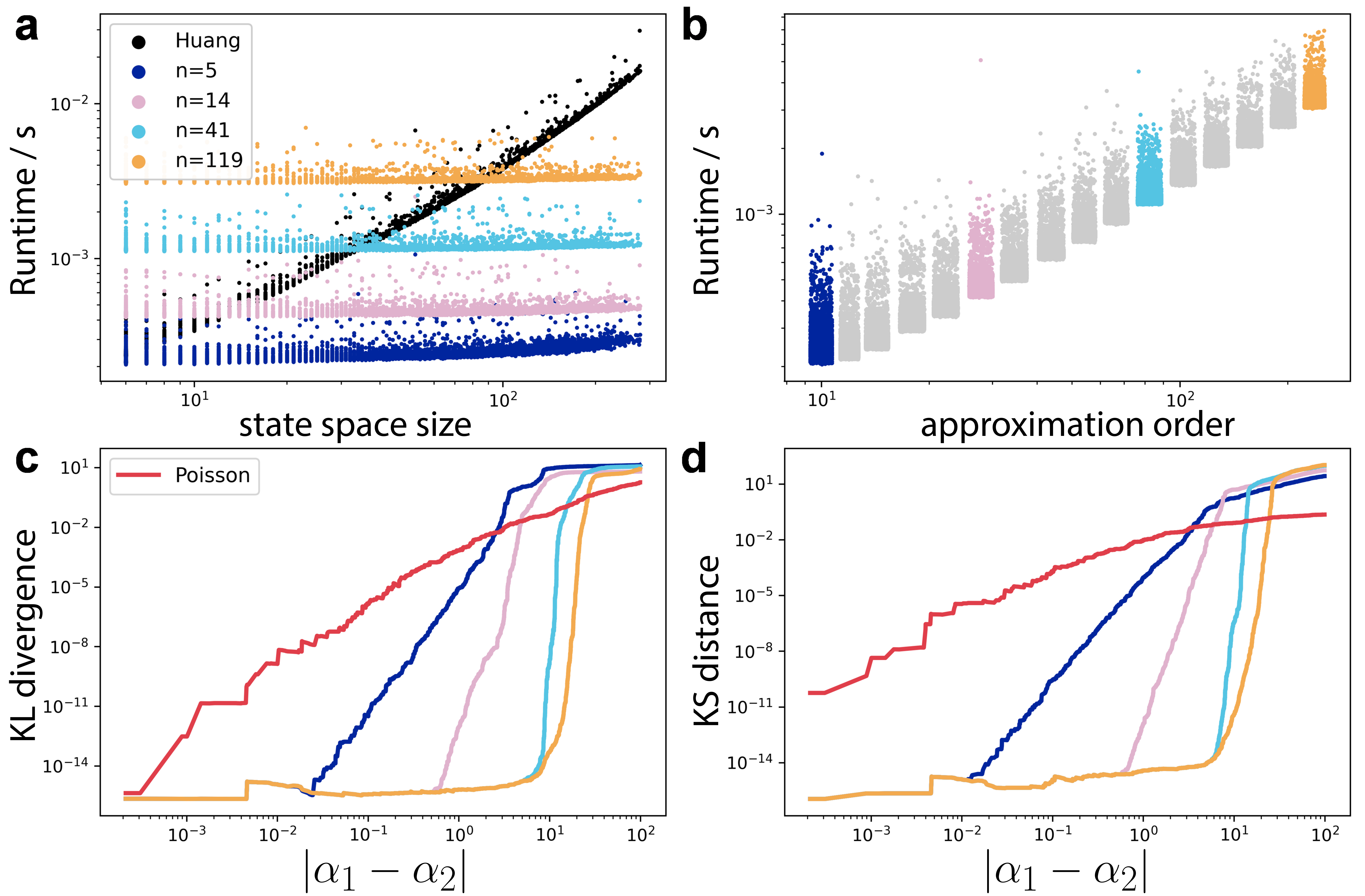}
\caption{Exploration of birth-death-switching solution performance and stability. For many randomly sampled parameter sets, we evaluated the effective state space size, runtime, and accuracy. Each dot corresponds to a different parameter set. See Appendix \ref{sec:app_validation} for details on the parameter sampling procedure and definition of state space size. (a) Runtime comparison between the ground truth hypergeometric solution from Huang et al. \cite{huang2015} and the diagrammatic estimates as a function of state space size, computing only the first $n$ terms of Eq. \ref{eq:bds_sln}. (b) Runtime comparison for the diagrammatic estimates as a function of approximation order. (c) Cumulative 98th percentile of the Kullback-Leibler divergence between estimates and ground truth as a function of $|\Delta \alpha|$ (colors correspond to colors in (b); red represents trivial Poisson solution). (d) Cumulative 98th percentile of the Kolmogorov-Smirnov divergence between estimates and ground truth as a function of $|\Delta \alpha|$ (colors correspond to colors in (b)). The last two panels confirm that increasing $\Delta \alpha$ makes convergence much slower, which in practice could mean inaccurate numerical results when computing Eq. \ref{eq:bds_sln} to fixed order. }
\label{fig:1sp_valid}
\end{figure}

This is essentially due to two reasons. The first reason is that our solutions were constructed as perturbative solutions in powers of $\Delta \alpha$; when this is not small---as in the case in most real situations---our implementations behave particularly poorly (Fig. \ref{fig:1sp_valid}a) because one must compute a very large number of series terms in order to closely approximate the correct answer. One is also likely to encounter overflow errors in taking large powers of $\Delta \alpha$.

One might wonder whether our formulas are asymptotic expansions of the true result, which would mean they formally diverge. This is not so. For example, in the birth-death-switching case we have that $C_n \sim (x/\mu)^n$, and that the Feynman diagram sums go like $1/(\gamma^n n!)$. Very roughly, we have
\begin{equation}
\frac{P_{ss}(x)}{\text{Poiss}(x, \mu)} \sim \sum_{n = 0}^{\infty} \left[ \frac{\Delta \alpha \cdot x}{\gamma \cdot \mu} \right]^n \frac{1}{n!} = \exp\left[ \frac{\Delta \alpha \cdot x}{\alpha_{eff}}  \right] \ .
\end{equation}
The analysis is similar in the multistep case, where we instead obtain
\begin{equation}
\frac{P_{ss}(\mathbf{x})}{\text{Poiss}(\mathbf{x}, \boldsymbol{\mu})} \sim \exp\left[ \sum_{j=0}^N \frac{\Delta \alpha \cdot q_j x_j}{\alpha_{eff}} \right] \ .
\end{equation}
The second reason is that additional instability arises due to the proliferation of time scales (i.e. the $\beta_i$) in the multistep problem. When the $\beta_i$ are all comparable, all of the Feynman diagrams contribute, leading to a large number of terms being required to converge to the correct answer. 

Despite these instabilities, the formulas are still usable if their regime of validity is kept in mind. We have found that computing the analytic generating function (Eq. \ref{eq:1sp_genfunc} and Eq. \ref{eq:multi_analytic_genfunc}), and then inverse fast Fourier transforming, tends to produce more stable results than using the state space formulas directly. This has the added advantage that one can bypass numerically computing the orthogonal polynomials.


We explore runtime performance and stability in quantitative fashion for the birth-death-switching solution in Fig. \ref{fig:1sp_valid}. Because the birth-death-switching problem has been solved before (in particular in \cite{huang2015}, by Huang et al.), we can assess the accuracy of our solution formula (Eq. \ref{eq:bds_sln}) by comparing it to their hypergeometric solution formula. By sampling physiologically plausible parameters and modulating the approximation order, we can compute the precision and computational complexity of the solution; the details of the sampling procedure are described in Appendix \ref{sec:app_validation}.

The diagrammatic solution exhibits substantially weaker state space size dependence than the hypergeometric solution (Fig. \ref{fig:1sp_valid}a), with nearly constant time complexity compared to the $\mathcal{O}(N^{1.00})$ complexity of the Huang solution in the relevant domain. The runtime dependence on order of approximation is approximately linear (Fig. \ref{fig:1sp_valid}b). In the low-$\Delta\alpha$ region, the diagrammatic solution recapitulates the analytical solution quite well, as quantified using the Kullback-Leibler and Kolmogorov-Smirnov divergence measures (Fig. \ref{fig:1sp_valid}c-d). Therefore, the algorithm performance suggests natural applications to the design of adaptive solvers, with the diagrammatic solution used as much as possible in the low-$\Delta\alpha$ regime. Finally, we note that the slow-switching Poisson mixture approximation given in Eq. \ref{eq:my_pmix} generally underperforms the diagrammatic solution.

But, as we might have anticipated, the discrepancy between the `ground truth' Huang result and our solution formula increases sharply as $\Delta \alpha$ gets large. This is clear from the large Kullback-Leibler divergence (Fig. \ref{fig:1sp_valid}c) and Kolmogorov-Smirnov distance (Fig. \ref{fig:1sp_valid}d) values in this regime.

In conclusion, the solution formulas have proven useful tools for examining the qualitative behavior of the models in various regimes (e.g. as we did in Sec. \ref{sec:special} and Sec. \ref{sec:multistep_special}), and can be used to obtain reliable theoretical results for observables like moments. But the series themselves are numerically poorly conditioned, and can be somewhat slow in the multistep case (where one must sum over many `paths'), and so are not appropriate for tasks like parameter inference on RNA counts data. In the second part of this two part article \cite{vastolaP2}, we present an alternative approach that has better numerical properties.

\clearpage

\section{Discussion}
\label{sec:discussion}

Using a novel theoretical methodology inspired by quantum mechanics, we have solved several complicated problems in stochastic chemical kinetics exactly: the chemical birth-death process coupled to a switching gene, the chemical birth-death process with additive noise coupled to a switching gene (the continuous limit of the former model), and a more realistic model of transcription involving a switching gene and an arbitrary number of downstream splicing steps. We also uncovered tantalizing formal connections between chemical kinetics and quantum physics: the dynamics of a system involving RNA production and degradation coupled to a switching gene looks much like the dynamics of a system involving non-interacting bosons coupled to a fermion, for example. Unfortunately, the instability of these solution formulas renders them inadequate for practical purposes like parameter inference. We will address this deficiency further in part II \cite{vastolaP2}.

There are various potential directions for generalizing this approach, some of which we have pointed out earlier in the paper. Most obviously, the idea of expressing a system's Hamiltonian in terms of `natural' ladder operators, and rewriting it in terms of `free' Hamiltonians and an interaction term, is broadly applicable---whether or not a switching gene is involved. In the context of stochastic chemical kinetics and gene regulation, birth-death-like processes like the ones we have studied here may be an appropriate choice of free Hamiltonian, with the interaction terms corresponding to dynamics we currently do not know how to solve exactly. Notably, in most cases we cannot exactly solve dynamics involving molecular binding/unbinding \cite{schnoerr2017}. One expects nonlinear operator products like $\hat{A}_1 \hat{A}_2 \hat{A}_3^+$ to appear in the interaction term in such cases.

Closer in spirit to our current results, one expects the ability to generalize to (i) more than two gene states, and (ii) more realistic transcription and splicing dynamics. Multistate models have been solved in at least one special case \cite{zhou_analytical_2012}, but it is expected that such a generalization would be highly nontrivial. It is not immediately obvious how to generalize to more than two gene states while maintaining nice anti-commutation properties for the switching-associated ladder operators; we offer some reflection on this problem towards the end of Appendix \ref{sec:app_moreswitch}. On the other hand, generalizing to more complicated models of transcription and splicing, involving branching splicing topologies and even alternative splicing, seems more straightforward. The main difficulty would be understanding the representation theory of this more general problem, which in some sense means reckoning with the representation theory of so-called monomolecular reaction networks \cite{jahnke2007, vastolaDP} (the natural multi-species generalization of the birth-death process).

One unexpected strength of our method is that the problem it treats is fairly abstract. Although we took our main CME (Eq. \ref{eq:CME_multistep}) to represent RNA dynamics coupled to a switching gene, almost the exact same equation can be used to describe a more granular model of transcription involving RNA polymerase molecules that hop sequentially along a gene \cite{filatova2020}. Also, as pointed out at the end of Sec. \ref{sec:continuous}, the continuous limit of our model corresponds exactly to a model of active matter run-and-tumble motion \cite{gm2020}. This suggests our techniques may be useful for solving a broad range of stochastic physical and biological problems.

Lastly, we must emphasize the surprising applicability of quantum mechanics-like operator-based approaches, and the natural appearance of objects that resemble Feynman diagrams. There are probably many more results (in both stochastic gene regulation and studies of active matter dynamics) that can be derived by exploiting useful parallels between quantum mechanics and stochastic chemical reaction dynamics, as several pieces of past work have suggested \cite{baez2018, vastolaSTOCHPATH, vastolaADD, bressloff2020, bressloff2021}.



\begin{acknowledgments}
The DNA and mRNA used in Fig. \ref{fig:allmodels_cartoon} are derivatives of the DNA Twemoji by Twitter, Inc., used under CC-BY 4.0. The palette used in Fig. \ref{fig:1sp_valid} is derived from \href{https://github.com/lambdamoses/IslamicArt}{\color{blue}IslamicArt by lambdamoses}, used under the MIT license. G.G. and L.P. were partially funded by NIH U19MH114830. J.J.V. and W.R.H. were supported by NSF Grant \# DMS 1562078.
\end{acknowledgments}

\enlargethispage{\baselineskip}


\section*{Authors' contributions}

J.J.V. and G.G. conceived of the work, studied the numerical properties of the analytic solutions, and wrote the manuscript. J.J.V. came up with the solution method and worked out the associated mathematics. L.P. and W.R.H. reviewed and edited the manuscript. 

\clearpage

\appendix

\section{Charlier polynomials}
\label{sec:app_charlier}

In this appendix, we will define the Charlier polynomials and describe their properties. It is useful to view them as a discrete analogue to the more familiar Hermite polynomials; in particular, the Charlier polynomials relate to the Poisson distribution the same way that Hermite polynomials relate to the normal distribution. Moreover, they reduce to the Hermite polynomials in the same limit that the Poisson distribution reduces to a normal distribution.

Given a parameter $\mu \geq 0$ and a nonnegative integer $n$, define the $n$th Charlier polynomial $C_n(x, \mu)$ for all $x \geq 0$ via the Rodrigues formula
\begin{equation} \label{eq:rodrigues_charlier}
C_n(x, \mu) := \frac{1}{\text{Poiss}(x, \mu)} \left[ - \nabla \right]^n \text{Poiss}(x, \mu)
\end{equation}
where we recall that
\begin{equation}
\text{Poiss}(x, \mu) :=  \frac{\mu^x e^{- \mu}}{x!} \ ,
\end{equation}
and where $\nabla$ is the backward difference operator that acts on a discrete-valued function $f$ according to
\begin{equation}
\nabla f(x) := f(x) - f(x - 1) \ .
\end{equation}
For example, to obtain $C_1(x, \mu)$, we compute
\begin{equation}
C_1(x, \mu) = \frac{\text{Poiss}(x-1,\mu) - \text{Poiss}(x,\mu)}{\text{Poiss}(x, \mu)} = \frac{\frac{x}{\mu} \text{Poiss}(x,\mu) - \text{Poiss}(x,\mu)}{\text{Poiss}(x, \mu)} = \frac{x}{\mu} - 1 \ . 
\end{equation}
From the point of view of our ladder operator solution of the chemical birth-death process, this definition of $C_n(x, \mu)$ arises from taking 
\begin{equation} 
\braket{x}{n} = \bra{x} \frac{( \hat{A}^+ )^n}{\sqrt{n!}} \ \ket{0} =  \sqrt{\frac{\mu^n}{n!}} \bra{x} \left( \hat{\pi} - 1  \right)^n \ \ket{0} 
\end{equation}
and defining
\begin{equation}
C_n(x, \mu) := \bra{x} \left( \hat{\pi} - 1  \right)^n \ \ket{0} \ .
\end{equation}
\clearpage
\noindent In our convention, the first few Charlier polynomials are
\begin{equation}
\begin{split}
C_0(x, \mu) &= 1 \\
C_1(x, \mu) &= \frac{x}{\mu} - 1 \\
C_2(x, \mu) &= \frac{x^2}{\mu^2} - \left[ \frac{1}{\mu^2} + \frac{2}{\mu} \right] x + 1 \\
C_3(x, \mu) &= \frac{x^3}{\mu^3} - \frac{3}{\mu^2} \left[ 1 + \frac{1}{\mu} \right] x^2 + \frac{1}{\mu} \left[ 3 + \frac{3}{\mu} + \frac{2}{\mu^2} \right] x - 1 \ .
\end{split}
\end{equation}
In general, the $n$th Charlier polynomial will be an $n$th order polynomial in $x$ which asymptotically goes like $(x/\mu)^n$. 

\begin{figure}[!htb]
\center{\includegraphics[scale=0.75]{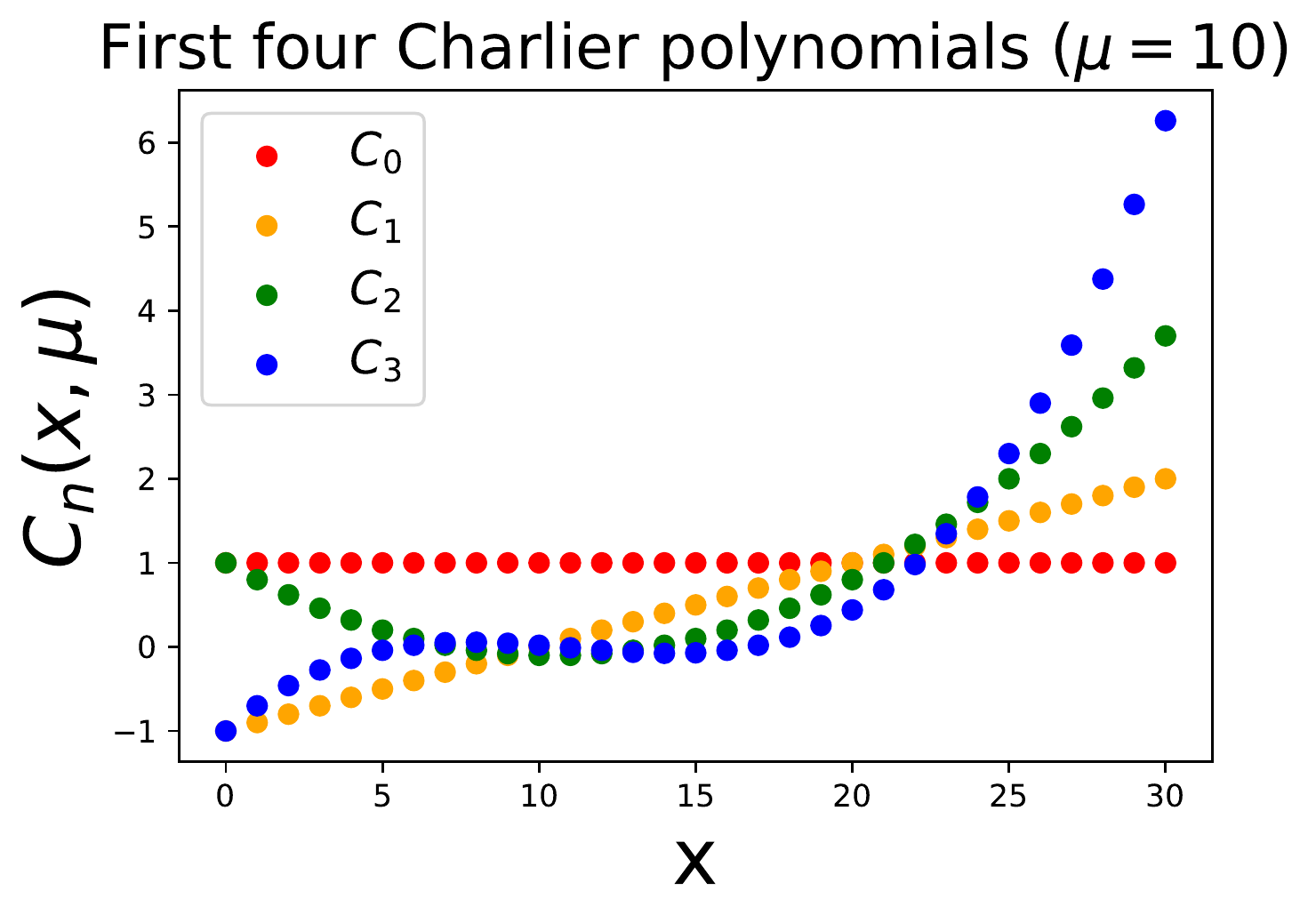}}
\caption{\label{fig:charlier_polys} 
The first few Charlier polynomials ($C_0, C_1, C_2, C_3$) for $\mu = 10$. For $x$ somewhat larger than $\mu$, we can see that $C_n \sim (x/\mu)^n$.}
\end{figure}

Two results are particularly useful for deriving various properties of the Charlier polynomials. One is its generating function $G_C(t, x)$ (not to be confused with probability generating function seen elsewhere in this paper), defined via
\begin{equation}
G_C(t, x) := \sum_{n = 0}^{\infty} \frac{C_n(x, \mu)}{n!} \ t^n \ .
\end{equation}
\noindent Using the Rodrigues formula (Eq. \ref{eq:rodrigues_charlier}), we can show that
\begin{equation} \label{eq:charlier_genfunc}
\begin{split}
G_C(t, x) &= \frac{1}{\text{Poiss}(x, \mu)}   \sum_{n = 0}^{\infty} \frac{\left[ - \nabla \ t \right]^n}{n!} \ \text{Poiss}(x, \mu) \\
&= \frac{1}{\text{Poiss}(x, \mu)}   e^{ - \nabla \ t} \ \text{Poiss}(x, \mu) \\
&= \frac{1}{\text{Poiss}(x, \mu)}   e^{ - t} \ \sum_{n = 0}^x \frac{\text{Poiss}(x - n, \mu) \ t^n}{n!} \\
&= \frac{1}{\text{Poiss}(x, \mu)}   e^{ - t} \ \left( 1 + \frac{t}{\mu} \right)^x \text{Poiss}(x, \mu) \\
&=  \left( 1 + \frac{t}{\mu} \right)^x  e^{ - t} \ .
\end{split}
\end{equation}
\noindent Another useful result comes from a novel integral representation of the Poisson distribution\footnote{See Gradshteyn and Ryzhik \cite{gradshteyn2014} (ET I 118(3), in section 3.382, on pg. 365).}
\begin{equation}
\text{Poiss}(x, \mu) = \int_{-\infty}^{\infty} \frac{dz}{2 \pi} \ \frac{e^{- i \mu z}}{(1 - i z)^{x + 1}} \ .
\end{equation}
Applying the Rodrigues formula to it, we can derive the integral representation
\begin{equation} \label{eq:charlier_intrep}
C_n(x, \mu) = \frac{1}{\text{Poiss}(x, \mu)} \int_{-\infty}^{\infty} \frac{dz}{2 \pi} \ \frac{(- iz)^n e^{- i \mu z}}{(1 - i z)^{x + 1}}
\end{equation}
which, incidentally, offers another method for explicitly computing $G_C(t, x)$. 

\noindent One property of the Charlier polynomials that we will use is that
\begin{equation}
\sum_{x = 0}^{\infty} C_n(x, \mu) g^x \ \frac{\mu^x e^{- \mu}}{x!} = (g - 1)^n e^{\mu (g - 1)} \ ,
\end{equation}
a result that can be derived using our expression for $G_C(t, x)$. To do so, note that
\begin{equation}
\begin{split}
\sum_{n = 0}^{\infty} \sum_{x = 0}^{\infty} C_n(x, \mu) g^x \ \frac{\mu^x e^{- \mu}}{x!} \frac{t^n}{n!} &= \sum_{x = 0}^{\infty}  g^x \ \frac{\left( \mu + t \right)^x}{x!}   e^{ - t - \mu} \\
&= \exp\left\{ g \left( \mu + t \right) - t - \mu \right\} \\
&= e^{\mu (g - 1)} \sum_{n = 0}^{\infty} \frac{\left( g - 1 \right)^n}{n!} t^n 
\end{split}
\end{equation}
and match the coefficients of $t^n$. This property, along with our formula for $G_C(t, x)$, can be used to derive what is probably the most important property of the Charlier polynomials: that they are orthogonal with respect to a Poisson weight function, i.e.
\begin{equation}
\sum_{x = 0}^{\infty} C_n(x, \mu) C_m(x, \mu) \ \text{Poiss}(x, \mu) = \frac{n!}{\mu^n} \ \delta_{n, m} \ .
\end{equation}
This orthogonality relation offers an alternative way to view the resolution of the identity described by Eq. \ref{eq:ROI_bd}.

Other properties that are useful to establish include that the Charlier polynomials obey a recurrence relation in $x$, and that they reduce to Hermite polynomials in the large $\mu$ limit. To establish the recurrence relation, it is easiest to solve the birth-death CME (Eq. \ref{eq:CME_bd}) via separation of variables. Given that we already know the allowed energies $E_n$ (c.f. Eq. \ref{eq:energies_bd}) and steady state distribution (c.f. Eq. \ref{eq:ground_bd}), we can substitute the ansatz
\begin{equation}
P(x, t) = C_n(x, \mu) \text{Poiss}(x, \mu) e^{- \gamma n t}
\end{equation}
into the time-dependent CME to derive (after simplifying)
\begin{equation}
\begin{split}
- n C_n(x, \mu) =& \ \mu \left[ C_n(x+1, \mu) - C_n(x, \mu) \right] \\
&+ x \left[ C_n(x-1, \mu) - C_n(x, \mu) \right] \ .
\end{split}
\end{equation}
In addition to offering another way to define the Charlier polynomials, this recurrence relation also offers one numerically efficient way to determine their values. Meanwhile, the large $\mu$ behavior of the Charlier polynomials is described by
\begin{equation} \label{eq:charlierlimit}
\begin{split}
\left( \sqrt{\mu} \right)^n C_n(x, \mu) &\xrightarrow{\mu \gg 1} \left( \frac{1}{\sqrt{2}} \right)^n H_n\left( \frac{x - \mu}{\sqrt{2 \mu}} \right) \\
G_C(\sqrt{\mu} \ t, x) &\xrightarrow{\mu \gg 1} G_H\left(\frac{t}{\sqrt{2}}, \frac{x - \mu}{\sqrt{2 \mu}} \right)  \ ,
\end{split}
\end{equation}
where $G_H(t, y)$ denotes the generating function of the (physicists') Hermite polynomials $H_n(x)$, which reads 
\begin{equation}
G_H(t, y) := \sum_{n = 0}^{\infty} \frac{H_n(y)}{n!} \ t^n = e^{2 y t - t^2} \ .
\end{equation}
This can be established by straightforwardly approximating $G_C(\sqrt{\mu} \ t, x)$. For large $\mu$, we have that
\begin{equation}
\begin{split}
G_C(\sqrt{\mu} \ t, x)  &= e^{- \sqrt{\mu} t} \left( 1 + \frac{t}{\sqrt{\mu}} \right)^x \\
&= e^{- \sqrt{\mu} t} e^{x \log\left[ 1 + \frac{t}{\sqrt{\mu}} \right]} \\
&\approx e^{- \sqrt{\mu} t} e^{x \left[ \frac{t}{\sqrt{\mu}} - \frac{t^2}{2 \mu} \right]}
\end{split}
\end{equation}
where we have expanded the logarithm to second order. Approximating $x \approx \mu$ in the $t^2$ term, we have
\begin{equation}
\begin{split}
e^{- \sqrt{\mu} t} e^{x \left[ \frac{t}{\sqrt{\mu}} - \frac{t^2}{2 \mu} \right]} &\approx e^{- \sqrt{\mu} t} e^{\frac{x t}{\sqrt{\mu}} - \frac{t^2}{2 } } \\
&= \exp\left\{ \frac{x - \mu}{\sqrt{\mu}} t - \frac{t^2}{2} \right\} \\
&= \exp\left\{ 2 \ \frac{x - \mu}{\sqrt{2 \mu}} \left( \frac{t}{\sqrt{2}} \right) - \left(\frac{t}{\sqrt{2}}\right)^2 \right\} \\
&= G_H\left(\frac{t}{\sqrt{2}}, \frac{x - \mu}{\sqrt{2 \mu}} \right)
\end{split}
\end{equation}
as desired. The reader may notice that the approximations that were necessary here are the same as the ones required to show that a Poisson distribution reduces to a normal distribution in the large $\mu$ limit, i.e. it is necessary in both cases to truncate the Taylor series of a logarithm and assume that $x \approx \mu$ in part of the formula.

\pagebreak

\section{More on switching gene representation theory} \label{sec:app_moreswitch}

Although the appearance of fermionic ladder operators in the solution of pure gene switching dynamics may at first seem mysterious, there is a systematic approach to finding these operators and deriving their properties. In this appendix, we discuss this approach, and point towards how it might be generalized for problems involving more than two gene states. 

The key insight is that the matrix\footnote{In this appendix, we will relax our convention of putting hats on matrices and arrows on vectors.} $H_s$ is infinitesimal stochastic, which means its columns sum to zero. This forces one of the eigenvalues of $H_s$ to be zero, and the other (assuming $H_s \neq 0$) to be strictly negative. Because its two eigenvalues are distinct, it is diagonalizable, and can be written in the form
\begin{equation}
H_s = - s \ Q D Q^{-1}
\end{equation} 
where we are using $- s$ to denote the negative eigenvalue of $H_s$, and where $D$ is the diagonal matrix
\begin{equation}
D := \begin{pmatrix} 0 & 0 \\ 0 & 1 \end{pmatrix} \ .
\end{equation}
Trivially, we can observe that
\begin{equation}
D = \begin{pmatrix} 0 & 0 \\ 0 & 1 \end{pmatrix} = \begin{pmatrix} 0 & 0 \\ 1 & 0 \end{pmatrix} \begin{pmatrix} 0 & 1 \\ 0 & 0 \end{pmatrix} = \tilde{B}^+ \tilde{B}
\end{equation}
where we have defined the matrices $\tilde{B}$ and $\tilde{B}^+$. Various properties of $\tilde{B}$ and $\tilde{B}^+$, including their commutation relations, are obvious:
\begin{equation} \label{eq:tilde_properties}
\begin{split}
\tilde{B} e_0 &= 0 \hspace{1in} \tilde{B} e_1 = e_0 \\
\tilde{B}^+ e_0 &= e_1 \hspace{1in} \tilde{B}^+ e_1 = 0 \\
\{ \tilde{B}, \tilde{B} \} &= \tilde{B}^2 + \tilde{B}^2 = 0 \\
\{ \tilde{B}^+, \tilde{B}^+ \} &= (\tilde{B}^+)^2 + (\tilde{B}^+)^2 = 0 \\
\{ \tilde{B}, \tilde{B}^+ \} &= \tilde{B} \tilde{B}^+ + \tilde{B}^+ \tilde{B} = I 
\end{split}
\end{equation}
where $e_0 := (1, 0)^T$ and $e_1 := (0, 1)^T$. These anticommutation properties are identical to those satisfied by $B$ and $B^+$, and the action of $\tilde{B}$ and $\tilde{B}^+$ on $e_0$ and $e_1$ looks like the action of $B$ and $B^+$ on the eigenvectors of $H_s$ (c.f. Eq. \ref{eq:B_effect} and Eq. \ref{eq:Bplus_effect}). This is no accident; noting that
\begin{equation}
H_s = - s \ Q \tilde{B}^+ \tilde{B} Q^{-1} = - s \ (Q \tilde{B}^+ Q^{-1})  (Q \tilde{B} Q^{-1}) \ ,
\end{equation}
we can define $B := Q \tilde{B} Q^{-1}$ and $B^+ := Q \tilde{B}^+ Q^{-1}$, i.e. $B$ and $B^+$ are just a similarity transformation away from $\tilde{B}$ and $\tilde{B}^+$. Various properties of $B$ and $B^+$ follow trivially from the properties of $\tilde{B}$ and $\tilde{B}^+$ listed in Eq. \ref{eq:tilde_properties}. For example,
\begin{equation}
Q \tilde{B} Q^{-1} Q e_1  = Q e_0 \ \implies \ B v_1 = v_0  \ ,
\end{equation}
since $Q$ constitutes a change of basis, e.g. $Q e_0 = v_0$. 

A `natural' weight matrix $W$ can be motivated by this idea that our system is simple up to a change of basis. To force the eigenvectors $v_i$ of $H_s$ to be orthogonal, we can note that the standard basis vectors $e_i$ are orthogonal, and that $Q$ changes from one basis set to the other. In particular,
\begin{equation}
\delta_{i, j} = e_i^T e_j = (Q^{-1} v_i)^T (Q^{-1} v_j) = v_i^T (Q^{-T} Q^{-1}) v_j \ ,
\end{equation} 
so we should define $W$ to be the symmetric matrix
\begin{equation}
W := Q^{-T} Q^{-1} \ .
\end{equation}
It is easy to see that $B$ and $B^+$ are Hermitian conjugates with respect to the inner product induced by this weight matrix, e.g.
\begin{equation}
W B = Q^{-T} Q^{-1} Q \tilde{B} Q^{-1} = Q^{-T} \tilde{B} Q^T Q^{-T} Q^{-1} = Q^{-T} (\tilde{B}^+)^T Q^T Q^{-T} Q^{-1} = (B^+)^T W \ .
\end{equation}
In principle, it is straightforward to extend this idea to represent gene switching dynamics involving more than two gene states. For example, for a three state gene switching problem we might generically expect $H_s$ to have three distinct eigenvalues: zero, and two distinct strictly negative eigenvalues (because, as before, $H_s$ is infinitesimal stochastic). In that case, $H_s$ can be diagonalized so that
\begin{equation}
\begin{split}
H_s &= Q \begin{pmatrix} 0 & 0 & 0 \\ 0 & - \lambda_1 & 0 \\ 0 & 0 & - \lambda_2 \end{pmatrix} Q^{-1} \\
&= - \lambda_1 \  Q \begin{pmatrix} 0 & 0 & 0 \\ 0 & 1 & 0 \\ 0 & 0 & 0 \end{pmatrix} Q^{-1} - \lambda_2 \ Q \begin{pmatrix} 0 & 0 & 0 \\ 0 & 0 & 0 \\ 0 & 0 & 1 \end{pmatrix} Q^{-1} \\
&= - \lambda_1 \  Q \begin{pmatrix} 0 & 0 & 0 \\ 1 & 0 & 0 \\ 0 & 0 & 0 \end{pmatrix} \begin{pmatrix} 0 & 1 & 0 \\ 0 & 0 & 0 \\ 0 & 0 & 0 \end{pmatrix} Q^{-1} - \lambda_2 \ Q \begin{pmatrix} 0 & 0 & 0 \\ 0 & 0 & 0 \\ 1 & 0 & 0 \end{pmatrix} \begin{pmatrix} 0 & 0 & 1 \\ 0 & 0 & 0 \\ 0 & 0 & 0 \end{pmatrix} Q^{-1} \\
&= - \lambda_1 \  Q \tilde{B}_1^+ \tilde{B}_1 Q^{-1} - \lambda_2 \ Q \tilde{B}_2^+ \tilde{B}_2 Q^{-1} \\
\end{split}
\end{equation}
and we obtain matrices $\tilde{B}_i$ and $\tilde{B}_i^+$ (for $i = 1, 2$) with similar properties as before. If we use $\delta_{ij}$ to denote the matrix with a $1$ in the $(i, j)$th place and zeros elsewhere, we can write
\begin{equation}
\begin{split}
\tilde{B}_1 &:= \delta_{01} \hspace{1in} \tilde{B}_1^+ := \delta_{10} \\
\tilde{B}_2 &:= \delta_{02} \hspace{1in} \tilde{B}_2^+ := \delta_{20} \ ,
\end{split}
\end{equation}
and note that the generalization to $(N+1)$ gene states is clear. 

If constructed in this way, one should be careful to note that the two sets anticommute with themselves, but not necessarily with \textit{each other}. For example, while $\tilde{B}_1 \tilde{B}_1^+ + \tilde{B}_1^+ \tilde{B}_1 = I$, we have
\begin{equation}
\tilde{B}_1 \tilde{B}_2^+ + \tilde{B}_2^+ \tilde{B}_1 = \delta_{01} \delta_{20} + \delta_{20} \delta_{01} = \delta_{21} \neq 0 
\end{equation}
which is naively not what we would expect from `independent' fermionic operators. Using the same similarity transformation idea as before, we can define matrices
\begin{equation}
\begin{split}
B_i &:= Q \tilde{B}_i Q^{-1} \\
B_i^+ &:= Q \tilde{B}_i^+ Q^{-1}
\end{split}
\end{equation}
that inherit the properties of the $\tilde{B}_i$ and $\tilde{B}_i^+$, so that
\begin{equation}
H_s = - \lambda_1 \ B_1^+ B_1 - \lambda_2 \ B_2^+ B_2 \ ,
\end{equation}
which looks like a decomposition of $H_s$ in terms of two `independent' fermions.  Just like the set $\{ I, B, B^+, B^+ B \}$ formed a basis for the set of all $2 \times 2$ matrices, the set
\begin{equation}
\begin{split}
\{ I, B_1, B_2, B_1^+, B_2^+, B_1^+ B_1, B_2^+ B_2, B_2^+ B_1, B_1^+ B_2 \}
\end{split}
\end{equation}
forms a basis for the set of all $3 \times 3$ matrices. Supposing there are $(N+1)$ gene states and that $H_s$ has distinct eigenvalues, we could write
\begin{equation}
H_s = - \lambda_1 \ B_1^+ B_1 - \cdots - \lambda_N \ B_N^+ B_N
\end{equation}
and note that a basis for the space of all $(N+1) \times (N+1)$ matrices can be constructed using $B_i$, $B_i^+$, and $B_i^+ B_j$ (for all $i, j$), for a total of $N^2 + 2 N + 1 = (N + 1)^2$ basis elements.

\clearpage

\section{Tables of low order Feynman diagrams}
\label{sec:app_tables}

\begin{table}[ht!]
\centering
{\renewcommand{\arraystretch}{1.9}
 \begin{tabular}{| c | c | c |} 
 \hline
Coeff. & Values & Relevant diagrams \\ 
 \hline \hline
$\displaystyle q^0_{1,0}$ & $\displaystyle 0$ & no diagrams \\     \hline
$\displaystyle q^0_{1,1}$ & $\displaystyle \frac{c_3}{s + \gamma \Bstrut}$ &  \raisebox{-0.3\height}{\includegraphics[scale=0.1]{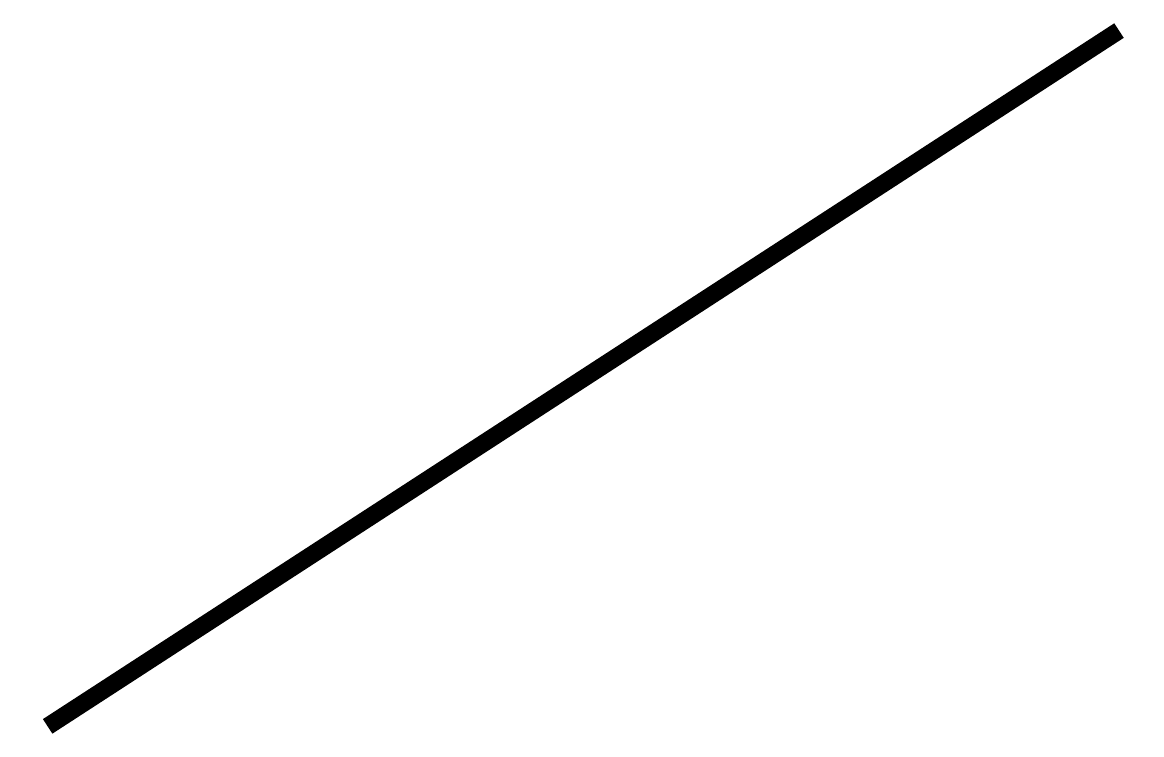}} \\ \hline
$\displaystyle q^0_{2,0}$ &  $\displaystyle \frac{c_3}{(s + \gamma) 2 \gamma \Bstrut}$ & \raisebox{-0.3\height}{\includegraphics[scale=0.1]{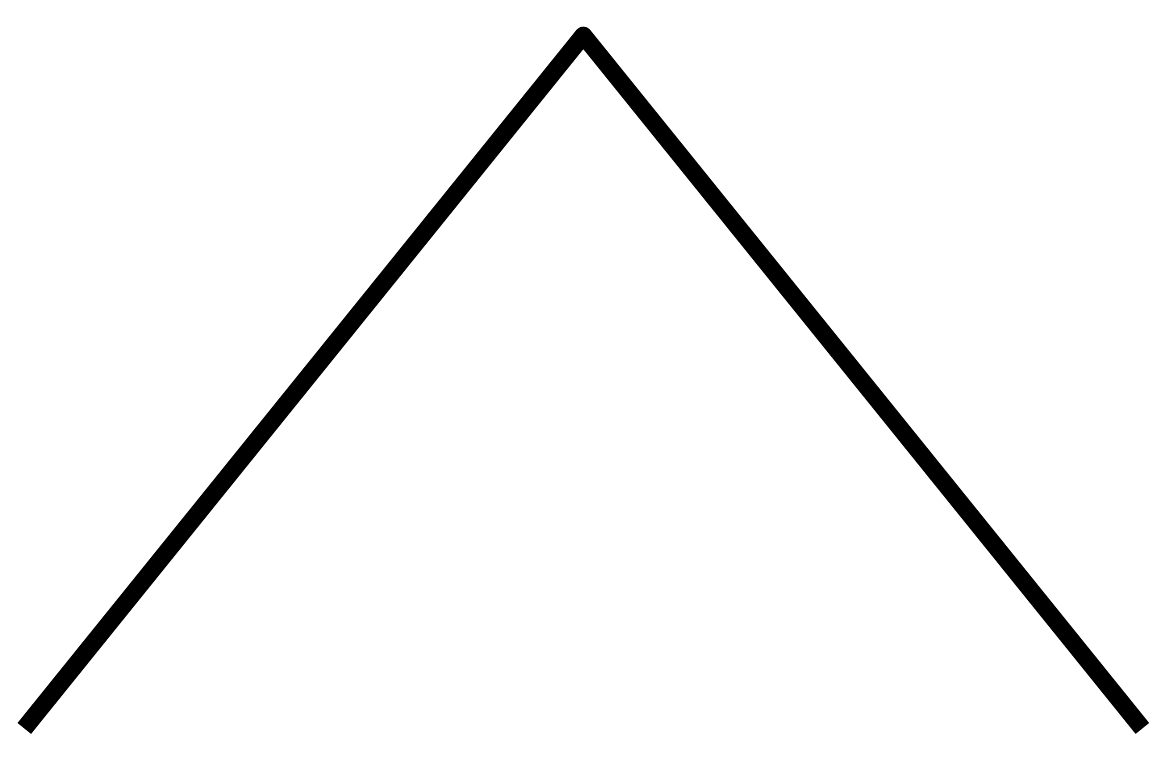}} \\ \hline
$\displaystyle q^0_{2,1}$  &   $\displaystyle \frac{c_3 c_4}{(s + \gamma) (s + 2 \gamma)\Bstrut}$ & \raisebox{-0.3\height}{\includegraphics[scale=0.1]{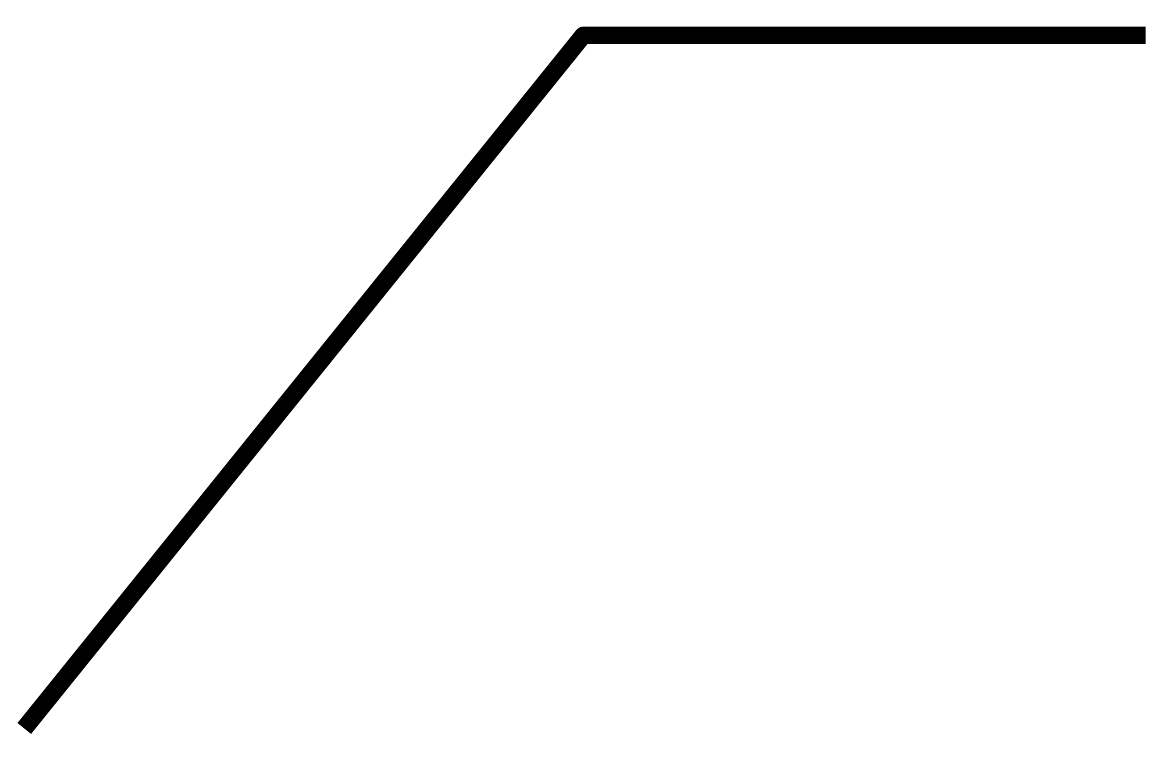}}  \\ \hline
$\displaystyle q^0_{3,0}$  &  $\displaystyle \frac{c_3 c_4}{(s + \gamma) (s + 2\gamma) 3 \gamma \Bstrut}$
 &   \raisebox{-0.3\height}{\includegraphics[scale=0.1]{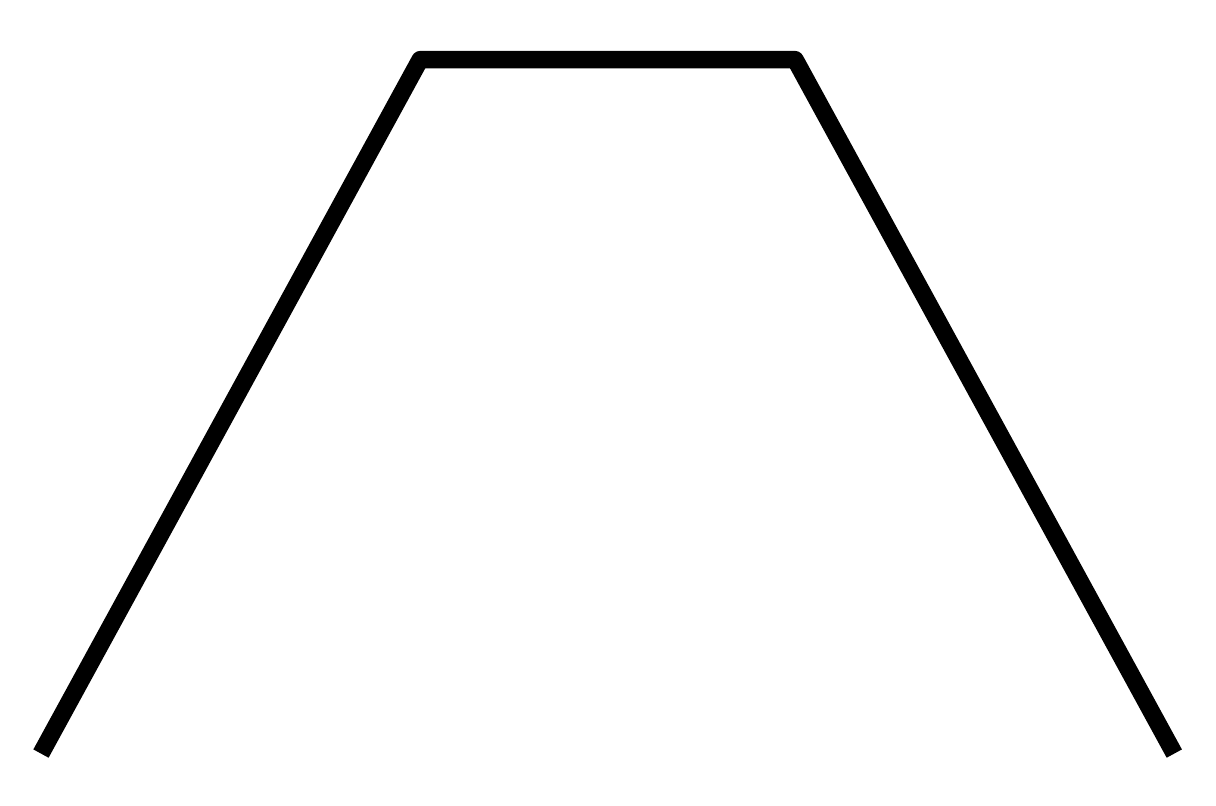}}  \\ \hline
$\displaystyle q^0_{3,1}$  &  $\displaystyle \frac{c_3 c_4^2}{(s + \gamma) (s + 2 \gamma) (s + 3 \gamma)} + \frac{c_3^2}{(s + \gamma) 2 \gamma (s + 3 \gamma) \Bstrut} $
 &  \raisebox{-0.3\height}{\includegraphics[scale=0.1]{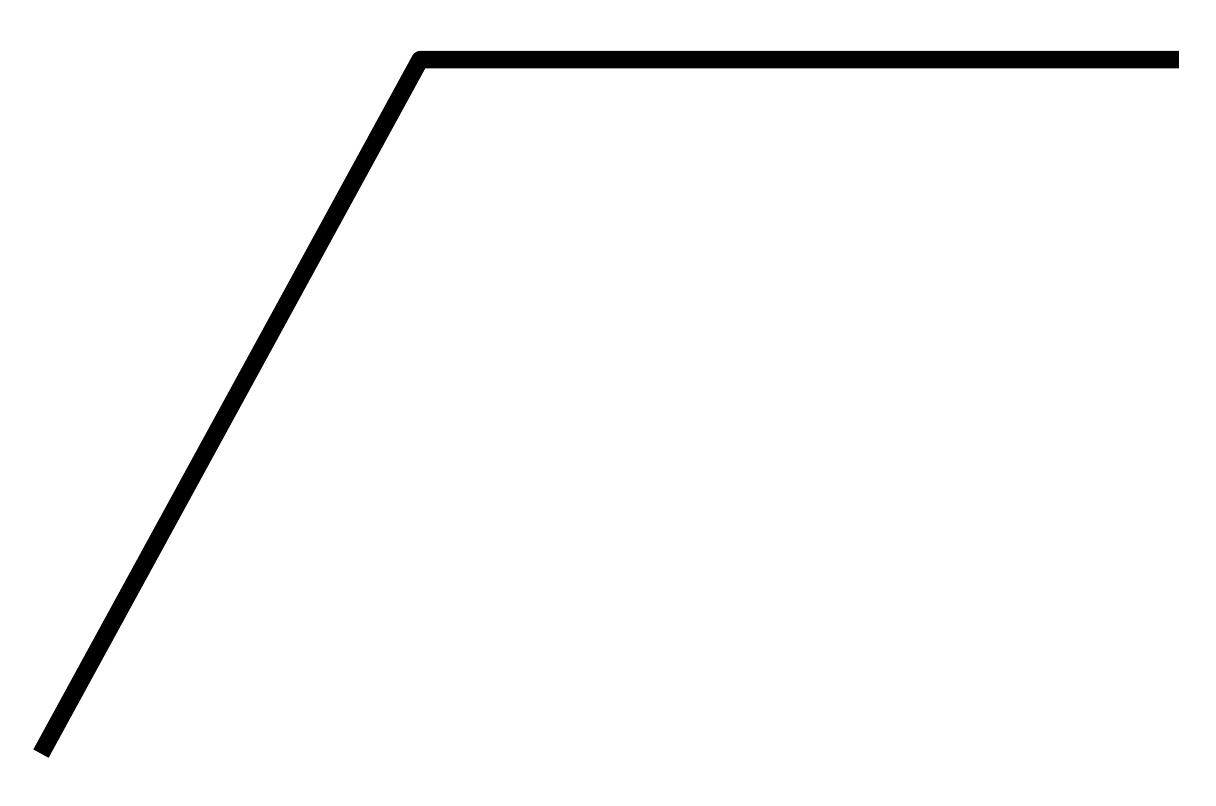}  \includegraphics[scale=0.1]{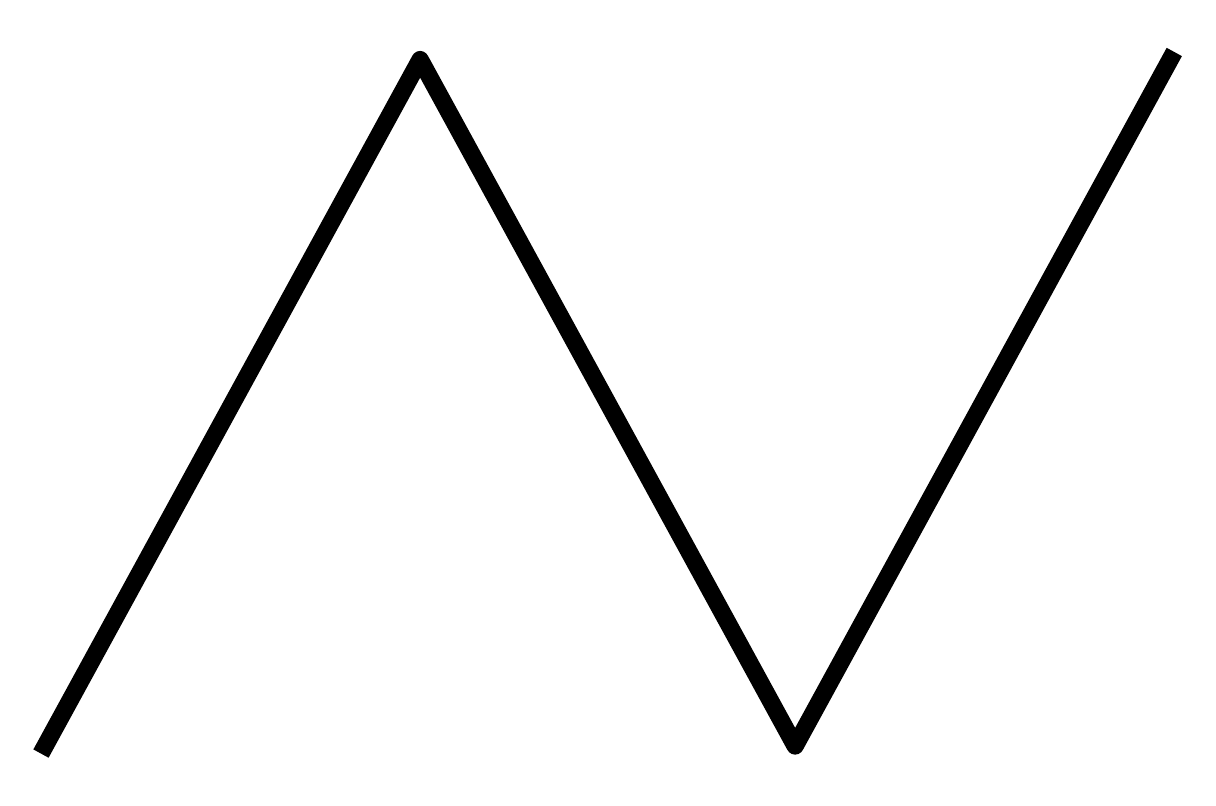}} \\ \hline
$\displaystyle q^0_{4,0}$  &  $\displaystyle \frac{c_3 c_4^2}{(s + \gamma) (s + 2 \gamma) (s + 3 \gamma) 4 \gamma} + \frac{c_3^2}{(s + \gamma) 2 \gamma (s + 3 \gamma) 4 \gamma \Bstrut} $
& \raisebox{-0.3\height}{\includegraphics[scale=0.1]{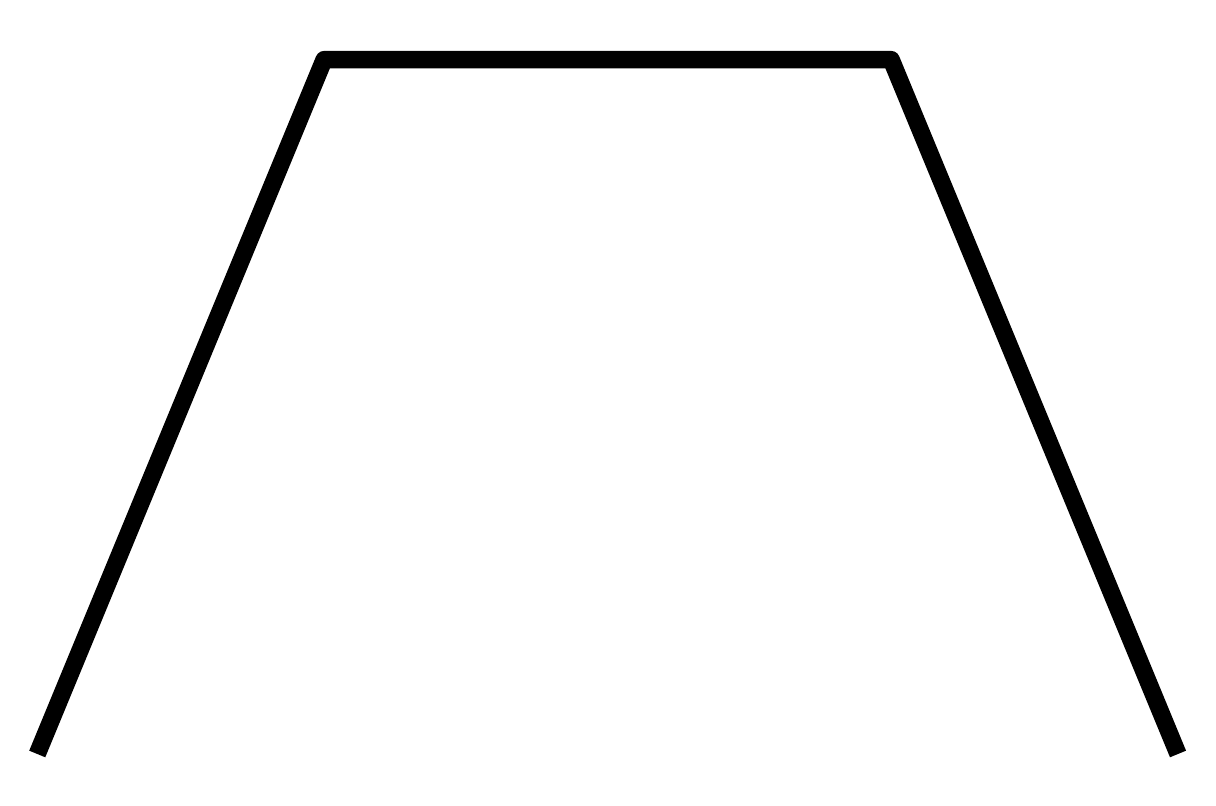} \includegraphics[scale=0.1]{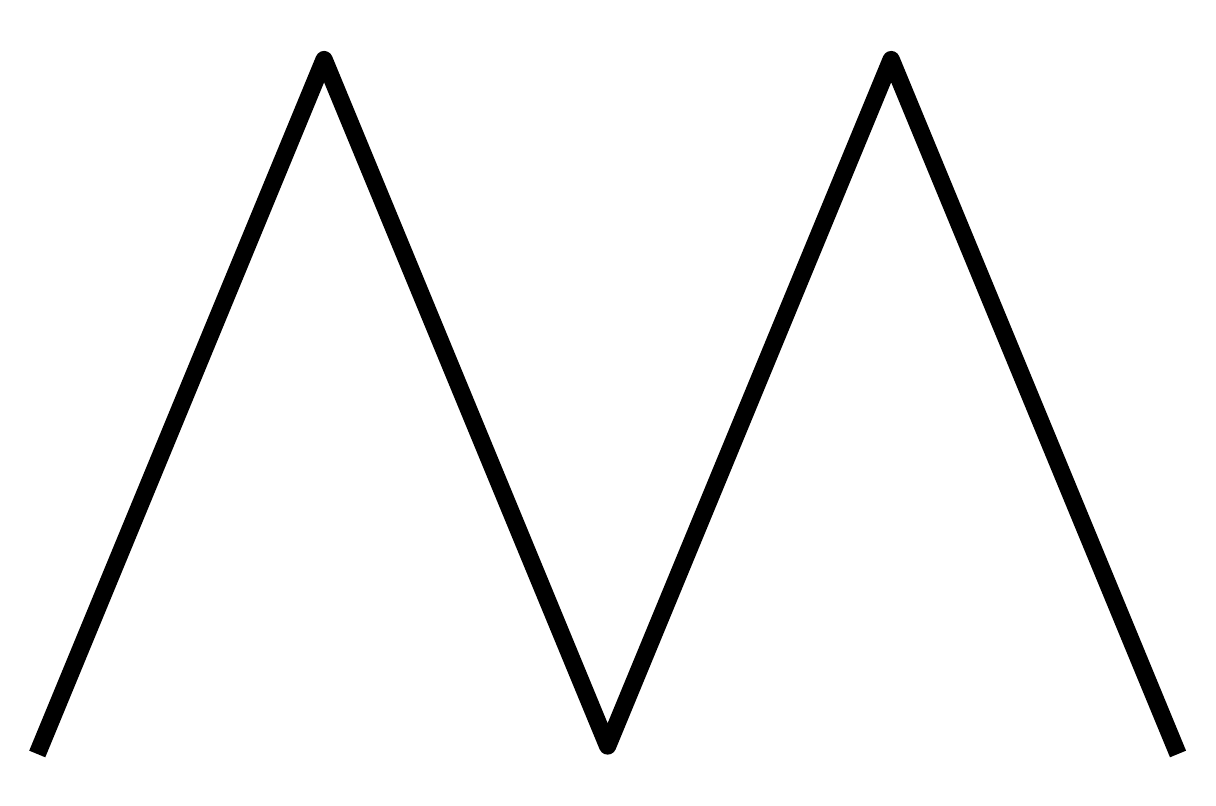}}   \\ \hline
$\displaystyle q^0_{4,1}$  &  $\displaystyle \frac{c_3 c_4^3}{(s + \gamma)(s + 2\gamma)(s + 3\gamma)(s + 4\gamma) \Bstrut} $ 
& \raisebox{-0.5\height}{\includegraphics[scale=0.1]{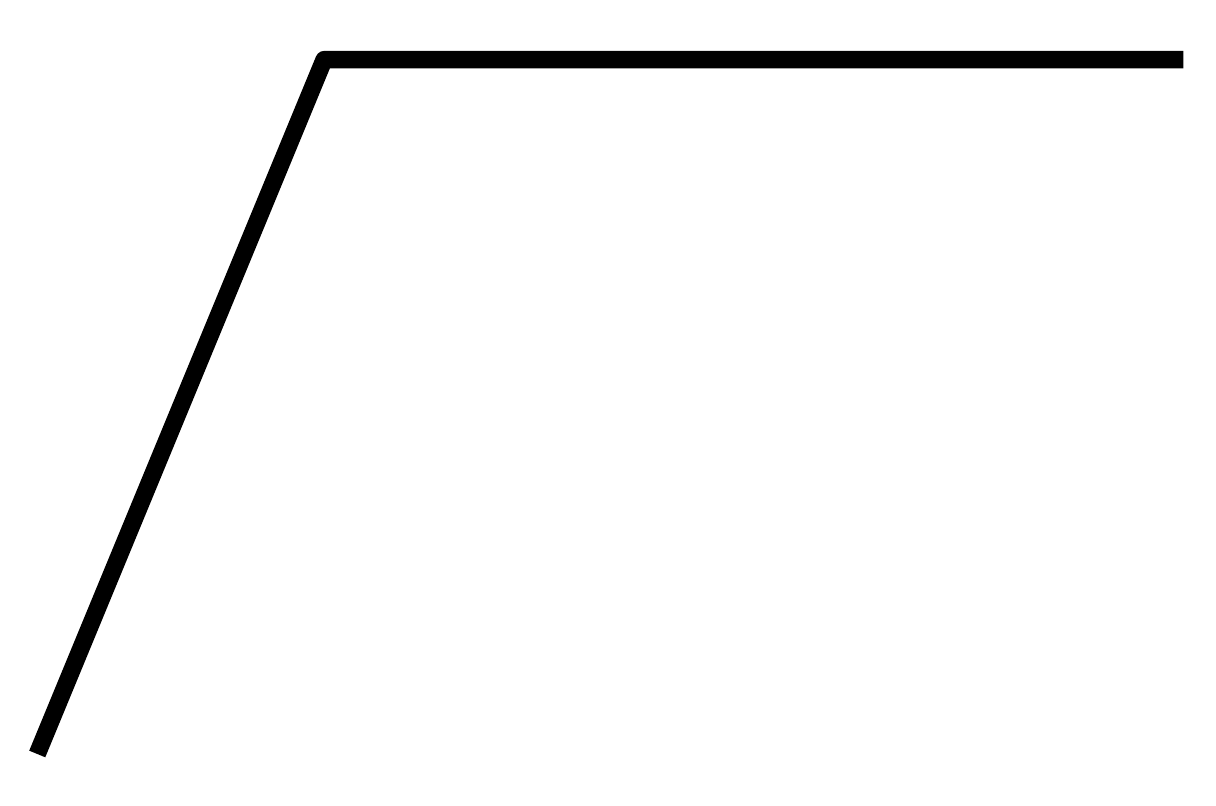}}  \\ 
 &  $\displaystyle + \frac{c_3^2 c_4}{(s + \gamma) 2\gamma (s + 3\gamma)(s + 4\gamma)}  + \frac{c_3^2 c_4}{(s + \gamma)(s + 2\gamma) 3\gamma(s + 4\gamma) \Bstrut} $ 
& \includegraphics[scale=0.1]{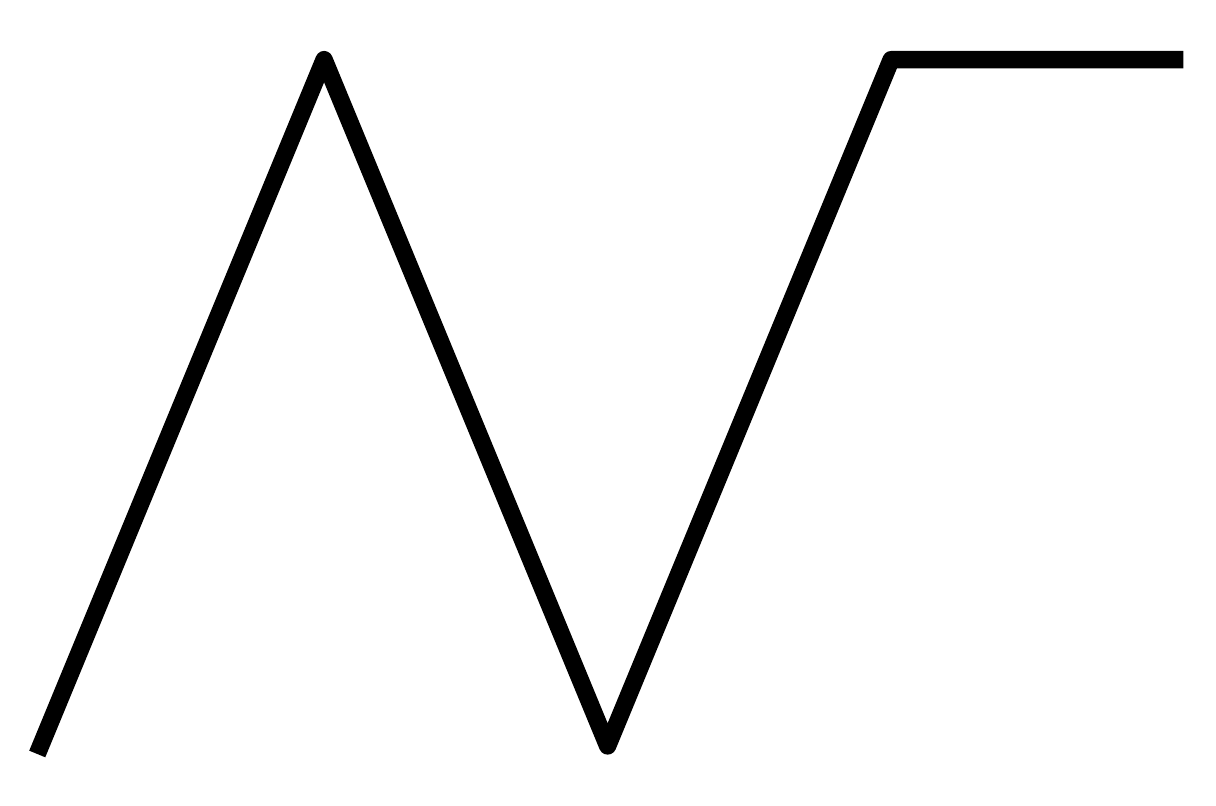} \includegraphics[scale=0.1]{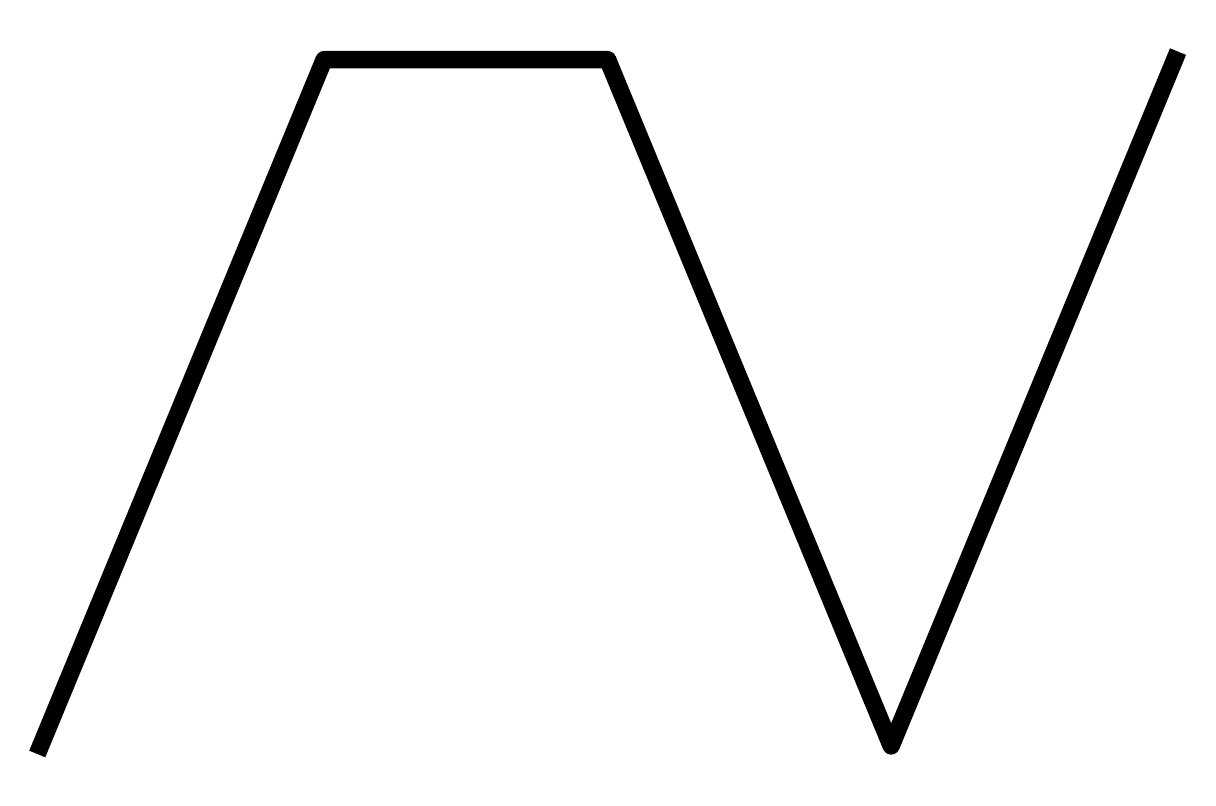}  \\
 \hline
 \end{tabular}}
 \caption{The first few coefficients $q^0_{k, g}$ and the corresponding diagrams.}
\label{table:feynman_examples0}
\end{table}

\begin{table}[ht!]
\centering
{\renewcommand{\arraystretch}{1.9}
 \begin{tabular}{| c | c | c |} 
 \hline
Coeff. & Values & Relevant diagrams \\ 
 \hline \hline
$\displaystyle q^1_{1,0}$ & $\displaystyle \frac{1}{\gamma - s \Bstrut}$ &  \raisebox{-0.3\height}{\includegraphics[scale=0.1]{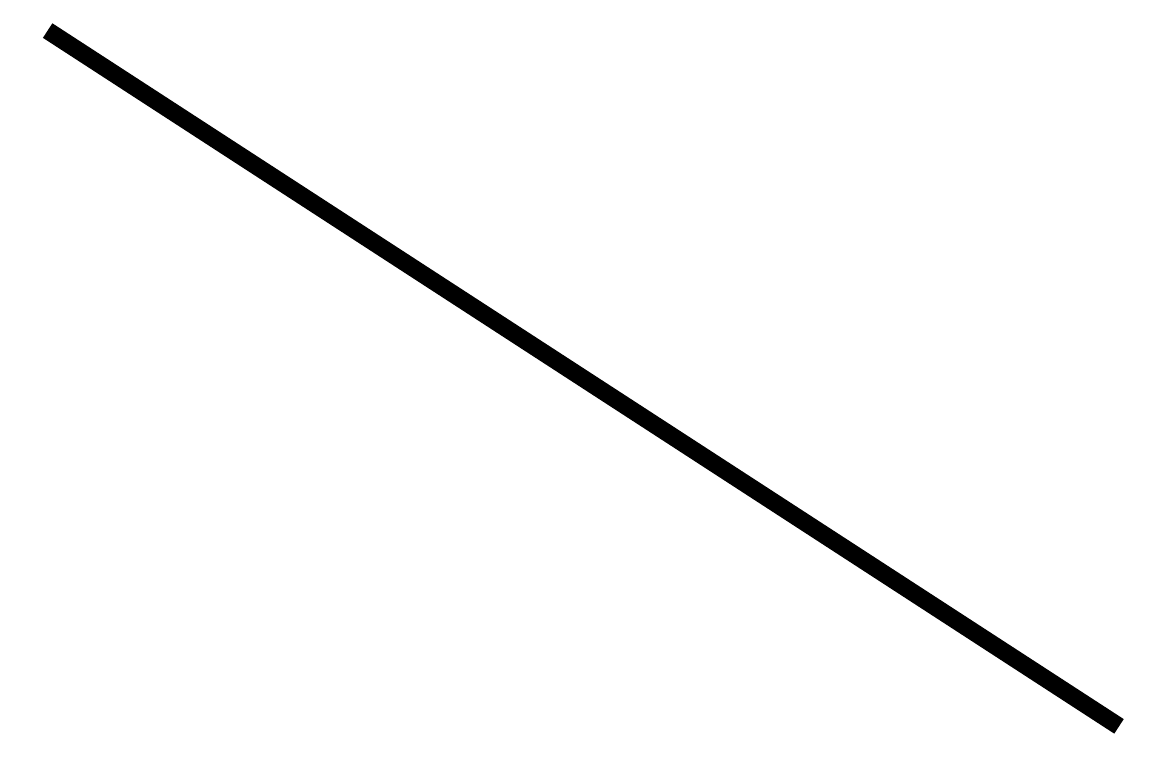}} \\ \hline
$\displaystyle q^1_{1,1}$ & $\displaystyle \frac{c_4}{\gamma \Bstrut}$ &  \raisebox{-0.3\height}{\includegraphics[scale=0.1]{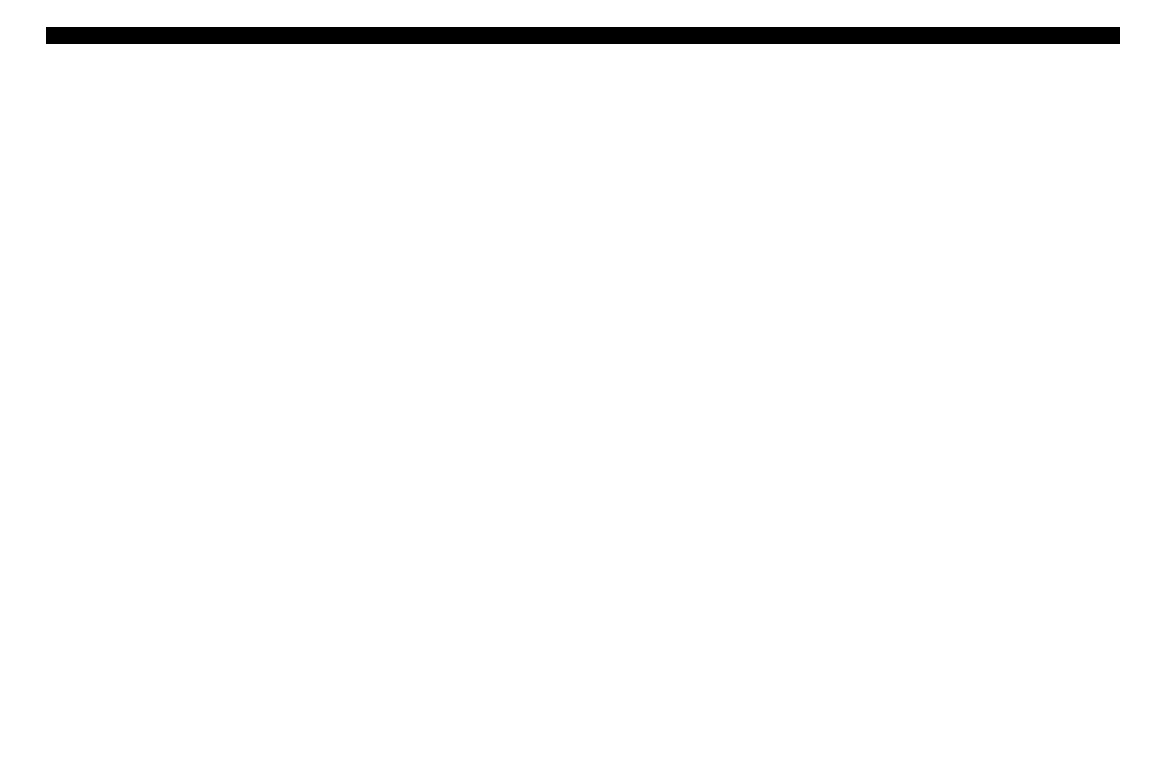}} \\ \hline
$\displaystyle q^1_{2,0}$ &  $\displaystyle \frac{c_4}{\gamma (2 \gamma - s) \Bstrut}$ & \raisebox{-0.3\height}{\includegraphics[scale=0.1]{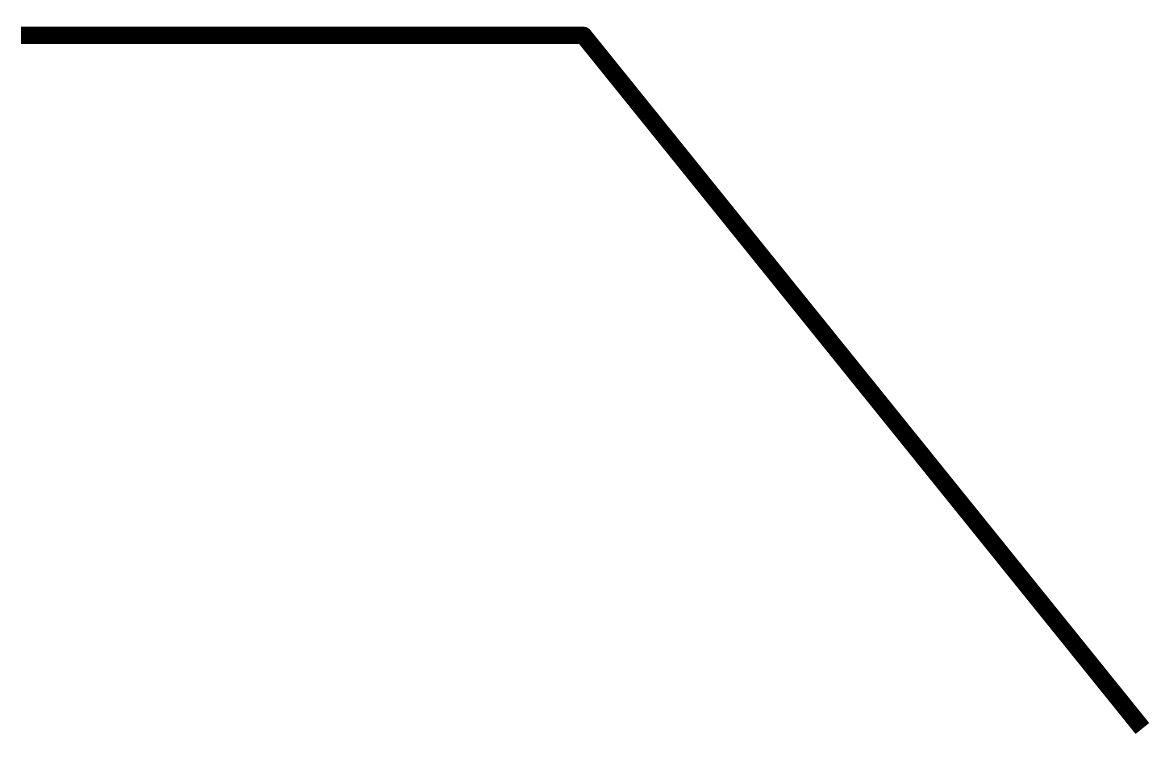}} \\ \hline
$\displaystyle q^1_{2,1}$  &   $\displaystyle \frac{c_3}{(\gamma - s) 2 \gamma} + \frac{c_4^2}{\gamma \cdot 2 \gamma \Bstrut}$ & \raisebox{-0.3\height}{\includegraphics[scale=0.1]{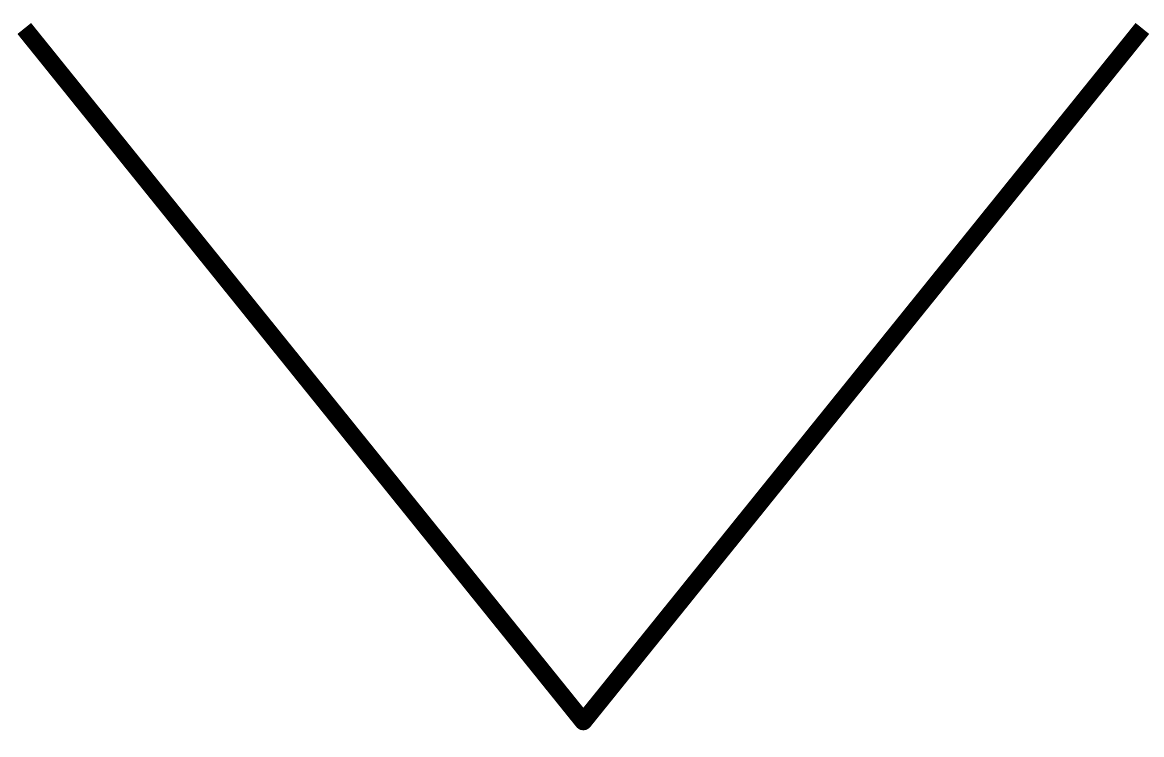}  \includegraphics[scale=0.1]{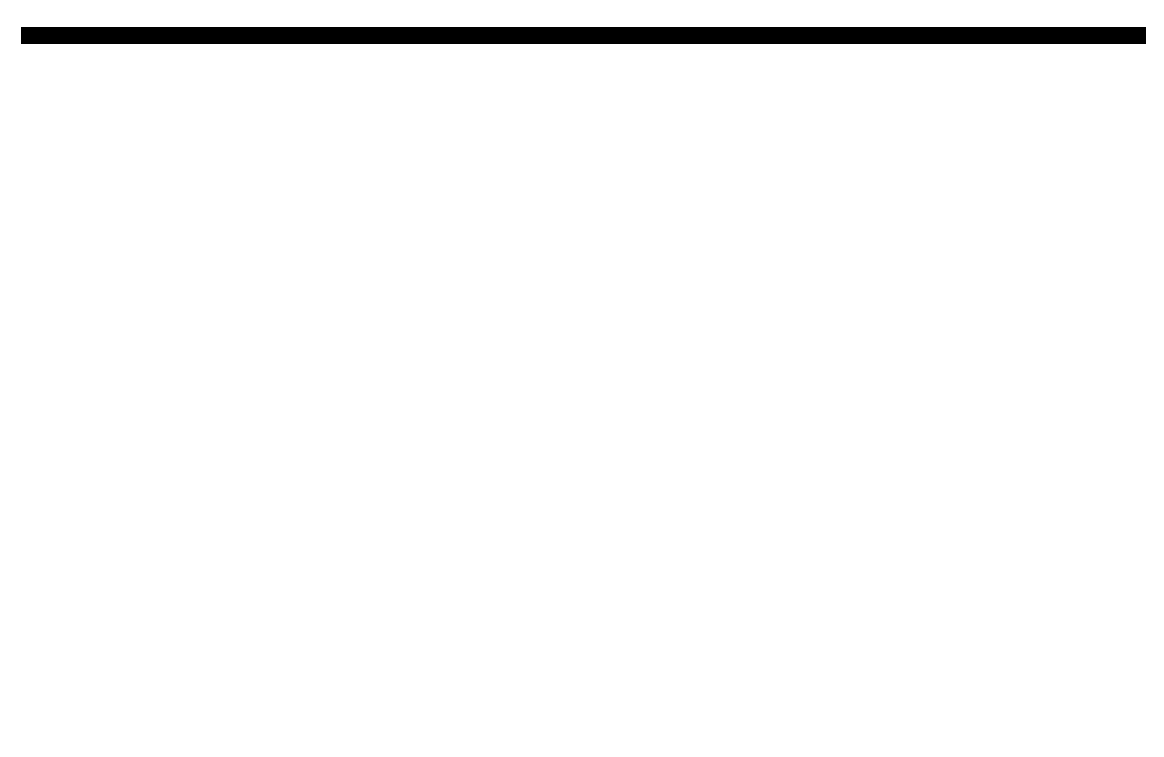}}  \\ \hline
$\displaystyle q^1_{3,0}$  &  $\displaystyle \frac{c_3}{(\gamma - s) 2 \gamma (3 \gamma - s)} + \frac{c_4^2}{\gamma \cdot 2 \gamma (3 \gamma - s) \Bstrut} $
 &   \raisebox{-0.3\height}{\includegraphics[scale=0.1]{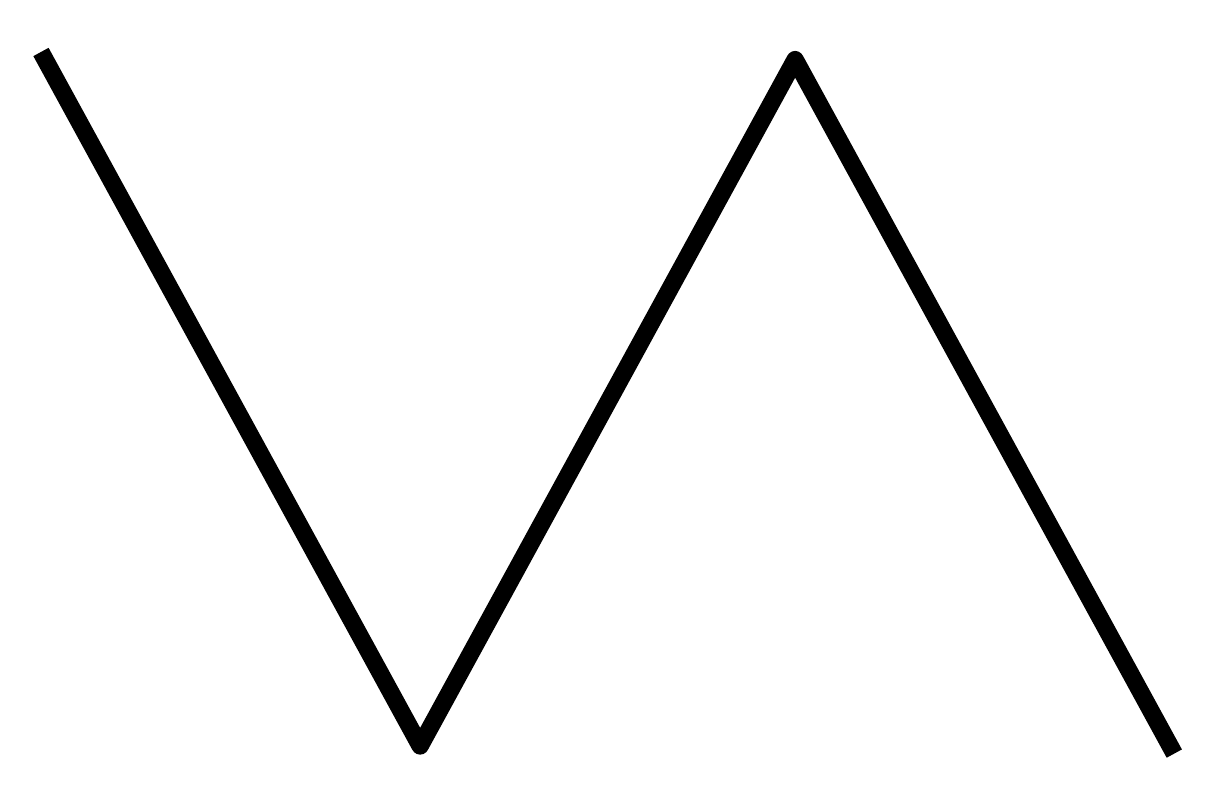} \includegraphics[scale=0.1]{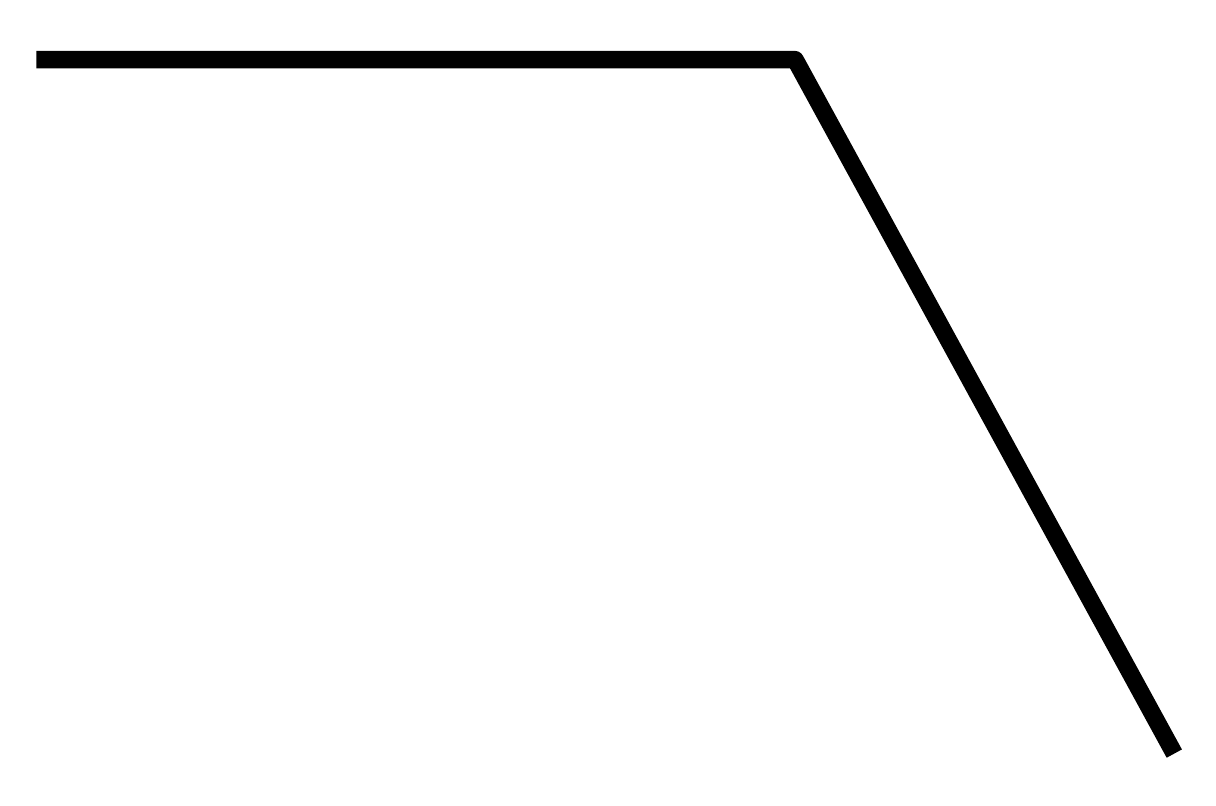}}  \\ \hline
$\displaystyle q^1_{3,1}$  &  $\displaystyle  \frac{c_3 c_4}{(\gamma - s) 2 \gamma \cdot 3 \gamma} + \frac{c_3 c_4}{\gamma (2\gamma - s) 3 \gamma} + \frac{c_4^3}{\gamma \cdot 2 \gamma \cdot 3 \gamma \Bstrut}$
 &   \raisebox{-0.3\height}{\includegraphics[scale=0.1]{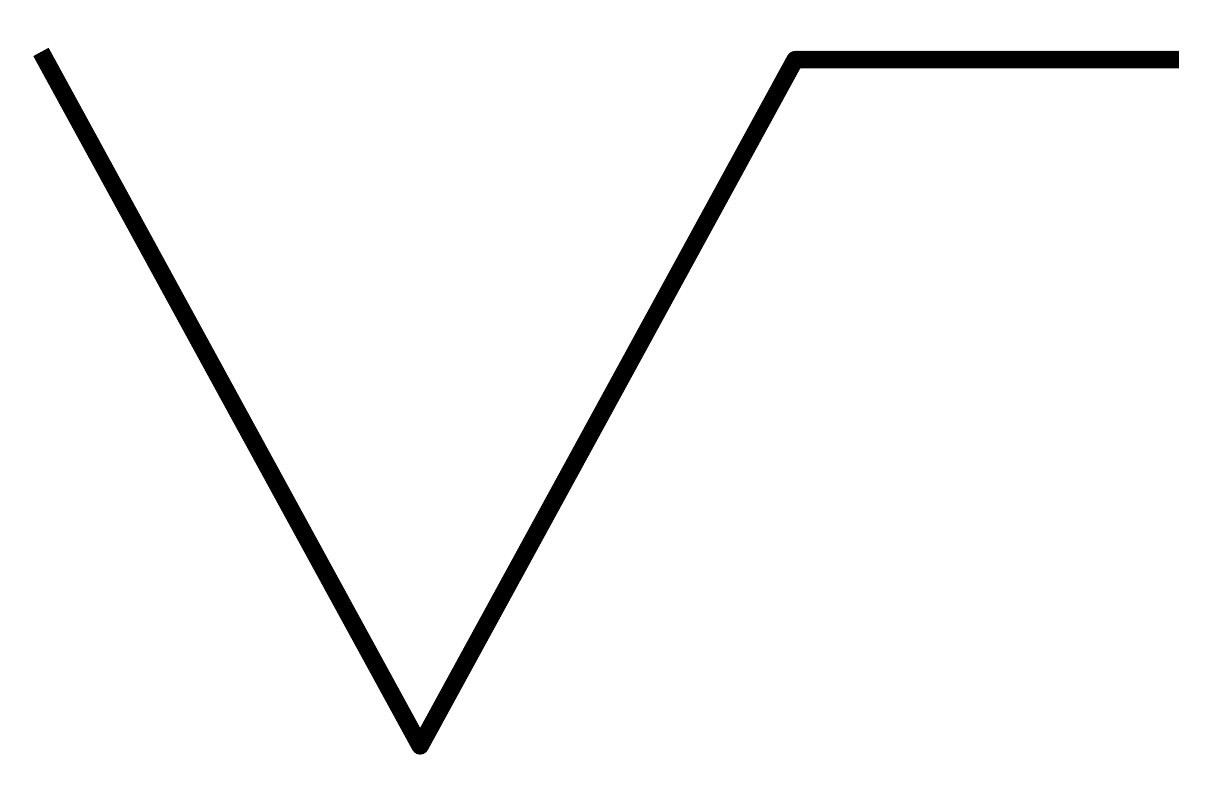} \includegraphics[scale=0.1]{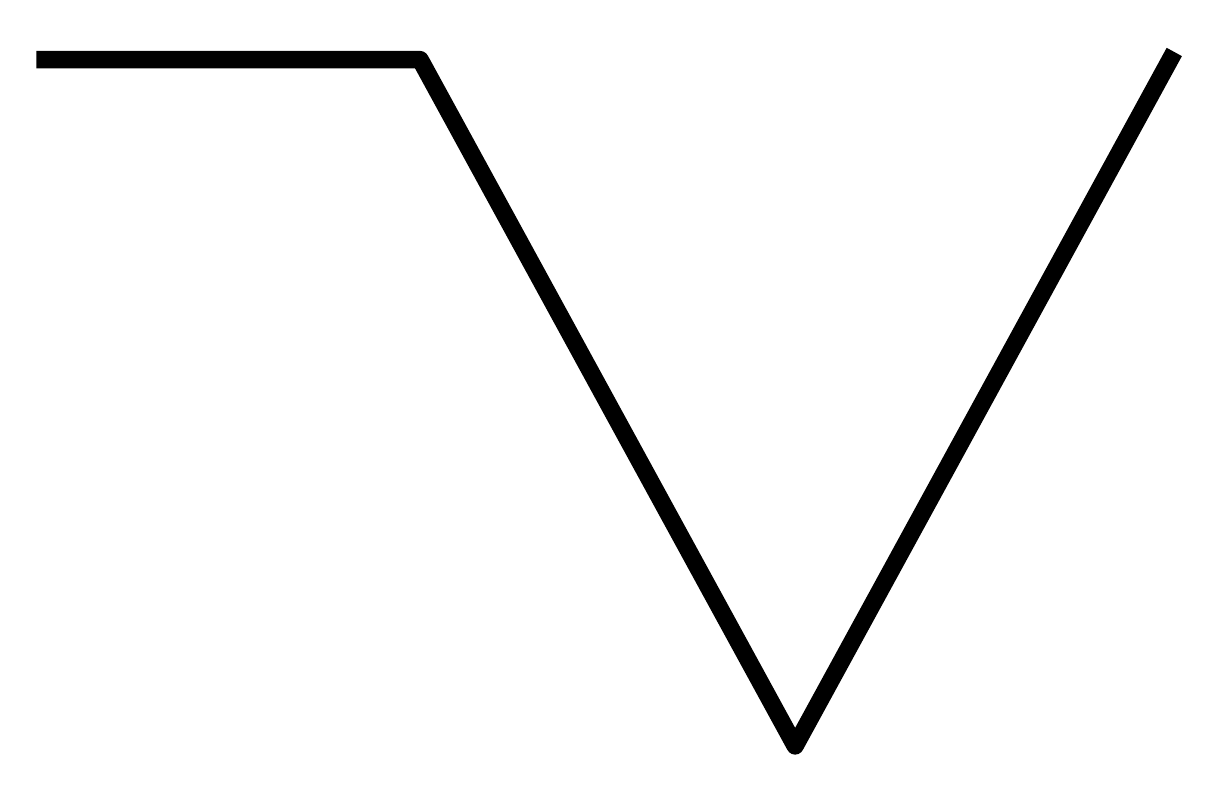}  \includegraphics[scale=0.1]{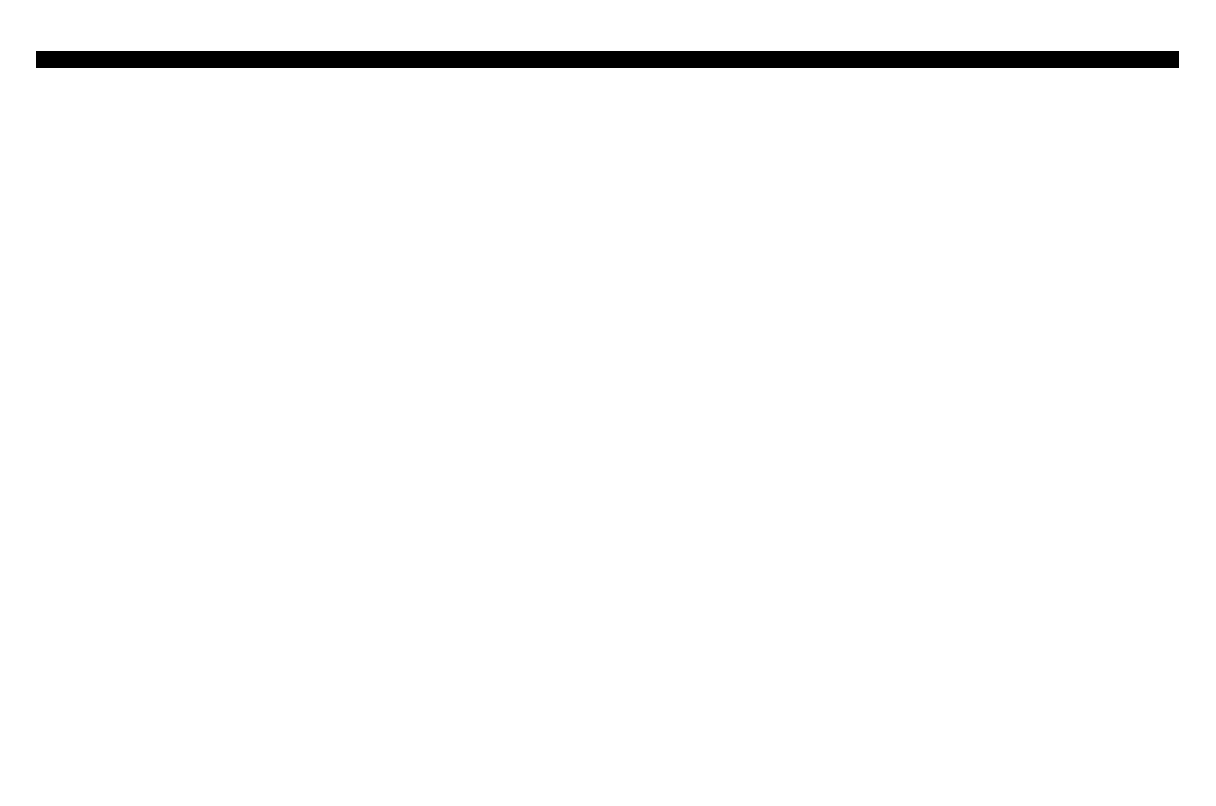}}  \\ \hline
$\displaystyle q^1_{4,0}$  &  $\displaystyle  \frac{c_3 c_4}{\gamma (2\gamma - s) 3 \gamma (4\gamma - s) \Bstrut}$
& \raisebox{-0.3\height}{\includegraphics[scale=0.1]{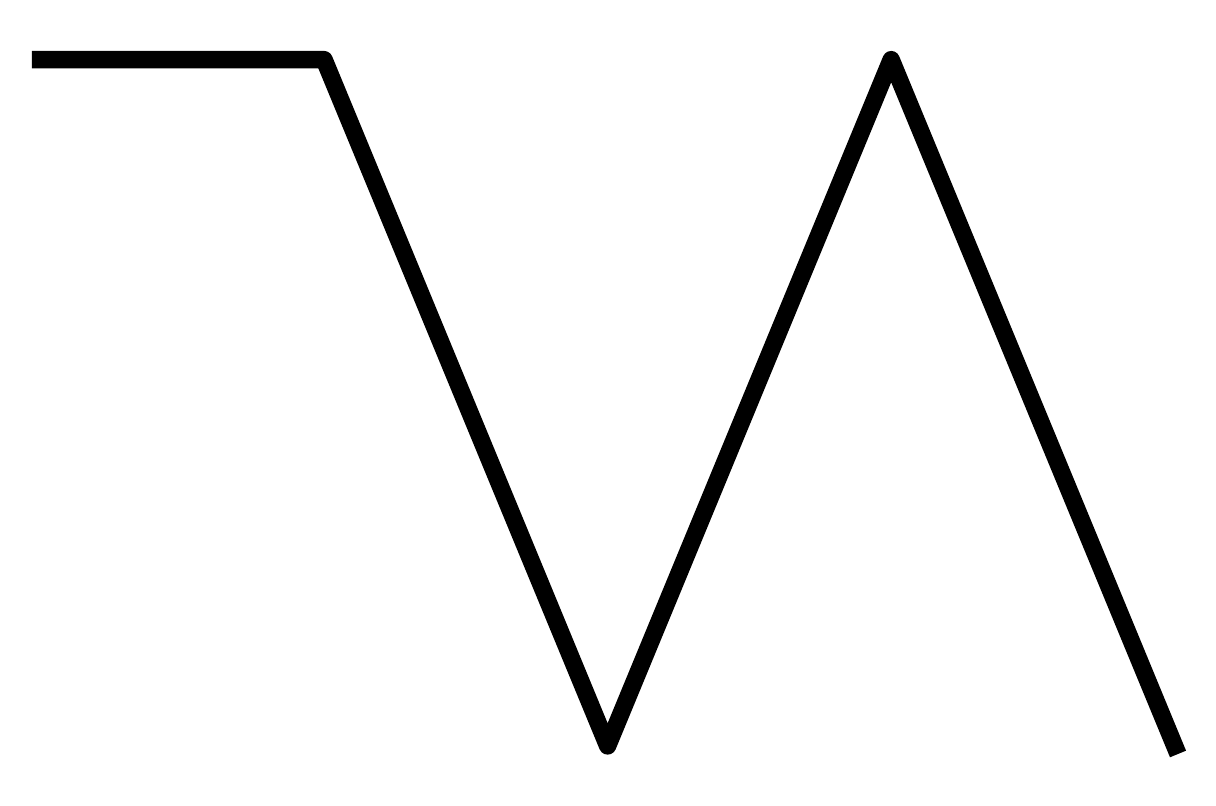}}    \\ 
  &  $\displaystyle  + \frac{c_3 c_4}{(\gamma - s) 2 \gamma \cdot 3 \gamma (4\gamma - s)} + \frac{c_4^3}{\gamma \cdot 2 \gamma \cdot 3 \gamma (4\gamma - s)\Bstrut}$
& \raisebox{-0.3\height}{\includegraphics[scale=0.1]{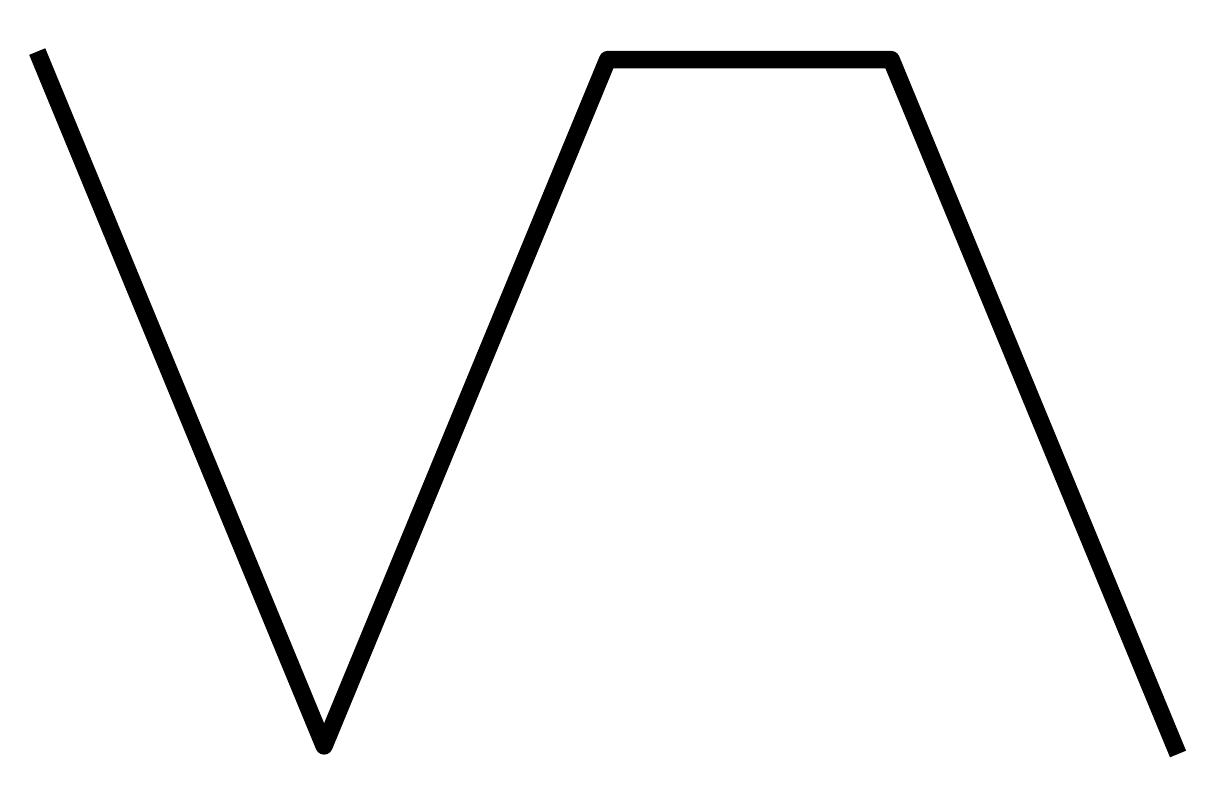} \includegraphics[scale=0.1]{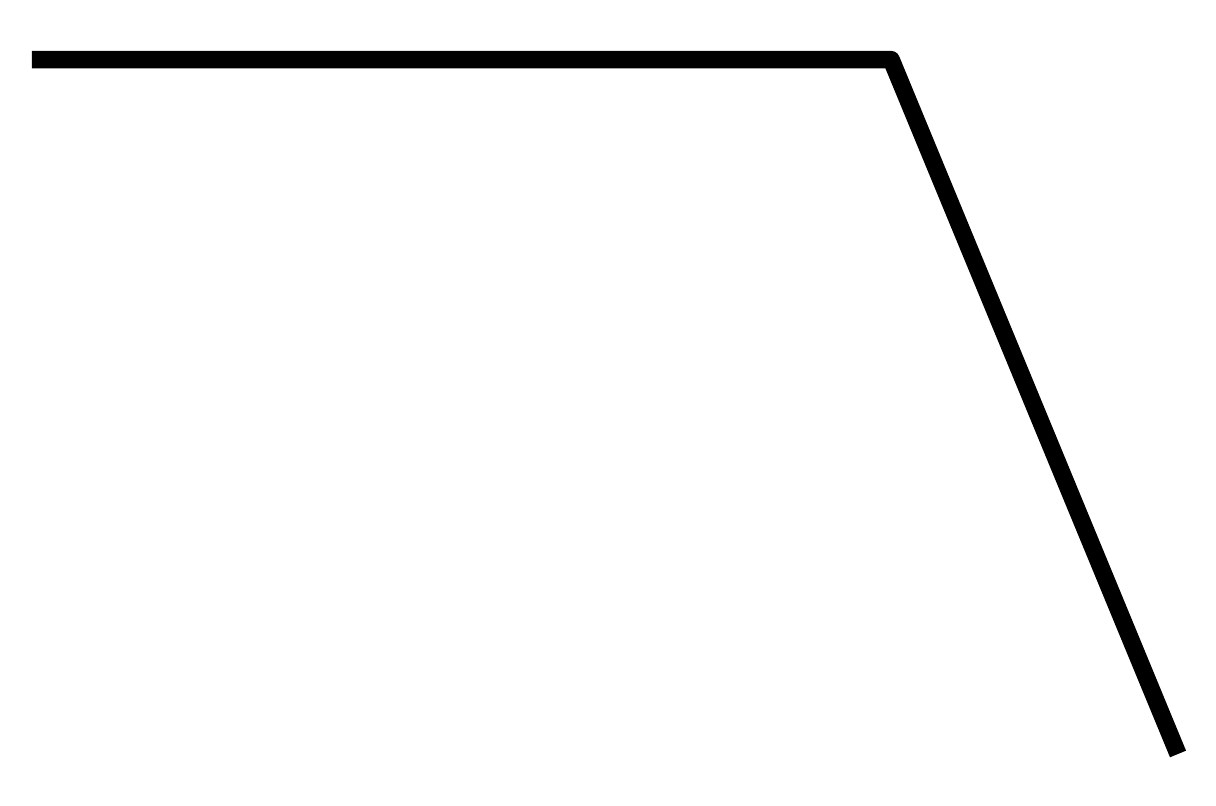}} \\ \hline
$\displaystyle q^1_{4,1}$  & $\displaystyle \frac{c_4^4}{\gamma \cdot 2 \gamma \cdot 3 \gamma \cdot 4 \gamma} + \frac{c_3 c_4^2}{(\gamma - s) 2\gamma \cdot 3 \gamma \cdot 4 \gamma\Bstrut} $   
& \raisebox{-0.3\height}{\includegraphics[scale=0.1]{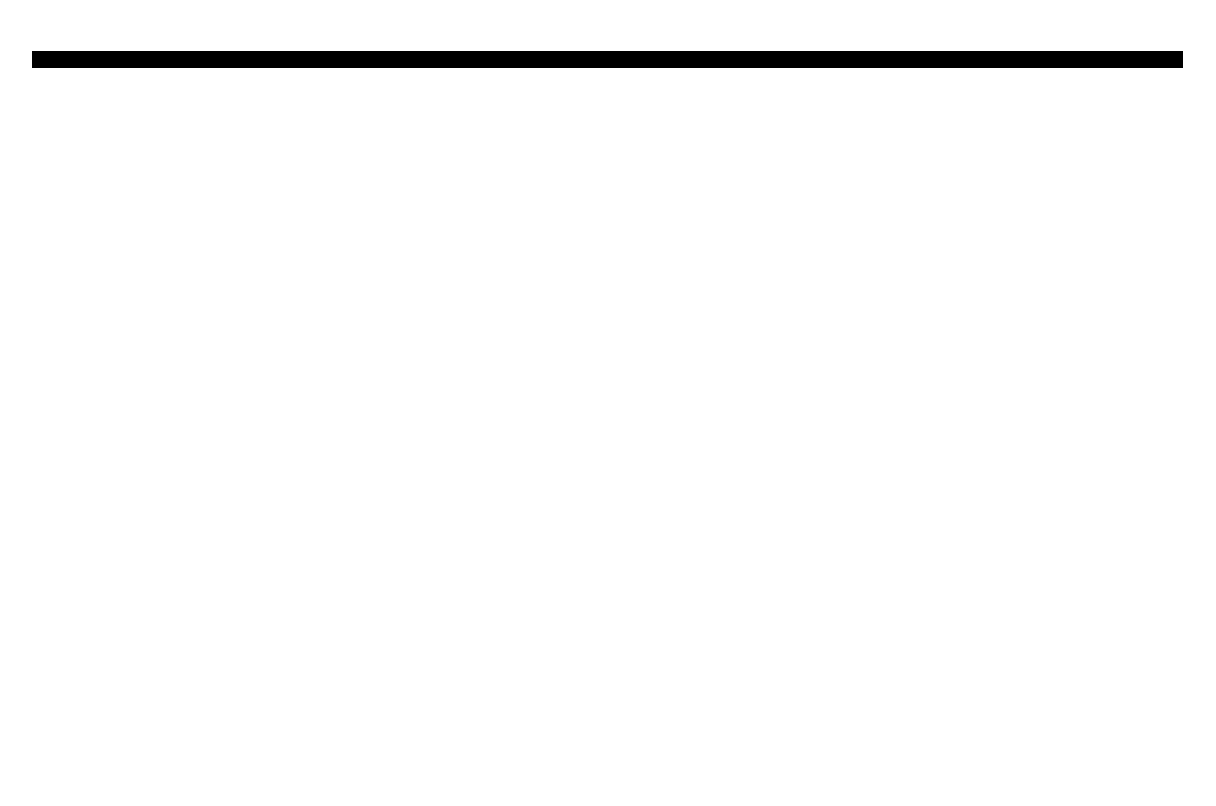} \includegraphics[scale=0.1]{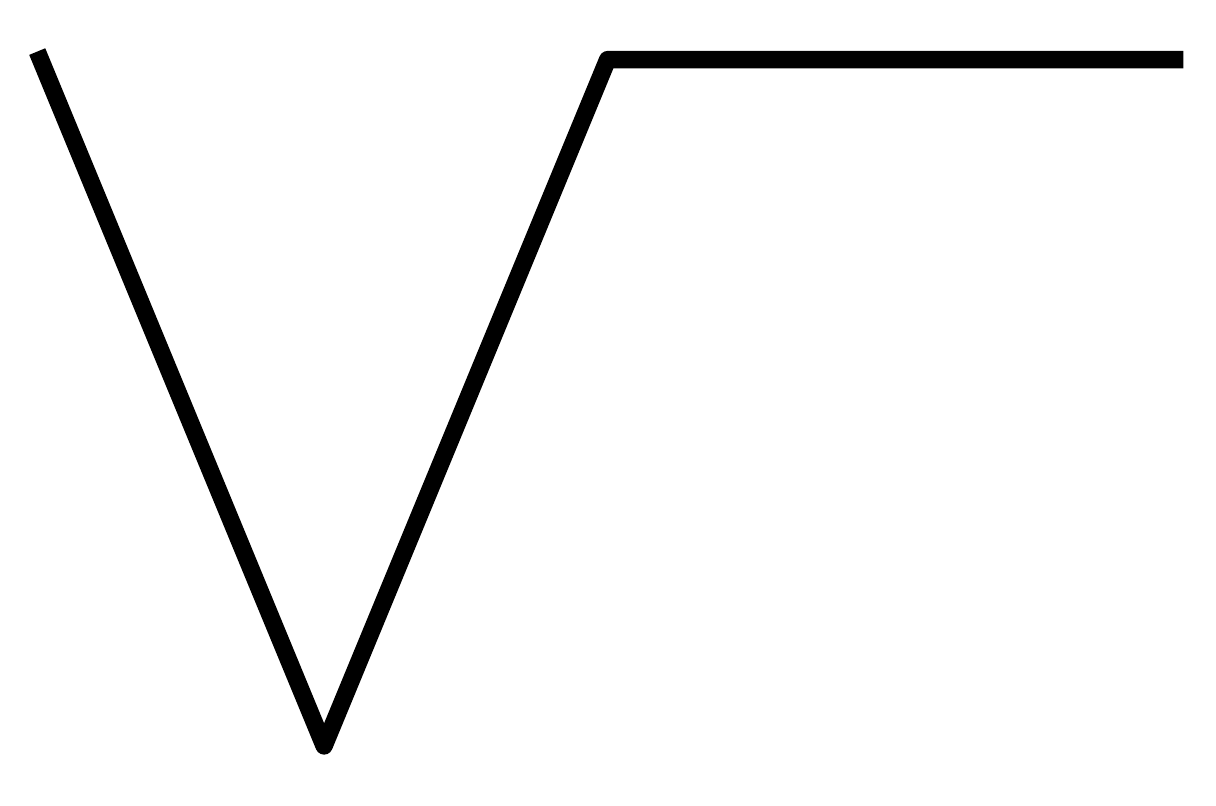}}   \\
 &  $\displaystyle + \frac{c_3 c_4^2}{\gamma (2\gamma - s) 3 \gamma \cdot 4 \gamma \Bstrut} $ 
&  \raisebox{-0.3\height}{\includegraphics[scale=0.1]{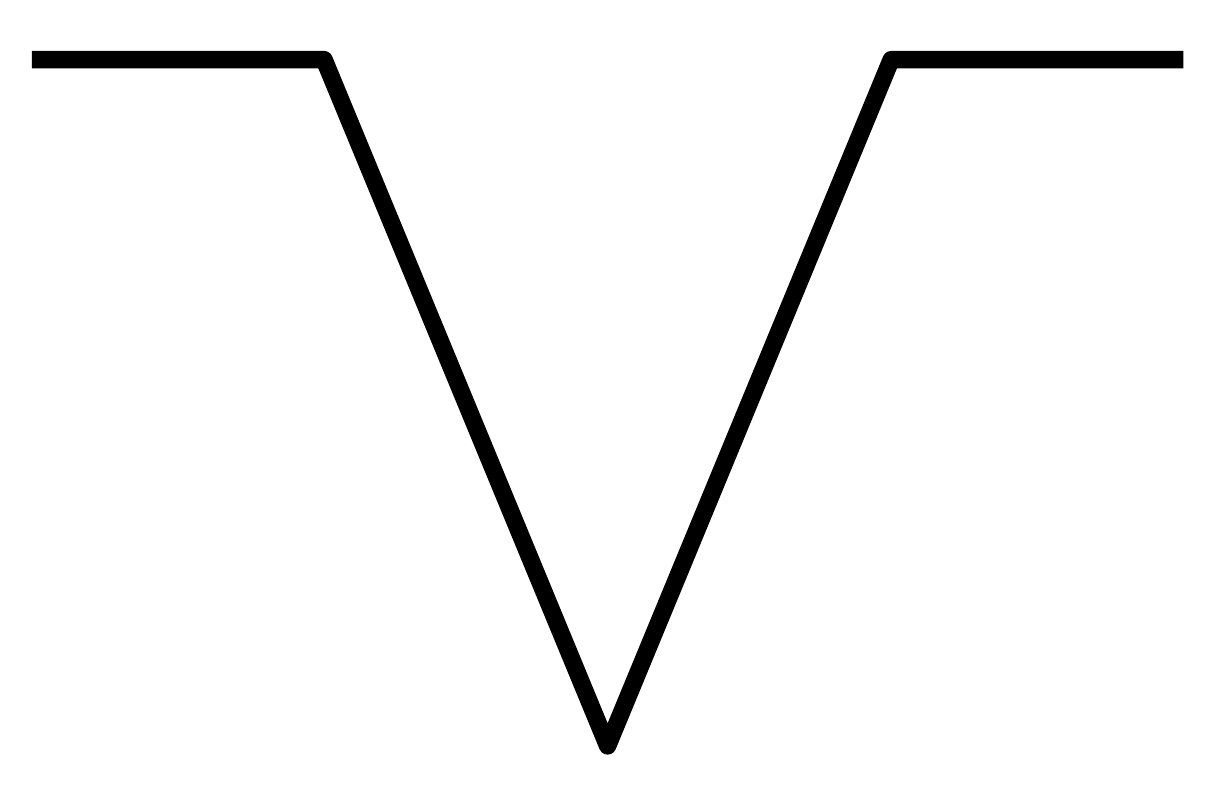}}  \\
 &  $\displaystyle + \frac{c_3^2}{(\gamma - s) 2\gamma (3 \gamma - s) 4 \gamma} + \frac{c_3 c_4^2}{\gamma \cdot 2 \gamma (3 \gamma - s) 4 \gamma \Bstrut} $ 
& \raisebox{-0.3\height}{\includegraphics[scale=0.1]{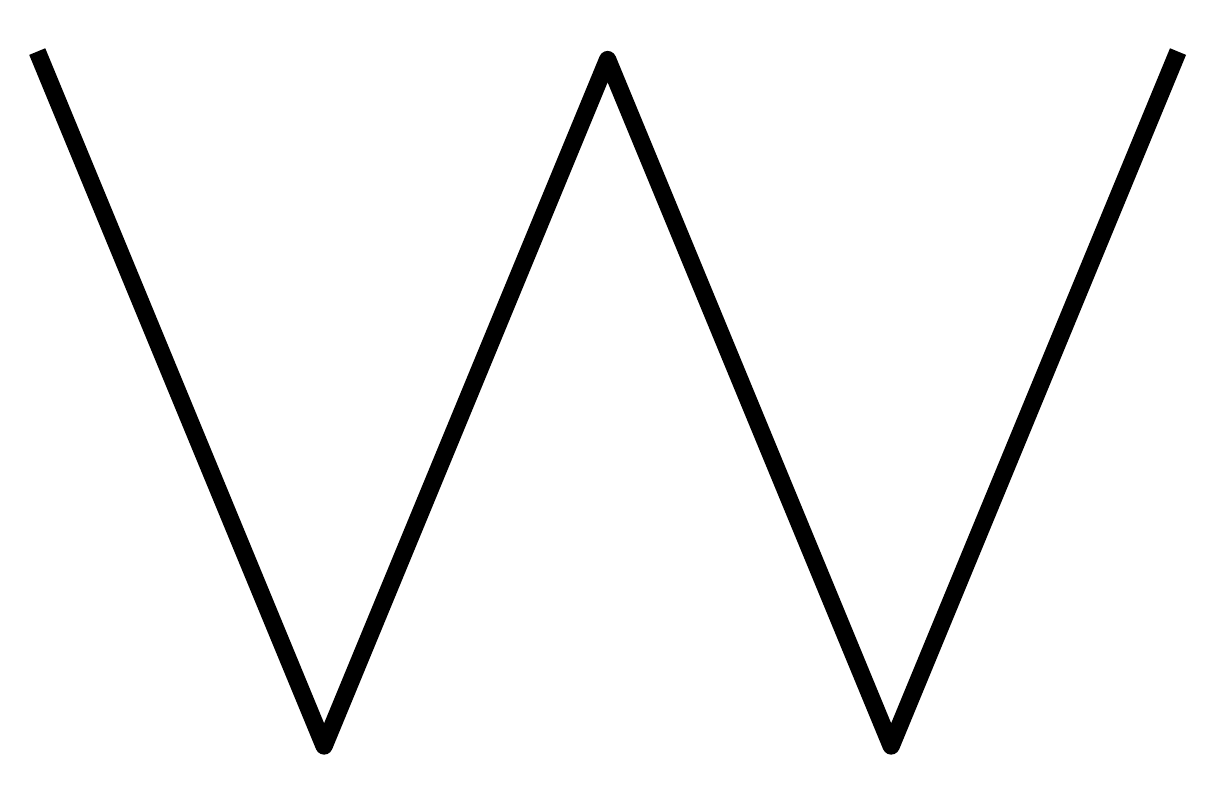} \includegraphics[scale=0.1]{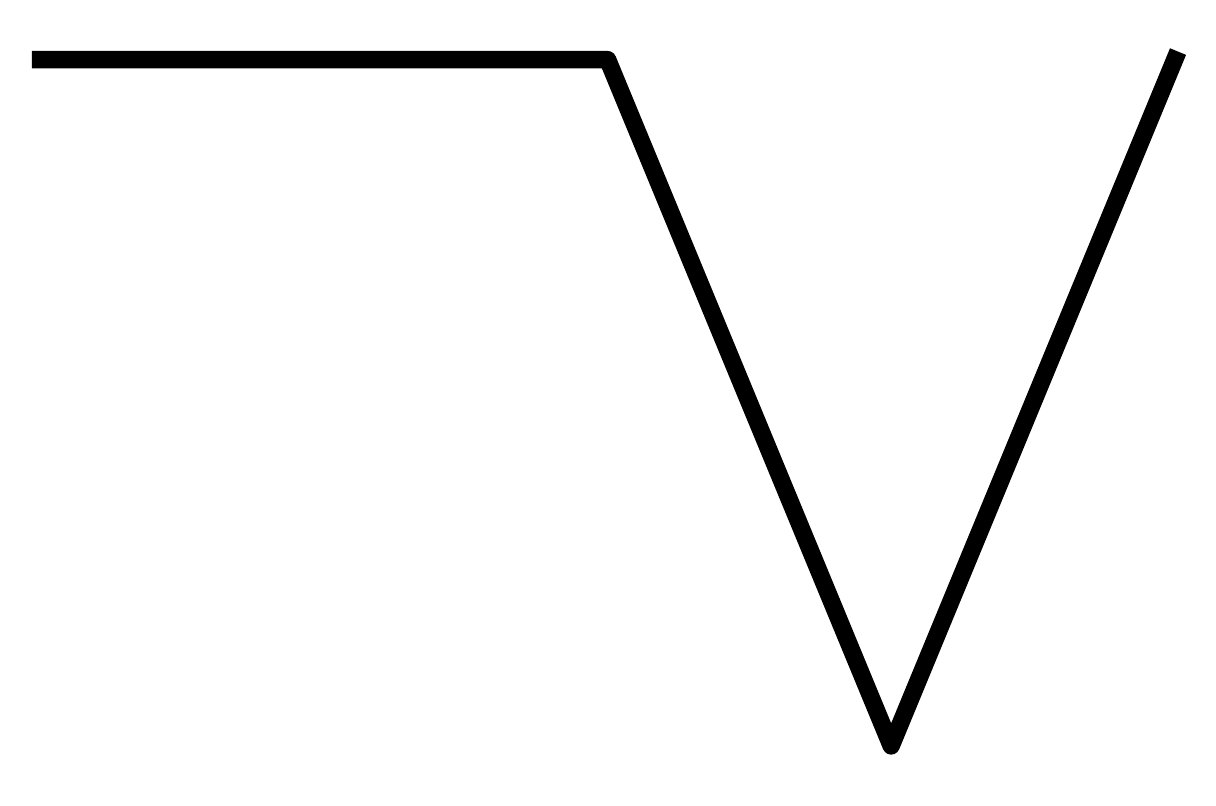}}  \\
 \hline
 \end{tabular}}
 \caption{The first few coefficients $q^1_{k, g}$ and the corresponding diagrams.}
\label{table:feynman_examples1}
\end{table}

\clearpage

\section{Multistep orthogonal polynomials}
\label{sec:app_multistep}

In this appendix, we define a new family of orthogonal polynomials that appear in the solution of the multistep splicing dynamics problem, and discuss their properties. They generalize the Charlier polynomials in a natural way, and many Charlier polynomial properties have analogues in this more general context. 

Fix a natural number $N \geq 0$ and mean parameter $\boldsymbol{\mu} := (\mu_0, ..., \mu_N) \in \mathbb{R}^{N+1}$. These polynomials will be functions of $\mathbf{x} := (x_0, ..., x_N) \in \mathbb{N}^{N+1}$. Denote the polynomial associated with the integer vector $\mathbf{n} := (n_0, n_1, ..., n_N) \in \mathbb{N}^{N+1}$ by $V_{\mathbf{n}}(\mathbf{x},\boldsymbol{\mu})$. We assume we have `ladder operators' made up of linear combinations of finite difference operators
\begin{equation}
\begin{split}
\hat{L}_i^+ &:= \sum_{j = 0}^N v^{(i)}_j \left( - \nabla_j  \right) \ \hspace{1in} i = 0, 1, ..., N 
\end{split}
\end{equation}
where $\mathbf{v}^{(i)}$ contains the coefficients of the $i$th ladder operator, e.g. 
\begin{equation}
\begin{split}
\hat{A}_{0}^+ =& \ \hat{a}_{0}^+ + \frac{\beta_{0}}{\beta_{1} - \beta_{0}} \ \hat{a}_{1}^+ + \frac{\beta_{0} \ \beta_{1}}{(\beta_{1} - \beta_{0}) (\beta_2 - \beta_{0})} \ \hat{a}_2^+ \\
& \implies \ \mathbf{v}^{(0)} = \begin{pmatrix} 1 & \frac{\beta_{0}}{\beta_{1} - \beta_{0}} & \frac{\beta_{0} \ \beta_{1}}{(\beta_{1} - \beta_{0}) (\beta_2 - \beta_{0})} \end{pmatrix}^T \ ,
\end{split}
\end{equation}
and where the backward difference operator $\nabla_i$ acts on a function $f(\mathbf{x})$ according to
\begin{equation}
\nabla_i f(\mathbf{x}) := f(\mathbf{x}) - f(x_0, ..., x_i - 1, ..., x_N) \ .
\end{equation}
The simplest way to define these polynomials is via the Rodrigues formula
\begin{equation} \label{eq:rodrigues_multi}
V_{\mathbf{n}}(\mathbf{x},\boldsymbol{\mu}) := \frac{1}{\text{Poiss}(\mathbf{x}, \boldsymbol{\mu})} \left[ \prod_{k = 0}^N  \left( \hat{L}_k^+ \right)^{n_k} \right] \text{Poiss}(\mathbf{x}, \boldsymbol{\mu}) \ .
\end{equation}
This definition exactly corresponds to the fact that the eigenstates of the multistep splicing problem can be constructed by repeatedly applying the up ladder operators:
\begin{equation}
\ket{\mathbf{n}} = \frac{( \hat{A}_0^+ )^{n_0}}{\sqrt{n_0!}} \cdots \frac{( \hat{A}_N^+ )^{n_N}}{\sqrt{n_N!}}  \ket{\mathbf{0}}
\end{equation}
where we have used a mostly arbitrary normalization convention. In the $N = 0$ case, these polynomials are just the Charlier polynomials. The first few polynomials in the $N = 1$ case are given by
\begin{equation}
\begin{split}
V_{00} =& \ 1 \\
V_{0,1} =& \ \frac{x_1}{\mu_1} - 1 \\
V_{1,0} =& \ \frac{x_0}{\mu_0} + \frac{\beta_0}{\beta_1 - \beta_0} \frac{x_1}{\mu_1} - \frac{\beta_1}{\beta_1 - \beta_0} \\
V_{0,2} =& \ \frac{x_1^2}{\mu_1^2} - \left[ \frac{1}{\mu_1^2} + \frac{2}{\mu_1} \right] x_1 + 1 \\
V_{1,1} =& \ \frac{x_0 x_1}{\mu_0 \mu_1} + \frac{\beta_0}{\beta_1 - \beta_0} \frac{x_1 (x_1 - 1)}{\mu_1^2} - \frac{x_0}{\mu_0} - \frac{(\beta_1 + \beta_0)}{\beta_1 - \beta_0} \frac{x_1}{\mu_1} + \frac{\beta_1}{\beta_1 - \beta_0} \\
V_{2,0} =& \ \frac{x_0 (x_0 - 1)}{\mu_0^2} + \left( \frac{\beta_0}{\beta_1 - \beta_0} \right)^2 \frac{x_1 (x_1 - 1)}{\mu_1^2} + \frac{2\beta_0}{\beta_1 - \beta_0} \frac{x_0}{\mu_0} \frac{x_1}{\mu_1} \\
 & - \frac{2 \beta_1}{\beta_1 - \beta_0} \frac{x_0}{\mu_0} -  \frac{2\beta_0 \beta_1}{(\beta_1 - \beta_0)^2} \frac{x_1}{\mu_1} + \left( \frac{\beta_1}{\beta_1 - \beta_0} \right)^2 \ .
\end{split}
\end{equation}
In general, $V_{\mathbf{n}}$ will be a polynomial of order $n_i$ with respect to the variable $x_i$, and of order $|\mathbf{n}| = n_0 + \cdots + n_N$ overall. The leading term asymptotically goes like 
\begin{equation}
\left( \frac{x_0}{\mu_0} \right)^{n_0} \cdots \left( \frac{x_N}{\mu_N} \right)^{n_N} 
\end{equation}
but other terms of the same order can appear (e.g. a term that goes like $x_1^2$ appears in $V_{11}$). As with the Charlier polynomials, two helpful results are the generating function and a particular integral representation. Noting the representation of a multivariate Poisson distribution via
\begin{equation}
\begin{split}
\text{Poiss}(\mathbf{x},\boldsymbol{\mu}) &= 
\int \frac{d\mathbf{z}}{(2\pi)^{N+1}} \frac{e^{- i \boldsymbol{\mu} \cdot \mathbf{z}}}{(\mathbf{1} - i \mathbf{z})^{\mathbf{x} + \mathbf{1}}} \\
&= \int \frac{dz_0 \cdots dz_N}{(2\pi)^{N+1}} \frac{e^{- i \mu_0 z_0 - \cdots - i \mu_N z_N}}{(1 - i z_0)^{x_0 + 1} \cdots (1 - i z_N)^{x_N + 1}} \ ,
\end{split}
\end{equation}
we can apply the Rodrigues formula to find the integral representation
\begin{equation}
\begin{split}
V_{\mathbf{n}}(\mathbf{x}, \boldsymbol{\mu}) &= \frac{1}{\text{Poiss}(\mathbf{x},\boldsymbol{\mu})}  \int \frac{d\mathbf{z}}{(2\pi)^{N+1}} \frac{\prod_{k = 0}^N \left[ - i \sum_{j = 0}^N v^{(k)}_j z_j  \right]^{n_k} e^{- i \boldsymbol{\mu} \cdot \mathbf{z}}}{(\mathbf{1} - i \mathbf{z})^{\mathbf{x} + \mathbf{1}}} \\
&= \frac{1}{\text{Poiss}(\mathbf{x},\boldsymbol{\mu})}  \int \frac{dz_0 \cdots dz_N}{(2\pi)^{N+1}} \frac{\prod_{k = 0}^N \left[ - i \sum_{j = 0}^N v^{(k)}_j z_j  \right]^{n_k} e^{- i \mu_0 z_0 - \cdots - i \mu_N z_N}}{(1 - i z_0)^{x_0 + 1} \cdots (1 - i z_N)^{x_N + 1}} \ .
\end{split}
\end{equation}
This can be used to derive the generating function
\begin{equation} \label{eq:multi_genfunc}
\begin{split}
G(\mathbf{z}) &=  \sum_{\mathbf{n}} \frac{V_{\mathbf{n}}(\mathbf{x}, \boldsymbol{\mu})}{\mathbf{n}!} \mathbf{z}^{\mathbf{n}} \\
&=  \prod_{j = 0}^N \left[ 1 + \frac{1}{\mu_j} \sum_k v^{(k)}_j z_k  \right]^{x_j} e^{- \sum_k v^{(k)}_j z_k } \\
&=  \prod_{j = 0}^N \left[ 1 + \frac{1}{\mu_j} \sum_k v^{(k)}_j z_k  \right]^{x_j} e^{- \sum_k v^{(k)}_j z_k } \ .
\end{split}
\end{equation}
This, in turn, can be used to derive the helpful identity
\begin{equation} \label{eq:multi_V_sum_x}
\begin{split}
\sum_{\mathbf{x}} V_{\mathbf{n}}(\mathbf{x}, \boldsymbol{\mu}) \ \mathbf{g}^{\mathbf{x}} \  \text{Poiss}(\mathbf{x},\boldsymbol{\mu}) &= \prod_{k = 0}^N \left[ \sum_{j = 0}^N (g_j - 1) v^{(k)}_j  \right]^{n_k} e^{\mu_k (g_k - 1)} \\
&= e^{\boldsymbol{\mu} \cdot (\mathbf{g} - \mathbf{1})}  \prod_{k = 0}^N \left[ (\mathbf{g} - \mathbf{1}) \cdot \mathbf{v}^{(k)}  \right]^{n_k} 
\end{split}
\end{equation}
which can be used both to find the (analytic) generating function in the multistep problem (c.f. Eq. \ref{eq:multi_analytic_genfunc}) and to marginalize over variables by setting different $g_i = 1$. To derive marginal distributions in Sec. \ref{sec:multistep_special}, we use this fact to obtain the formulas
\begin{equation}
\sum_{x_1, ..., x_N} V_{\mathbf{n}}(\mathbf{x}, \boldsymbol{\mu}) \ \text{Poiss}(\mathbf{x},\boldsymbol{\mu}) = (\delta_{n_1, 0} \cdots \delta_{n_N, 0}) C_{n_0}(x_0, \mu_0) \ \text{Poiss}(x_0, \mu_0)
\end{equation}
and
\begin{equation}
\sum_{x_0, ..., x_{N-1}} V_{\mathbf{n}}(\mathbf{x}, \boldsymbol{\mu}) \ \text{Poiss}(\mathbf{x},\boldsymbol{\mu}) = \prod_{k = 0}^N \left[ v^{(k)}_N  \right]^{n_k} \  C_{n_N}(x_N, \mu_N) \ \text{Poiss}(x_N, \mu_N) \ .
\end{equation}
The formulas we have derived so far can be used to determine various special values of these polynomials. For example, using the generating function (Eq. \ref{eq:multi_genfunc}) we can find that
\begin{equation}
V_{\mathbf{n}}(\mathbf{0}, \boldsymbol{\mu}) = \left( - \sum_{j = 0}^N v^{(0)}_j \right)^{n_0} \cdots \left( - \sum_{j = 0}^N v^{(N)}_j \right)^{n_N} \ .
\end{equation}
Using Eq. \ref{eq:multi_V_sum_x}, we can find that
\begin{equation}
V_{0, 0, ..., 0, n_N}(\mathbf{x}, \boldsymbol{\mu}) = C_{n_N}(x_N, \mu_N) \ .
\end{equation} 
What about orthogonality? Are $V_{\mathbf{n}}$ and $V_{\mathbf{n}'}$ orthogonal for $\mathbf{n} \neq \mathbf{n}'$? Perhaps surprisingly the answer is no in general; for example, in the $N = 1$ case, $V_{10}$ and $V_{01}$ are not orthogonal with respect to a Poisson weight function:
\begin{equation}
\sum_{x_0, x_1} V_{10}(\mathbf{x}, \boldsymbol{\mu}) V_{01}(\mathbf{x}, \boldsymbol{\mu}) \text{Poiss}(\mathbf{x}, \boldsymbol{\mu}) = \frac{\beta_0}{\beta_1 - \beta_0} \frac{1}{\mu_1} \neq 0 \ .
\end{equation}
It \textit{is} true, however, that $V_{\mathbf{n}}$ and $V_{\mathbf{n}'}$ are orthogonal when $|\mathbf{n}| \neq |\mathbf{n}'|$ (which is not true in the above case, where we have $0 + 1 = 1 + 0$). This can be argued using the generating function (Eq. \ref{eq:multi_genfunc}). 


\clearpage

\section{Consistency of current results with previous results}
\label{sec:app_consistency}

Previous work on the birth-death-switching problem (Iyer-Biswas et al. \cite{iyer2009}, Huang et al. \cite{huang2015}, and Cao and Grima \cite{cao2018}) found an exact analytic solution for the steady state generating function $\psi_{ss}(g)$ given by
\begin{equation} \label{eq:huang_sln}
\psi_{ss}(g) = e^{\frac{\alpha_2}{\gamma} (g - 1)} \ _1F_1\left(\frac{k_{12}}{\gamma}, \frac{s}{\gamma}; \frac{\Delta \alpha}{\gamma} (g-1) \right) = e^{\frac{\alpha_2}{\gamma} (g - 1)} \sum_{n = 0}^{\infty} \frac{\left[ \frac{\Delta \alpha}{\gamma} (g-1) \right]^n}{n!} \frac{\left( \frac{k_{12}}{\gamma} \right)_n}{\left( \frac{s}{\gamma} \right)_n}
\end{equation}
where $_1 F_1$ denotes the confluent hypergeometric function of the first kind, and $(a)_n := a (a + 1) \cdots (a + n - 1)$ denotes the Pochhammer symbol/rising factorial. Naively, this result looks quite different from ours, and does not involve Charlier polynomials or Feynman diagrams at all. We have checked the numerical consistency of our work with previous work elsewhere in this paper; in this appendix, we show that our result is mathematically identical to this one. 

First, we recall that our result for the probability generating function is
\begin{equation}
\begin{split}
\psi_{ss}(g) &= e^{\mu (g - 1)} \left\{ 1 + \sum_{k = 1}^{\infty} \left[ \Delta \alpha (g-1) \right]^k  \ \sum_{i_1, ..., i_{k-1} = 0, 1} \ \frac{T_{0 i_{k-1}}}{(k \gamma)} \cdots \frac{T_{i_1 0}}{(\gamma + i_1 s)}    \right\} \ .
\end{split}
\end{equation}
The Feynman diagram part of this formula can be rewritten in terms of the transfer matrix $T$ and another matrix, which we arbitrarily denote using $R$:
\begin{equation}
R := \begin{pmatrix} 1 & 0 \\ 0 & 0  \end{pmatrix} \ .
\end{equation}
Specifically, it is the $00$ (upper left) entry of a certain product of $2 \times 2$ matrices, i.e.
\begin{equation}
\sum_{i_1, ..., i_{k-1} = 0, 1} \ \frac{T_{0 i_{k-1}}}{(k \gamma)} \cdots \frac{T_{i_1 0}}{(\gamma + i_1 s)}   \stackrel{00}{=}   \frac{\left[ k I + r R \right] T \cdots \left[ I + r R \right] T}{\gamma^k \ k! \ (1 + r)_k} 
\end{equation}
where $r := s/\gamma$ and $I$ is the $2 \times 2$ identity matrix. Also note that
\begin{equation}
\begin{split}
\mu &= \frac{\alpha_1 k_{12} + \alpha_2 k_{21}}{\gamma s} = \frac{\alpha_1 k_{12} - \alpha_2 k_{12} + \alpha_2 k_{12} + \alpha_2 k_{21}}{\gamma s} = \frac{\Delta \alpha}{\gamma} x + \frac{\alpha_2}{\gamma} 
\end{split}
\end{equation}
where in this context we define $x := k_{12}/s$. Now we can write
\begin{equation}
\begin{split}
\frac{\psi_{ss}(g)}{e^{\frac{\alpha_2}{\gamma} (g - 1)}}   &\stackrel{00}{=}  e^{\frac{\Delta \alpha}{\gamma} x (g-1)} \sum_{k = 0}^{\infty} \frac{\left[ \frac{\Delta \alpha (g-1)}{\gamma} \right]^k}{k!}  \ \frac{\left[ k I + r R \right] T \cdots \left[ I + r R \right] T}{(1 + r)_k} \ .
\end{split}
\end{equation}
Let us rewrite the right-hand side by multiplying out the two infinite series involved term by term (in powers of $(g - 1)$). The $k$th term $d_k$ of the (Cauchy) product reads
\begin{equation}
\begin{split}
d_k &= \sum_{\ell = 0}^k \frac{\left[ \frac{\Delta \alpha}{\gamma} x (g-1) \right]^{k - \ell}}{(k - \ell)!} \frac{\left[ \frac{\Delta \alpha (g-1)}{\gamma} \right]^{\ell}}{\ell!}  \ \frac{\left[ \ell +  r R \right] T \cdots \left[ 1 + r R \right] T}{(\ell + r) \cdots (1 + r)}  \\
&= \left[ \frac{\Delta \alpha}{\gamma} (g-1) \right]^k \sum_{\ell = 0}^k \frac{x^{k - \ell}}{(k - \ell)!} \frac{1}{\ell!}  \ \frac{\left[ \ell +  r R \right] T \cdots \left[ 1 + r R \right] T}{(\ell + r) \cdots (1 + r)}  \\
&= \frac{\left[ \frac{\Delta \alpha}{\gamma} (g-1) \right]^k}{k!} \sum_{\ell = 0}^k \binom{k}{\ell} \ x^{k - \ell}  \ \frac{\left[ \ell +  r R \right] T \cdots \left[ 1 + r R \right] T}{(\ell + r) \cdots (1 + r)}  \ .
\end{split}
\end{equation}
We can simplify this using the surprising identity
\begin{equation}
\frac{r x (r x + 1) \cdots (rx + k - 1)}{r (r + 1) \cdots (r + k - 1)} \stackrel{00}{=} \sum_{\ell = 0}^k \binom{k}{\ell} \ x^{k - \ell}  \ \frac{\left[ \ell +  r R \right] T \cdots \left[ 1 + r R \right] T}{(\ell + r) \cdots (1 + r)} \ .
\end{equation}
Doing so, we have
\begin{equation}
\begin{split}
\psi_{ss}(g) &= e^{\frac{\alpha_2}{\gamma} (g - 1)} \sum_{k = 0}^{\infty} \frac{\left[ \frac{\Delta \alpha}{\gamma} (g-1) \right]^k}{k!} \frac{r x (r x + 1) \cdots (rx + k - 1)}{r (r + 1) \cdots (r + k - 1)} \\
&= e^{\frac{\alpha_2}{\gamma} (g - 1)} \ _1F_1\left(r x, r; \frac{\Delta \alpha}{\gamma} (g-1) \right)
\end{split}
\end{equation}
which matches the previously published result. The remainder of this appendix is dedicated to proving the unusual identity that allows us to make this simplification.

\noindent \textbf{Lemma.} Let $f(\ell)$ and $g(\ell)$ denote the $00$ and $10$ entries of the $\ell$-fold matrix product
\begin{equation}
\frac{\left[ \ell +  r R \right] T \cdots \left[ 1 + r R \right] T}{(\ell + r) \cdots (1 + r)}  \ ,
\end{equation}
where
\begin{equation}
\begin{split}
T &:= \begin{pmatrix} 0 & 1 \\ x (1 - x) & 1 - 2 x  \end{pmatrix} \\
R &:= \begin{pmatrix} 1 & 0 \\ 0 & 0  \end{pmatrix}  \ .
\end{split}
\end{equation}
Define
\begin{equation}
\begin{split}
S_{00}(k) &:= \sum_{\ell = 0}^k \binom{k}{\ell} \ x^{k-\ell} f(\ell) \\
S_{10}(k) &:= \sum_{\ell = 0}^k \binom{k}{\ell} \ x^{k-\ell} g(\ell) \ .
\end{split}
\end{equation}
We have that
\begin{equation}
\begin{split}
S_{00}(k) &:= \frac{r x (r x + 1) \cdots (rx + k - 1)}{r (r + 1) \cdots (r + k - 1)} \\
S_{10}(k) &:= \frac{k (1 - x)}{r + k} S_{00}(k) 
\end{split}
\end{equation}
for all integers $k \geq 0$ (using the convention that the empty product is $1$, so that $S_{00}(0) = 1$). 
\begin{proof}
First, by definition, we note that $f(\ell)$ and $g(\ell)$ satisfy the recurrence relations
\begin{equation}
\begin{split}
f(\ell) &= g(\ell - 1) \\
g(\ell) &= \frac{\ell}{\ell + r} \left[ x (1 - x) f(\ell - 1) + (1 - 2 x) g(\ell - 1) \right]
\end{split}
\end{equation}
for all $\ell \geq 1$. Next, note that
\begin{equation}
\begin{split}
f(0) &= 1 \\
f(1) &= 0 \\
f(2) &= \frac{x (1 - x)}{r + 1} \\
g(0) &= 0 \\
g(1) &= \frac{x (1 - x)}{r + 1} \\
g(2) &= \frac{2 x (1 - x) (1 - 2 x)}{(r + 1) (r + 2)} \ .
\end{split}
\end{equation}
We will proceed by induction. The $k = 0$ case is trivially true. The $k = 1$ case is easy to show:
\begin{equation}
\begin{split}
S_{00}(1) &= x f(0) + f(1) = x \\
S_{10}(1) &= x g(0) + g(1) = \frac{x (1 - x)}{r + 1} \ .
\end{split}
\end{equation}
Assume our formulas for $S_{00}$ and $S_{10}$ hold up to some $k$. Now we have
\begin{equation}
\begin{split}
S_{00}(k + 1) &= \sum_{\ell = 0}^{k+1} \binom{k+1}{\ell} \ x^{k + 1 - \ell} \ f(\ell) \\
&= x^{k+1} + f(k+1) + \sum_{\ell = 1}^{k} \left[ \binom{k}{\ell} + \binom{k}{\ell - 1} \right] \ x^{k + 1 - \ell} \ f(\ell) 
\end{split}
\end{equation}
where we have used Pascal's identity. Rearranging, this becomes
\begin{equation}
\begin{split}
S_{00}(k + 1) &= x \sum_{\ell = 0}^{k} \binom{k}{\ell}  \ x^{k - \ell} \ f(\ell) + \sum_{\ell = 1}^{k+1}  \binom{k}{\ell - 1}  \ x^{k + 1 - \ell} \ f(\ell) \\
&= x S_{00}(k) + \sum_{\ell = 0}^{k}  \binom{k}{\ell}  \ x^{k - \ell} \ f(\ell + 1) \\
&= x S_{00}(k) + \sum_{\ell = 0}^{k}  \binom{k}{\ell}  \ x^{k - \ell} \ g(\ell) \\
&= x S_{00}(k) + S_{10}(k) \\
&= \left[ x + \frac{k (1 - x)}{r + k}  \right] S_{00}(k)  \\
&= \left[ \frac{rx + k}{r + k}  \right] S_{00}(k)  \\
\end{split}
\end{equation}
which is exactly what we wanted. Now we will perform the induction step for $S_{10}$; it is somewhat more involved. It begins as we argued for $S_{00}$:
\begin{equation}
\begin{split}
S_{10}(k + 1) &= \sum_{\ell = 0}^{k+1} \binom{k+1}{\ell} \ x^{k + 1 - \ell} \ g(\ell) \\
&= x^{k+1} + g(k+1) + \sum_{\ell = 1}^{k} \left[ \binom{k}{\ell} + \binom{k}{\ell - 1} \right] \ x^{k + 1 - \ell} \ g(\ell)  \\
&= x \sum_{\ell = 0}^{k} \binom{k}{\ell}  \ x^{k - \ell} \ g(\ell) + \sum_{\ell = 1}^{k+1}  \binom{k}{\ell - 1}  \ x^{k + 1 - \ell} \ g(\ell) \\
&= x S_{10}(k) + \sum_{\ell = 0}^{k}  \binom{k}{\ell}  \ x^{k - \ell} \ g(\ell + 1)
\end{split}
\end{equation}
where we have again used Pascal's identity. The sum on the right is somewhat tricky to evaluate; denote it by $S'$. Using the recurrence relation relating $g(\ell + 1)$ to $f(\ell)$ and $g(\ell)$,
\begin{equation}
\begin{split}
S' =& \ \sum_{\ell = 0}^{k}  \binom{k}{\ell}  \ x^{k - \ell} \ \frac{\ell + 1}{\ell + 1 + r} \left[ x (1 - x) f(\ell) + (1 - 2 x) g(\ell) \right] \\
=& \ \sum_{\ell = 0}^{k}  \binom{k}{\ell}  \ x^{k - \ell} \ \left[ 1 - \frac{r}{\ell + 1 + r} \right] \left[ x (1 - x) f(\ell) + (1 - 2 x) g(\ell) \right] \\
=& \ x (1 - x) S_{00}(k) + (1 - 2 x) S_{10}(k) \\
& \ - \frac{r}{k + 1 + r} \sum_{\ell = 0}^{k}  \binom{k}{\ell}  \ x^{k - \ell} \ \frac{k + 1 + r}{\ell + 1 + r} \ \left[ x (1 - x) f(\ell) + (1 - 2 x) g(\ell) \right] \\
=& \ x (1 - x) S_{00}(k) + (1 - 2 x) S_{10}(k) \\
& \ - \frac{r}{k + 1 + r} \sum_{\ell = 0}^{k}  \binom{k}{\ell}  \ x^{k - \ell} \ \left[ \frac{k + 1 + r}{\ell + 1 + r} - 1 + 1 \right] \ \left[ x (1 - x) f(\ell) + (1 - 2 x) g(\ell) \right] \\
=& \ x (1 - x) S_{00}(k) + (1 - 2 x) S_{10}(k) \\
& \ - \frac{r}{k + 1 + r} \sum_{\ell = 0}^{k}  \binom{k}{\ell}  \ x^{k - \ell} \ \left[ \frac{k - \ell}{\ell + 1} \frac{\ell + 1}{\ell + 1 + r} + 1 \right] \ \left[ x (1 - x) f(\ell) + (1 - 2 x) g(\ell) \right] \\
=& \ \left[ x (1 - x) S_{00}(k) + (1 - 2 x) S_{10}(k) \right] \left[ 1 - \frac{r}{k + 1 + r} \right] \\
& \ - \frac{r}{k + 1 + r} \sum_{\ell = 0}^{k}  \binom{k}{\ell}  \ x^{k - \ell} \ \frac{k - \ell}{\ell + 1} \frac{\ell + 1}{\ell + 1 + r}  \ \left[ x (1 - x) f(\ell) + (1 - 2 x) g(\ell) \right] \\
=& \ \left[ x (1 - x) S_{00}(k) + (1 - 2 x) S_{10}(k) \right] \left[ 1 - \frac{r}{k + 1 + r} \right]  \\
& \ - \frac{r}{k + 1 + r} \sum_{\ell = -1}^{k-1}  \frac{k!}{(k - \ell)! \ \ell!}  \ x^{k - \ell} \ \frac{k - \ell}{\ell + 1} g(\ell + 1) \\
=& \ \left[ x (1 - x) S_{00}(k) + (1 - 2 x) S_{10}(k) \right] \left[ 1 - \frac{r}{k + 1 + r} \right]  \\
& \ - \frac{r}{k + 1 + r} \sum_{\ell = -1}^{k-1}  \frac{k!}{(k - \ell - 1)! \ (\ell + 1)!}  \ x^{k - \ell} \  g(\ell + 1) \\
=& \ \left[ x (1 - x) S_{00}(k) + (1 - 2 x) S_{10}(k) \right] \left[ 1 - \frac{r}{k + 1 + r} \right] - \frac{r x}{k + 1 + r} S_{10}(k) \\
=& \ x (1 - x) S_{00}(k) + (1 - 2 x) S_{10}(k)  - \frac{r}{k + 1 + r} (1 - x) \left[ x S_{00}(k) + S_{10}(k) \right] \ .
\end{split}
\end{equation}
Adding all the terms together yields that
\begin{equation}
\begin{split}
S_{10}(k+1) &= x S_{10}(k) + x (1 - x) S_{00}(k) + (1 - 2 x) S_{10}(k)  - \frac{r}{r + k + 1} (1 - x) \left[ x S_{00}(k) + S_{10}(k) \right] \\
&= (1- x) \left[ x S_{00}(k) + S_{10}(k) \right]  - \frac{r}{r + k + 1} (1 - x) \left[ x S_{00}(k) + S_{10}(k) \right] \\
&= \frac{k + 1}{r + k + 1} (1- x)  \left[ x S_{00}(k) + S_{10}(k) \right] \\
&= \frac{k + 1}{r + k + 1} (1- x)  \left[ x + \frac{k (1 - x)}{r + k} \right] S_{00}(k) \\
&= \frac{k + 1}{r + k + 1} (1- x)  \frac{r x + k}{r + k} S_{00}(k) 
\end{split}
\end{equation}
which suffices to establish our result. 
\end{proof}

\clearpage

\section{Numerical validation details}
\label{sec:app_validation}

The validation of a method for the solution of a given system requires its characterization under three criteria. Firstly, the runtime must be benchmarked against a known solution; we implement the solution due to Huang et al. \cite{huang2015} (specifically, the hypergeometric generating function solution given by Eq. \ref{eq:huang_sln}). Secondly, the bounds on the error between the estimate and the ground truth distribution must be estimated, particularly as a function of the parameter domain. Finally, the switching gene problem necessitates the criterion of nontriviality. Specifically, in parameter regimes with very slow gene state transition rates, the solution is given by a mixture of two Poisson distributions, given by Eq. \ref{eq:my_pmix}. Therefore, it is necessary to confirm that the method outperforms this trivial and computationally facile solution.

We explore the stationary solution of the chemical master equation. Since the time variable is immaterial, the five-parameter system reduces to a four-parameter one (i.e. we have $\alpha_1, \alpha_2, \gamma, k_{12}, k_{21}$). We set the degradation rate $\gamma$ to 1 with no loss of generality. Physically, the production timescale is assumed to be rather shorter than the degradation timescale, whereas the transition timescale is assumed to be longer. Therefore, we draw the $\alpha$ parameters from a log-uniform distribution on $[10^{-1},10^2]$ and the $k$ parameters from a log-uniform distribution on $[10^{-3},10^0]$. 

Aside from the parameter values, the evaluation of the distributions requires a domain, i.e., a distribution support $\{0,1,2,...,m\}$. One approach simulates the system for $N$ cells, considers the observations $Y_1,...,Y_N$ collected after equilibration, and uses $m\gets \max_j Y_j$. This method is rather natural, but problematic in practice, because simulation is substantially more computationally expensive than the computation of distributions by any method. 

\begin{figure}[h]
\centering
\includegraphics[width=0.55\columnwidth]{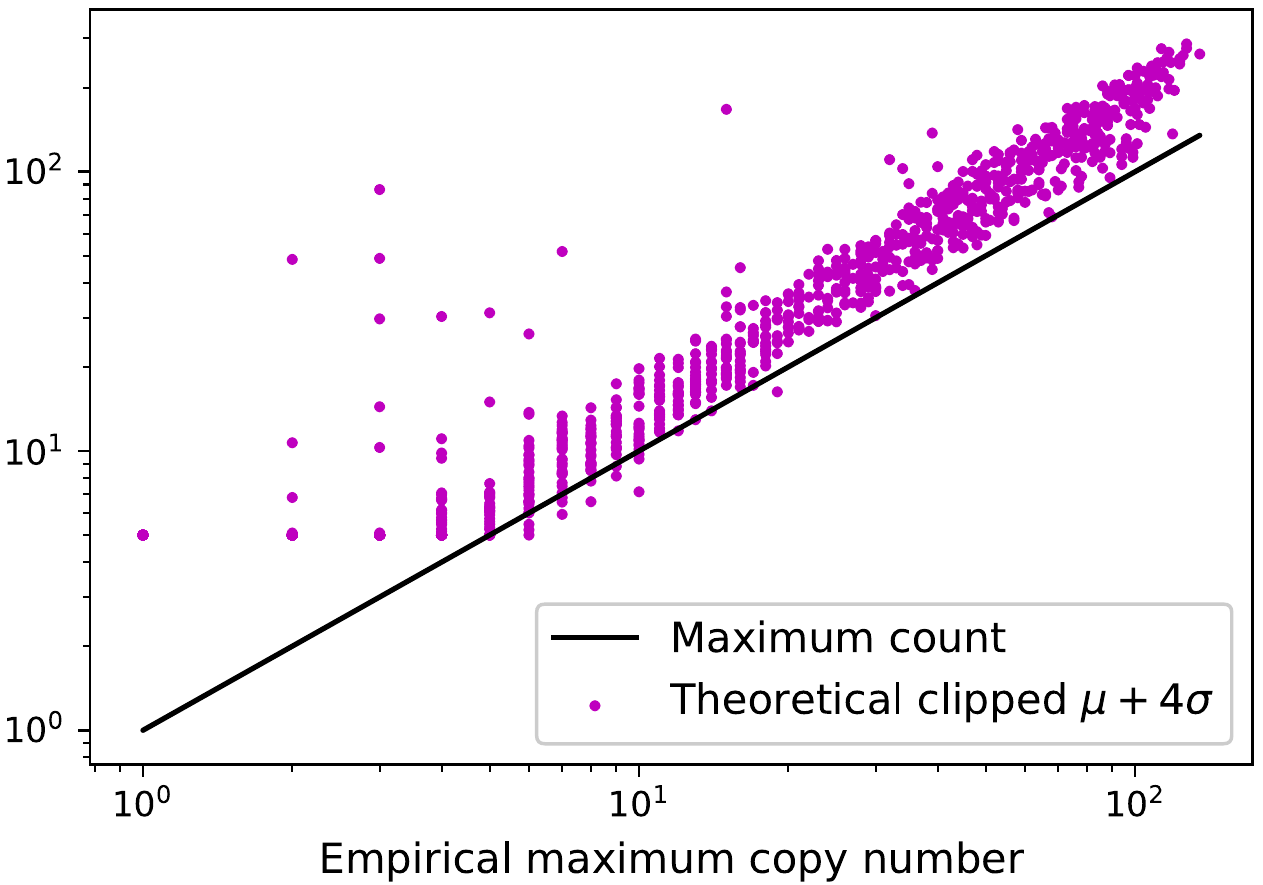}
\caption{Demonstration of the performance of the moment-based threshold against the simulation maximum-based threshold for state space definition.}
\label{fig:calibration}
\end{figure}

Instead, we adopt a heuristic method, inspired by previous literature in the field \cite{gupta2017}. Specifically, given a distribution with mean $\mu$ and standard deviation $\sigma$, we implement $m\gets \max(5,\mu+4\sigma)$. Since $\mu$ and $\sigma$ have simple analytical expressions for the two-state switch (see Sec. \ref{sec:results_bds}), this procedure is substantially less computationally intensive. To benchmark this choice of meta-parameters, we selected 1000 parameters and simulated $N=100$ cells for each. Only 1.7\% of $m$ produced by the moment-based estimate were lower than those produced by the simulation-based estimate, mostly restricted to low $m$ (Fig. \ref{fig:calibration}). Therefore, we adopt $\max(5,\mu+4\sigma)$ as a conservative estimate for the state space upper bound.

Thus, given a parameter vector and a state space $\{0,...,m\}$, we compute the solution due to Huang et al. as the non-trivial ground truth, as well as the Poisson solution as the trivial estimate, and compare them to the diagrammatic solution for various orders of approximation. The ground truth solution requires $m+1$ evaluations of the hypergeometric function; therefore, we expect the runtime to be a strong function of the state space size. Furthermore, the approximation runtime is a function of the approximation order. We compute and report both dependence trends.

The divergence from the ground truth affords a choice of possible measures. We use the Kullback-Leibler divergence and the Kolmogorov-Smirnov distance, with probabilities fixed to a minimum value of $10^{-16}$. However, raw divergences are somewhat uninformative for the purpose of determining robust domains. Given the trend of roughly monotonically increasing divergence as a function of $|\alpha_1-\alpha_2|$ (as seen in Fig. \ref{fig:1sp_valid}), a natural choice reports the maximum of all divergences up to a particular value of $|\alpha_1-\alpha_2|$. Intuitively, such a function of $|\Delta \alpha|$ reports an estimate for an upper bound on error. For a given approximation order and choice of acceptable error bound, the function immediately provides the upper bound of the $|\Delta \alpha|$ domain that meets this error threshold.

The maximum is not a particularly robust statistic, and is unstable to outliers. Instead, we report the 98th percentile of all divergences up to a particular value of $|\Delta \alpha|$. Therefore, such a trace is an estimate of the $98\%$ confidence interval for the maximum error in the domain. As the sampling density increases, this interval converges to the true one with $O(M^{-1/2})$ in the number of parameters sampled. We use 5000 parameter samples to estimate the intervals.

This procedure lends itself especially easily to quantitative comparison with the trivial Poisson solution. Specifically, if the divergence between ground truth and the diagrammatic estimate is lower than the divergence between ground truth and the Poisson mixture estimate, the diagrammatic estimate quantitatively outperforms the accuracy of the Poisson solution.

\end{document}